\documentclass{emulateapj}

\pdfoutput=1 
\newcommand{\figurepath}{./}
\newbox\grsign \setbox\grsign=\hbox{$>$}
\newdimen\grdimen \grdimen=\ht\grsign
\newbox\laxbox \newbox\gaxbox
\setbox\gaxbox=\hbox{\raise.5ex\hbox{$>$}\llap
     {\lower.5ex\hbox{$\sim$}}}\ht1=\grdimen\dp1=0pt
\setbox\laxbox=\hbox{\raise.5ex\hbox{$<$}\llap
     {\lower.5ex\hbox{$\sim$}}}\ht2=\grdimen\dp2=0pt

\shorttitle{Bipolar jet launching I}
\shortauthors{Sheikhnezami et al.}

\usepackage{amsmath}
\usepackage{float}
\usepackage{ctable} 
\usepackage{color}
\usepackage{natbib}
\usepackage{threeparttable}

\begin{document}

\title{Bipolar jets launched from magnetically diffusive accretion disks.\\
       I. Ejection efficiency vs field strength and d\pdfoutput=1 iffusivity}
\author{Somayeh Sheikhnezami\altaffilmark{1,2}, 
        Christian Fendt\altaffilmark{1}, 
        Oliver Porth\altaffilmark{1,3}, 
        Bhargav Vaidya\altaffilmark{1,4},
        Jamshid Ghanbari\altaffilmark{2} }
\altaffiltext{1}{Max Planck Institute for Astronomy, K\"onigstuhl 17, D-69117 Heidelberg, Germany}
\altaffiltext{2}{Department of Physics, Faculty of Sciences, Ferdowsi University of Mashhad, Iran}
\altaffiltext{3}{School of Mathematics,           University of Leeds, Leeds, LS2 9JT, UK}
\altaffiltext{4}{School of Physics and Astronomy, University of Leeds, Leeds, LS6 2EN, UK}
\email{nezami@mpia.de, fendt@mpia.de}

%///////////////////////////////////////////////////////////////////////

%      \date{\today}

%/////////////////////////////////////////////////////////////////////////
\begin{abstract}
 We investigate the launching of jets and outflows from magnetically diffusive accretion disks.
Using the PLUTO code we solve the time-dependent resistive MHD equations
taking into account the disk and jet evolution simultaneously.
The main question we address is {\em which kind of disks do launch jets and which kind of disks do not?}
In particular, we study how the magnitude and distribution of the (turbulent) magnetic diffusivity 
affect mass loading and jet acceleration. 
We have applied a turbulent magnetic diffusivity based on $\alpha$-prescription,
but have also investigate examples where the scale height of diffusivity is larger than that 
of the disk gas pressure.
We further investigate how the ejection efficiency is governed by the magnetic field strength.
Our simulations last for up to 5000 dynamical time scales corresponding to 900 orbital periods of
the inner disk.
As a general result we observe a continuous and robust outflow launched from the inner part of the disk,
expanding into a collimated jet of super fast magneto-sonic speed. 
For long time scales the disk internal dynamics changes, as due to outflow ejection
and disk accretion the disk mass decreases. 
For magneto-centrifugally driven jets we find that for 
i) less diffusive disks, 
ii) a stronger magnetic field, 
iii) a low poloidal diffusivity, or a 
iv) lower numerical diffusivity (resolution), 
 the  mass loading of the outflow is increased - resulting in 
more powerful jets with high mass flux.
For weak magnetization the (weak) outflow is driven by the magnetic pressure gradient.
We consider in detail the advection and diffusion of magnetic flux within the disk and we 
find that the disk and outflow magnetization may substantially change in time. % (by up to a factor 100).
This may have severe impact on the launching and formation process - an initially highly 
magnetized disk may evolve into a disk of weak magnetization which cannot drive strong outflows.
 We further investigate the jet asymptotic velocity and the jet rotational velocity
in respect of the different launching scenarios. 
We find a lower degree of jet collimation than previous studies,
most probably due to our revised outflow boundary condition.
\end{abstract}
%//////////////////////////////////////////////////////////////////////////////////////
\keywords{
   accretion, accretion disks --
   MHD -- 
   ISM: jets and outflows --
   stars: pre-main sequence --
   galaxies: jets --
   galaxies: active 
 }
%/////////////////////////////////////////////////////////////////////////////////////
\section{Introduction}
Jets as highly collimated beams of high velocity material and outflows of comparatively
lower degree of collimation and lower speed are an ubiquitous phenomenon among 
astrophysical objects.
Jets are powerful signs of activity and are observed over a wide range of luminosity and spatial scale.
Among the jet sources are young stellar objects (YSO), micro-quasars, active galactic nuclei (AGN), 
and most probably also gamma ray bursts
\citep{Fanaroff1974, Abell1979, Mundt1983, Rhoads1997,
Mirabel1994Na}.
The common models of launching, acceleration, and collimation work in the framework of 
magnetohydrodynamic (MHD) forces (see e.g.~\citealt{BP1982, Pudritz1983, Uchida1985}), 
although the details of the process are not fully understood.

Jets and outflows from YSO and AGN affect their environment, and, thus, the formation process
of the objects they are launching them.
Numerous studies investigate effects of such feedback mechanisms in star formation and galaxy formation
(see e.g. \citealt{Banerjee2007,Carroll2009, Gaibler2011}). 
However, a quantitative investigation of how much of mass, momentum, or energy from the infall is actually
recycled into a high speed outflow needs to resolve the innermost jet-launching region and to model the
physical process of launching directly. This is the major aim of the present paper.

According to the current understanding, accretion and ejection are related to each other.
One efficient way to remove angular momentum from a disk is to connect it to a magnetized outflow.
This has been motivated by the observed correlation between signatures of accretion and ejection 
in jet sources (see e.g. \citealt{Cabrit1990, Hartigan1995}).

The overall idea is that the energy and angular momentum are extracted from the disk by an efficient 
magnetic torque relying on a global, i.e. large-scale magnetic field threading the disk.
If the inclination of the field lines is sufficiently small, magneto-centrifugal forces can 
accelerate the matter along the field line. Beyond the Alfv\'en point also Lorentz forces 
contribute to the acceleration.
The collimation of the outflow is thought to be achieved by magnetic tension due to a toroidal 
component of the magnetic field. 
Still, we have to keep in mind the fact the the toroidal field pressure gradient acts de-collimating,
and an existing external gas over-pressure may contribute to jet collimation.

Before going into further details, we like to make clear that with jet {\em formation} we denote the process of 
accelerating and collimating an already existing slow disk wind or stellar wind in to a jet beam.
With jet {\em launching} we denote the process which conveys material from radial accretion into a vertical 
ejection, thereby lifting it from the disk plane into the corona, and thus establishing a disk wind.

A vast literature exists on magnetohydrodynamics jet modeling.
We may distinguish i) between steady-state models and time-dependent numerical simulations, or
ii) between simulations considering the jet formation only from a fixed-in-time disk surface and
simulations considering also the launching process, thus taking into account disk and jet
evolution together.

Steady-state modeling have mostly followed the self-similar Blandford \& Payne approach
(e.g. \citep{Sauty1994AA, Contopoulos1994C}), but also fully two-dimensional
models were proposed \citep{Pelletier1992, Li1993}, some of them taking even into account
the central stellar dipole \citep{Fendt1995, Paatz1996}.
Further, some numerical solutions  have been proposed by .\citep{konigl2010M, Salmeron2011S, Wardle1993}
in a weakly ionized protostellar accretion discs that are threaded by a large-scale magnetic field 
 as a wind-driving accretion disk.
They have studied the effects of different regimes for ambipolar diffusion or Hall and Ohm diffusivity dominance in
these disk.
(Self-similar) steady-state models have also been applied to the jet launching domain
\citep{Ferreira1995, Li1995,Casse2000}, connecting the collimating outflow with
the accretion disk structure.
In addition to the steady-state approaches, the magneto-centrifugal jet formation mechanism has been 
subject of a number of time-dependent numerical studies.
In particular, \citet{Ustyugova1995} and \citet{Ouyed1997} have demonstrated for the first time 
the feasibility of the MHD self-collimation property of jets.
We note, however, that it was already 1985 when the first jet formation simulations were published,
in that case for much shorter simulation time scale \citep{Shibata1985}.
Among these works, some studies investigated artificial collimation \citep{Ustyugova1999}, a more 
consistent disk boundary condition \citep{Krasnopolsky1999}, the effect of magnetic diffusivity on 
collimation \citep{Fendt2002}, or the impact of the disk magnetization profile on collimation
\citep{Fendt2006}.
In the aforementioned studies, the jet-launching accretion disk is taken into account as a boundary condition,
{\em prescribing} a certain mass flux or magnetic flux profile in the outflow.
This is a reasonable setup in order to investigate jet formation, i.e. the acceleration and collimation process
of a jet.
However, such simulations cannot tell the efficiency of mass loading or angular momentum loss from flux
from disk to jet, or cannot help answering the question which kind of disks do launch jets and under 
which circumstances.

It is therefore essential to extend the jet formation setup and include the launching process in the
simulations.
Numerical simulations of the MHD jet launching from accretion disks have been presented first 
by \citet{Kudoh1998} and \citet{Casse2002}, treating the ejection of a collimated outflow 
out of an evolving-in time resistive accretion disk.
\citet{Zanni2007} further developed this approach with emphasis on how resistive affect 
the dynamical evolution.
An additional central stellar wind was considered by \citet{Meliani2006}.
Further studies were concerned about the effects of the absolute field strength or the field geometry, in 
particular investigating field strengths around and below equipartition 
\citep{Kuwabara2005, Tzeferacos2009, Murphy2010}.
These latter were {\em long-term} simulations for several 100s of (inner) disk orbital periods, providing 
sufficient time evolution to also reach a (quasi) steady state for the fast jet flow. 

Finally, we like to emphasize the fact that jets and outflows are observed as {\em bipolar} streams.
Jet and counter jet appear typically asymmetric with only very few exceptions.
One exception is the protostellar jet HH\,212 showing an almost perfectly symmetric bipolar structure
\citep{Zinnecker1998}.
So far, only very few numerical simulations investigating the bipolar launching of disk jets have 
been performed.
Among them are the works of \citet{Rekowski2003} or \citet{Rekowski2004} which even 
included a disk dynamo action.
Recent publications consider asymmetric ejections of stellar wind components from an offset
multi-pole stellar magnetosphere \citep{Long2008M,Lovelace2010, Long2012}.
It is therefore interesting to investigate the evolution of both hemispheres of a {\em global} 
jet-disk system in order to see whether and how a global asymmetry in the large-scale outflow can 
be governed by the disk evolution. 
This has not been done so far.

In a series of two papers, we will address both the detailed physics of MHD jet launching (paper I)
and the bipolarity aspects of jets (paper II).
In the present paper (paper I) we investigate the details of the launching, acceleration, and collimation
of MHD jets from resistive magnetized accretion disks.
We investigate how the mass and angular momentum fluxes depend on the internal disk physics applying 
long-lasting global MHD simulations of the disk-jet system.
We hereby investigate different resistivity profiles of the disk.
This paper is organized as follows. 
Section 2 is dedicated to MHD equations and to describe the numerical setup, the initial and boundary 
conditions of our simulations.
The general evolution of jet launching and the physical processes involved are presented in 
section 3 with the help of a reference simulation.
Section 4 is then devoted to a parameter study comparing jets from different setups.

In paper II we will present the bipolar jets simulations, discussing their symmetry properties and 
how symmetry can be broken by the intrinsic disk evolution.
%
%////////////////////////////////////////////////////////////////////////////////////////////////////
\section{Model setup}
We model the launching of a MHD outflow from a slightly sub-Keplerian disk, initially in pressure equilibrium with 
a non-rotating disk-corona.

\begin{figure}
\centering
\includegraphics[width=0.8\columnwidth]{\figurepath/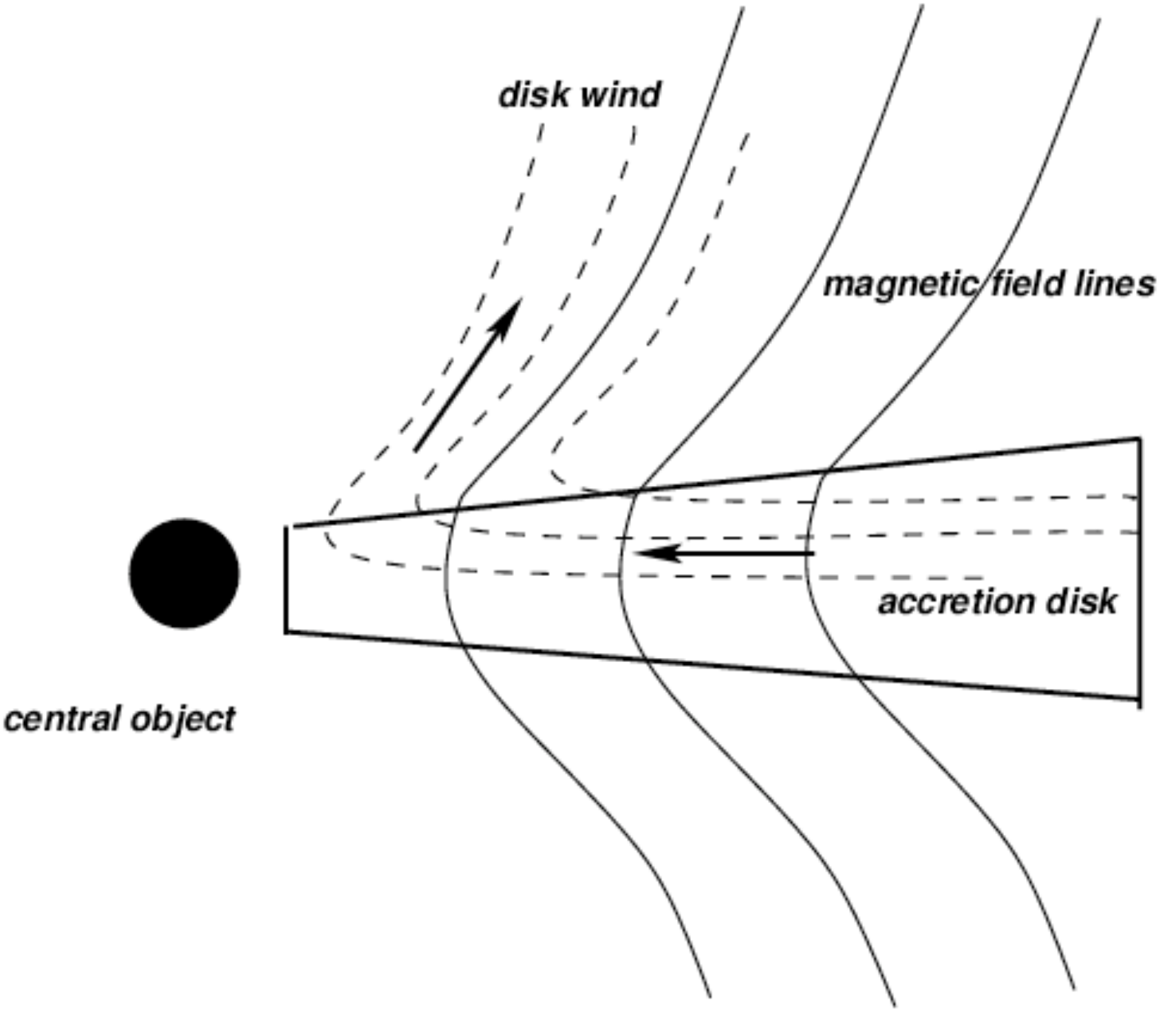}
\caption{Cartoon display of the outflow launching process from accretion disks. 
Matter {\it (dashed lines)} is accreted along the disk surrounding a central object and is 
loaded on to the magnetic field lines {\it (solid lines)}.
The emerging disk wind is further accelerated and collimated into a high velocity beam 
(jet formation).}
\smallskip
\label{fig:cartoonshow}
\end{figure}

As illustrated in Fig.~\ref{fig:cartoonshow}, matter is first accreted radially along a disk surrounding
a central object and is then loaded on to the magnetic field lines.
The large-scale magnetic field is threading the disk and thereby connects the accretion and ejection
processes.

Our main goals are
\begin{itemize}
\item[i)]
    to determine the relevant mass fluxes (mass ejection and accretion rate), and 
    to study the influence on them by the leading physical parameters such as
    magnetic diffusivity, magnetic field strength,
\item[ii)]
    to determine the resulting jet geometry - that is the asymptotic jet radius and opening angle, along with 
    the size of the jet launching area of the disk, and the asymptotic jet velocity.
\end{itemize}
We will in a follow-up paper further extend our setup into two hemispheres, to investigate iii) the launching 
of bipolar jets and their symmetry characteristics.
The relevant mass and energy fluxes for accretion and ejection are of essential interest
for feedback mechanisms in star formation or galaxy formation. 
We our highly resolved simulations of the innermost regions of these
objects we intend to quantify these properties for a range of possible
parameters.
%
%----------------------------------------------------------------------------------------------------
\subsection{MHD equations}
For our numerical simulations, we apply the MHD code PLUTO 3.01 \citep{Mignone2007}, solving the
conservative, time-dependent, resistive, inviscous MHD equations, namely for the
conservation of mass, momentum, and energy,
\begin{equation}
\frac{\partial\rho}{\partial t} + \nabla \cdot(\rho \vec v)=0,
\end{equation}
\begin{equation}
\frac{\partial(\rho \vec v)}{\partial t} + 
\nabla \cdot \left[\vec v \rho \vec v - \frac{\vec B \vec B}{4\pi} \right] + \nabla \left[ P + \frac{B^2}{8\pi} \right]
+ \rho \nabla \Phi = 0,
\end{equation}
\begin{multline}
 \frac{\partial e}{\partial t} + \nabla \cdot \left[ \left( e + P + \frac{B^2}{8\pi} \right)\vec v 
 - \left(\vec v \cdot \vec B \right)\frac{\vec B}{4\pi} + (\bar{\bar{\eta}} : \vec j ) \times \frac{\vec B}{4\pi}  \right]\\
 = - \Lambda_{\rm cool}. 
\end{multline}
Here, $\rho$ is the mass density, $\vec v$ the velocity, $P$ the thermal gas pressure,
$\vec B$ the magnetic field,
and $\Phi = - GM/R $ the gravitational potential of the central object of mass $M$,
withe the spherical radius $R = \sqrt{r^2 + z^2}$.
In general, the magnetic diffusivity is defined as a tensor $\bar{\bar{\eta}}$ (see \S 2.5).
The evolution of the magnetic field is described by the induction equation,
\begin{equation}
\frac{\partial \vec B}{\partial t} - \nabla\times (\vec v \times \vec B - \bar{\bar{\eta}}: \vec j) = 0,
\end{equation}
with the electric current density $\vec j$ given by Amp\'ere's law 
$\vec j = (\nabla \times \vec B) / 4\pi$.
The cooling term $\Lambda$ can be expressed in terms of Ohmic heating $\Lambda = g \Gamma$, with
$\Gamma = (\bar{\bar{\eta}}: \vec j) \cdot \vec j$, and with $g$ measuring the fraction of the 
magnetic energy that is radiated away instead of being dissipated locally. 
For simplicity, here we adopt $g=1$. The gas pressure follows an equation of state 
$P = (\gamma - 1) u $ with the polytropic index $\gamma$ and the internal energy density $u$.
The total energy density is
\begin{equation}
e = \frac{P}{\gamma - 1} + \frac{\rho v^2}{2} + \frac{B^2}{8\pi} + \rho \Phi.
\end{equation}

%--------------------------------------------
Our simulations are performed in axisymmetry 
applying cylindrical coordinates.
The CENO3 algorithm as $3^{rd}$ order interpolation scheme is used for spatial integration \citep{Zanna2002}
together with a $3^{rd}$ order Runge-Kutta scheme for time evolution
 and an HLL Riemann solver.
For the magnetic field evolution we apply the constrained transport method (FCT) 
ensuring solenodality $\nabla \cdot \vec B =0$.

\subsection{Units and normalization}
Throughout the paper distances are expressed in units of the inner disk radius $r_{\rm i}$,
while $p_{\rm d,i}$ and $\rho_{\rm d,i}$ denote the disk pressure and density at this radius, 
respectively.
The index 'i' refers to a number value at the inner disk radius at $z=0$ and time $t=0$.

In the Appendix \ref{app:units-normalization} we show for comparison the astrophysical scaling
for yound stellar (YSO) jets and AGN jets.
Typically, $r_{\rm i} < 0.1\,$ AU for YSO and $r_{\rm i} < 10$ Schwarzschild radii for AGNs.
Naturally, we cannot treat any relativistic effects of AGN jets with our non-relativistic setup.
Velocities are measured in units of the Keplerian speed $v_{\rm K,i}$ at the inner disk 
radius.
Time is measured in units of $t_{\rm i} = r_{\rm i} / v_{\rm K,i}$, which can be related to 
the Keplerian orbital period $\tau_{\rm K,i} = 2\pi t_{\rm i} $.
Pressure is given in units of  $p_{\rm d,i} = \epsilon^2 \rho_{\rm d,i} v_{\rm K,i}^2$.
The magnetic field is measured in units of $B_{\rm i} = B_{z,\rm i}$.
As usual we define the aspect ratio of the disk  $\epsilon$ as the ratio of the isothermal 
sound speed to the Keplerian speed, both evaluated at disk mid plane, 
$\epsilon  \equiv {c_{\rm s}}/{v_{\rm K}}$\footnote{In PLUTO the 
magnetic field is normalized considering $4\pi = 1$.}.
For a more details see appendix \ref{app:units-normalization}.
%
%//////////////////////////////////////////////////////////////////////////////////////////////

\subsection{Initial setup - disk and corona}
We define initial conditions following a setup applied by other authors before -
a magnetically diffusive accretion disk is prescribed in sub-Keplerian rotation,
above which a hydrostatic corona in pressure balance with the disk is located
\citep{Zanni2007, Murphy2010}.
The coronal density is chosen several orders of magnitude below the disk density, 
thus implying an entropy and a density jump from disk to corona.
However, contrary to previous authors
\citep{Zanni2007, Tzeferacos2009, Murphy2010}
we do not apply an initial vertical velocity profile in the disk, and only prescribe a radial velocity profile.
We have seen that for our long-term simulations the whole disk system does adjust to a new 
dynamical equilibrium which does not depend on the vertical profile of the initial velocity distribution.

\subsubsection{Initial disk structure}
We prescribe an initially geometrically thin disk with 
$\epsilon = H/r =  0.1$ which is itself in vertical equilibrium between thermal 
pressure and gravity.

We follow the standard setup employed in a number of previous papers 
\citep{Zanni2007,Murphy2010}.

As initial disk density distribution we prescribe
\begin{equation}
 \rho_{\rm d} = \rho_{\rm d,i}
                \left(  \frac{2}{5\epsilon^2}\frac{r_{\rm i}}{r} 
                \left[\frac{r}{R}-  
                \left(1-\frac{5\epsilon^2}{2}\right)
                \right] 
                \right)^{3/2}
\end{equation}

\citep{Murphy2010},

while the initial disk pressure distribution follows
\begin{equation}
P_{\rm d} = P_{\rm d,i} \left(\frac{\rho_{\rm d,i}}{\rho_{\rm d}}\right)^{5/3}.
\end{equation}
The disk is set into slightly sub-Keplerian rotation accounting for the radial 
gas pressure gradient and advection, 
\begin{equation}
v_{\phi, \rm d}(r) = \sqrt{1-\frac{5\epsilon^2}{2}}\,\sqrt{\frac{GM}{r}},
\end{equation}
following \citet{Murphy2010}, but neglecting viscosity.

In our setup the initial poloidal velocity is imposed by hand following
the prescription of \cite{Zanni2007}, 
\begin{equation}
v_{r,\rm d}(r) = \epsilon^2 \sqrt{2 \mu} \left(\frac{5}{4}+\frac{5}{3m^2}\right)\sqrt{\frac{G M}{r}} 
               = \frac{r}{z}\,v_{z, \rm d}.
\label{eq:disk_vp}
\end{equation}
Simulations with initially vanishing disk accretion resulted in the same asymptotic
inflow-outflow evolution. Starting with a disk inflow as in Eq.~\ref{eq:disk_vp},
the inflow-outflow structure is established on a shorter time scale.

\begin{figure}
\centering
\includegraphics[width=0.8\columnwidth]{\figurepath/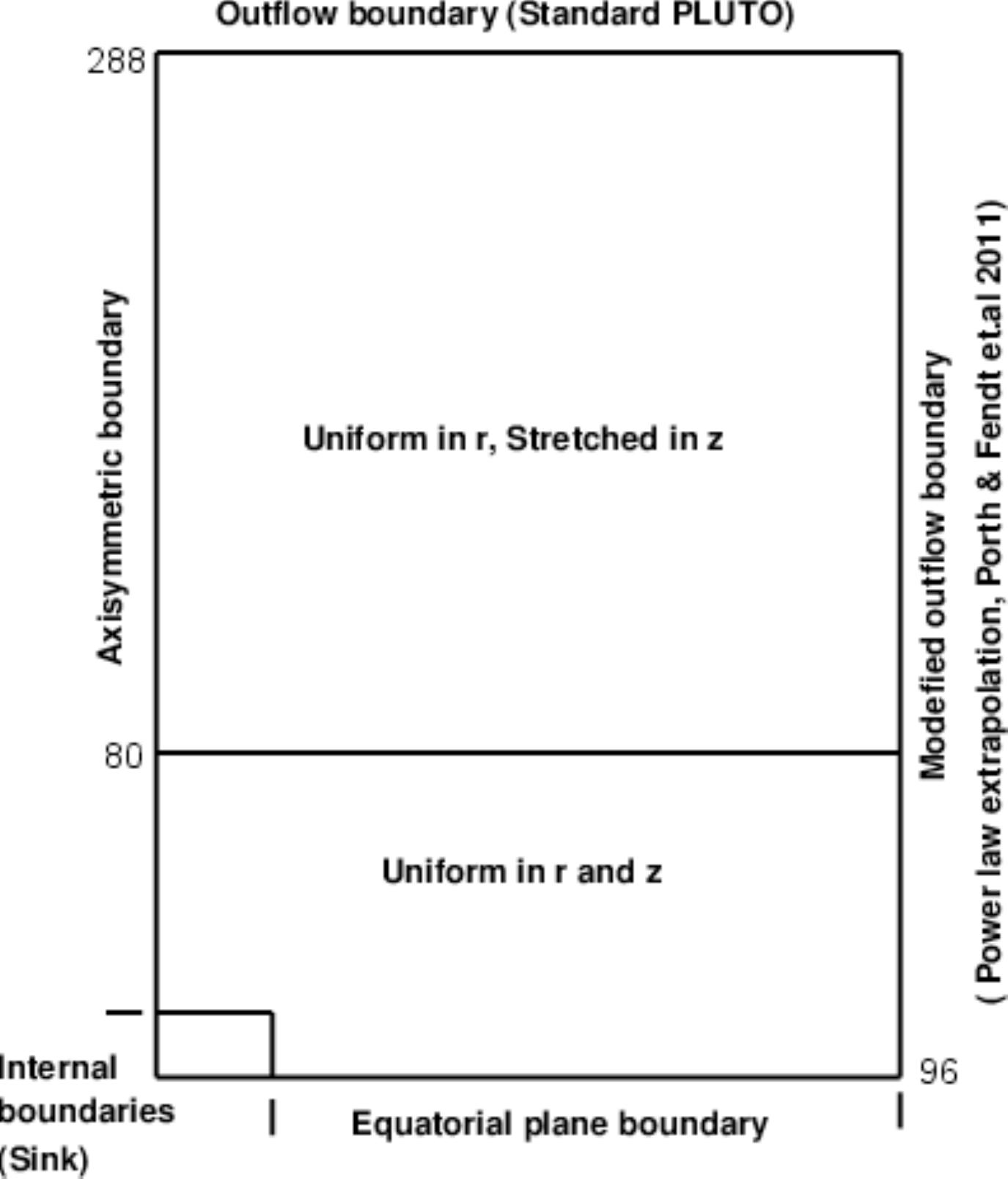}
\caption{Computational domain consisting of two grids and a set of different boundary conditions.
The computational domain covers $(1500 \times 1200)$ uniform grid cells on a physical
domain of $(r \times z) = (96.0 \times 80.0) r_{\rm i}$.
Another 2000 stretched grid cells are attached in vertical direction $80.0 < z < 288.0$.
 For the high resolution simulations a physical domain of $(50.0 \times 50.0) r_{\rm i}$
is covered with $(2000 \times 2000)$ grid cells to which another 2000 stretched grid cells are attached
in vertical direction between $50.0 < z < 180.0$.}
\label{fig:domain1}
\end{figure}

%------------------------------------------------------------------------------------------
\subsubsection{Structure of the coronal region}
Above the disk, we define an initially hydrostatic density and pressure stratification
(the so-called "corona"),
\begin{equation}
\rho_{\rm c}=\rho_{\rm a,i}\left(\frac{r_{\rm i}}{R}\right)^\frac{1}{\gamma-1}\!,\,
P_{\rm c}=\rho_{\rm a,i}\frac{\gamma-1}{\gamma}\frac{GM}{r_{\rm i}}\left(\frac{r_{\rm i}}{R}\right)^\frac{\gamma}{\gamma-1}\!.
\end{equation}
The parameter $\delta \equiv \rho_{\rm c} /\rho_{\rm d}$ quantifies the initial density contrast between disk 
and corona. 
We adopt $\delta = 10^{-4}$.

\subsubsection{Magnetic field distribution}
The initial magnetic field is prescribed by the magnetic flux function $\psi$ following \cite{Zanni2007},
\begin{equation}
\displaystyle
\psi(r,z) = \frac{3}{4} B_{z,i} r_{\rm i}^2 
            \left(\frac{r}{r_{\rm i}}\right)^{3/4}
            \frac{m^{5/4}}{\left( m^2 + \left(z/r\right)^2\right)^{5/8} }.
\label{eq;magpsi}
\end{equation}
Here $B_{z,0}$ measures the vertical field strength at $(r=r_{\rm i},z=0)$.
The magnetic field components are calculated by
$r B_z = \partial \Psi/\partial r$ and $r B_r = \partial \Psi/\partial z$.
\subsection{Numerical grid}
The computational domain spans a rectangular grid region applying a purely uniform spacing in radial 
direction and a uniform plus stretched spacing vertically (Fig.~\ref{fig:domain1}). 
We have run simulations in two different resolutions (see Tab.~\ref{tbl:cases}).
For the low resolution simulations the domain extends to $(96 \times 288)$ inner disk radii $r_{\rm i}$
on a grid of $(1500 \times 3200)$ cells, resulting in a resolution of $\Delta r = 0.064$.
For the high resolution simulations the resolution is increased to $0.025$
on the cost of being limited to a smaller domain of $(50 \times 180) r_{\rm i}$.
For all simulations presented here, a uniform grid is used for the magnetically diffusive 
disk\footnote{Although previous publications have applied a stretched grid also for the part of the domain  
enclosing the disk structure (see e.g. \citet{Murphy2010}),
we note that up to the latest version of the code, PLUTO 3.1., 
the treatment of magnetic diffusivity is limited to equidistant grids (see the PLUTO user manual)}.

For the typical disk model applied with $\epsilon = 0.1$ the low resolution provides
only 2 grid cells per disk scale height at the inner disk radius, while for somewhat
larger radii, say at $r=5$, we have about 10 cells per disk scale height.
In the high resolution runs the disk resolution is increased by a factor 2.5.
As discussed by \citet{Murphy2010} the low resolution barely resolves the jet launching
disk surface layer of the very inner disk, 
where steep gradients in density, pressure and diffusivity are present.
Here, numerical diffusivity is supposed to play a role and we have devoted a section to
investigate this effect.

In order to follow the jet outflow over long distances and in order to provide a sufficient
mass reservoir for disk accretion a large domain would be wishful.
In order to resolve the wind launching area, a high resolution would be required.
However, as mentioned before, PLUTO in its current version does not allow for diffusive MHD
simulations in a stretched grid. 
Furthermore, we experiences that for large domains together with stretched grids (implying an 
elongated shape of the outer grid cells) the code had difficulties to solve the conservative equation
and the simulations crashed.
\begin{figure}
\begin{center}
\includegraphics[width=0.8\columnwidth]{\figurepath/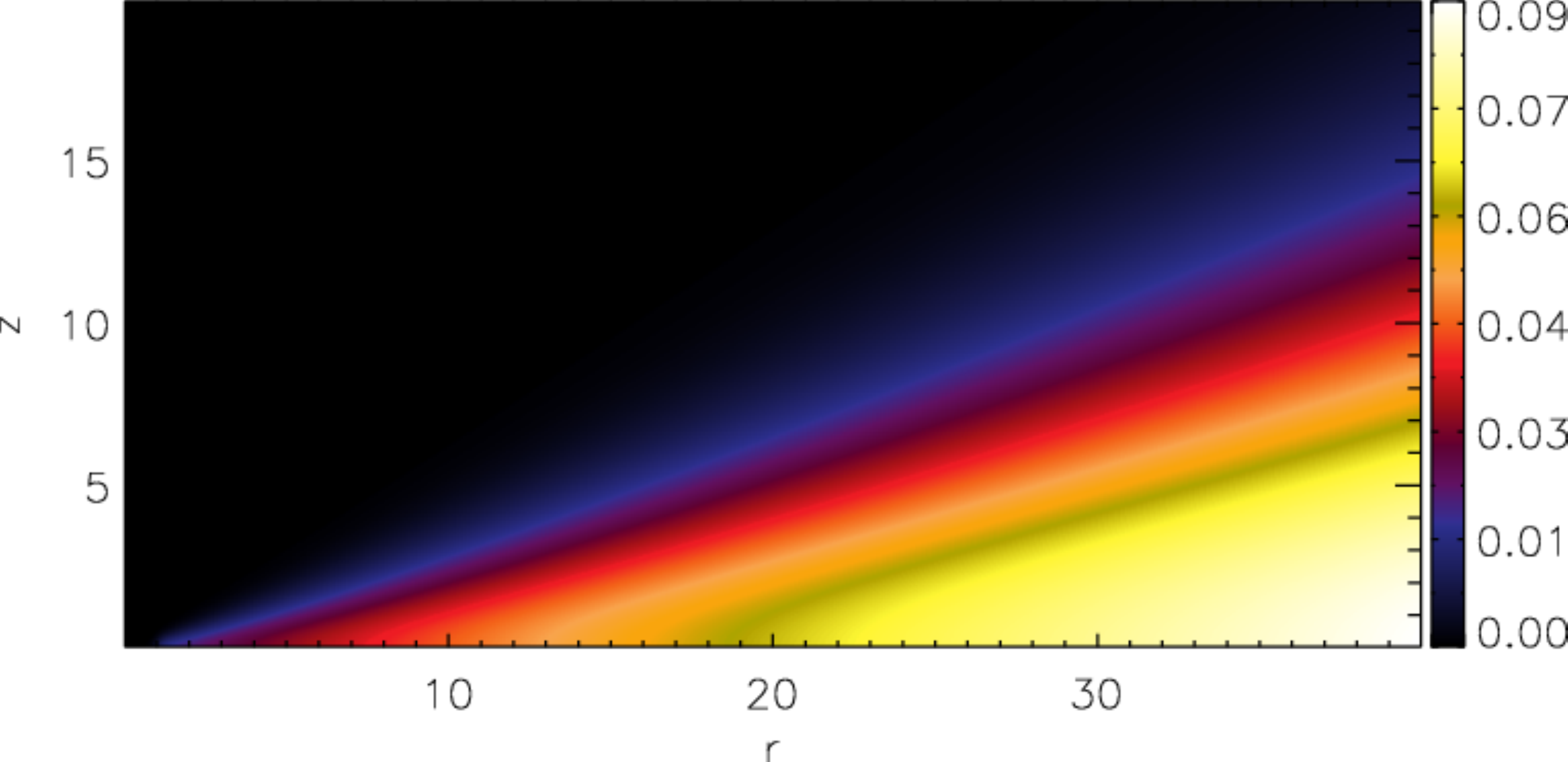}
\caption{Distribution of the turbulent magnetic diffusivity $\eta_{\rm p} = \eta_{\rm p,i} f(r,z)$ in the 
disk-jet system for the reference simulation (see Eq.~\ref{eq:magdiff}). 
Here, the  diffusive scale height is $H_{\eta} = 0.4 r$.}
\label{fig:diff_prof}
\end{center}
\end{figure}

\subsection{Boundary conditions}
For the boundary conditions, axisymmetry on rotation axis and equatorial symmetry for the disk mid plane 
are imposed.
At the upper $z$ boundary, we use the standard PLUTO outflow boundary condition (zero gradient), 
but at the outer radial boundary condition, we apply a modified outflow condition
as derived by \cite{Porth2011} in order to avoid artificial collimation.
Most essential is the internal boundary enclosing the origin which we call a sink.

\subsubsection{Internal boundary - a central sink}
We prescribe a sink for the mass flux in the very inner region of the domain.
The sink is essential for the following reasons.
First, the numerically problematic singularity at the origin can be hidden 
by this internal boundary.
Second, the central sink allows to absorb the accretion flow through the 
disk and thus, emulate a central accreting object.

The sink is numerically introduced as an internal boundary condition, located 
on the region $r < r_{\rm i}$ and $ z< z_s $.\footnote {$z_s$ is defined corresponding to 
the resolution, at least 4 cells is taken as a sink height in any run.}
The internal boundary conditions defined at the right and top sides of the sink 
have significant impact on the simulation.
If ill-defined, spurious effects arise during the early evolution already.
For the right side of the sink, we impose {\em accretion conditions}, implying a zero gradient 
for pressure, density, and the vertical velocity component.
For $v_{\rm \phi}$ and $B_{\rm \phi}$ a first order extrapolation (constant gradient) is 
imposed which ensures an angular momentum decrease across the boundary which is 
essential for accretion.
To allow for accretion further, and also to minimize feedback from the sink to the 
domain, we constrain min$(0, v_{\rm r_{\rm i}})$ in the ghost cells for the radial 
velocity component.
We assume that the magnetic flux is not advected into the central object, and we 
impose $ E_{\rm \phi}= 0 $.
The normal component of the magnetic field is calculated from the solenodality condition
along both sides of the sink .

On top of the sink, we prescribe the initial local value for the gas pressure.
In order to avoid evacuation of the regions close to the symmetry axis, we impose a 
density of 110\% the initial local density.
Effectively, this condition replenishes the mass into the domain near to the axis to
overcome numerically difficult low densities.

\subsubsection{Outer boundary conditions}
An essential point of our setup is to impose a proper outflow boundary at the outer boundary
of the domain in $r$-direction avoiding artificial collimation forces.
Here we have implemented a current-free outflow condition
(see Appendix \ref{app:outer-boundary} )
which avoids spurious collimation by Lorentz forces and which has been thoroughly tested in 
our previous papers \citep{Porth2010, Porth2011, Vaidya2011}.
In order to enable a long-term disk-jet simulation requiring a long-living disk accretion,
we need to provide a sufficiently long lasting mass reservoir.
This could be realized by either i) a prescribed mass inflow, by a ii) local mass replenishment, or
by iii) providing a large mass reservoir for jet launching part of the accretion disk by extending the 
outer disk radius to large radii. All these approaches have been chosen in the literature.
We decided to follow option iii) and provide a 
sufficiently high mass reservoir outside of the inner launching region  
by extending the computational grid (and thus the outer disk radius) up to about 100 inner disk radii.
Similar approaches haven been applied by \citet{Murphy2010, Long2012}.

We will see below (see Sect.~3.2) that our approach is working well, but has, however, its
limits if applied for a very long lasting simulations. 
During about 5000 dynamical time steps we loose about 30-40\% of disk the mass due to ejection 
and accretion and due to unwanted mass loss across the outer disk boundary.

%----------------------------------------------------------------------------------------
\subsection{The disk magnetic diffusivity}
A dissipative effect is required for steady-state accretion in order to allow matter to diffuse 
across through the magnetic field threading the disk. 
Magnetic diffusivity allows the mass flux to cross the field lines and thus to allow for accretion
along the disk.
We consider the magnetic diffusivity to be turbulent in nature, or "anomalous", 
and, thus, cannot be calculated self-consistently in our setup.
The origin of the turbulent magnetic diffusivity is usually referred to be as caused by the magneto-rotational 
instability (MRI) \citep{Balbus1991, Balbus1998}        %REFERENCES 
in a moderately magnetized disk\footnote{We will later show that we do not resolve
the MRI with the numerical resolution applied in our disk-jet setup (see \S 4)}.

Nevertheless, we argue that as we load  turbulently diffusive disk plasma into the outflow, 
it is natural to assume that the outflowing material is  initially turbulent initially as well.
Therefore, we have the option of applying a diffusive scale height $H_{\eta} = \epsilon_{\rm \eta} r$ 
larger than the thermal scale height $H = \epsilon r$ of the disk, supposing that the turbulent disk material 
lifted into the outflow will remain turbulent a few more scale lengths until turbulence decays.  
 It is clear that the strong magnetization in the upper layers of the disk
will prevent generation of turbulence by e.g. the MRI and will also quench the turbulence in 
the outflowing material.
Without going into detail we may estimate the time scale for the decay of the turbulence 
potentially existing in the outflowing material as follows.
The decay of (helical) MHD turbulence follows a power law $E_{\rm M} \sim t^{-n}$, where
$E_{\rm M}$ is the magnetic energy and the power law index depends on further details
but is in the range of $n = 1/2 ... 2/3$ \citep{Brandenburg2011}.
Simulations indicate that MHD turbulence decays as fast as the hydrodynamic 
turbulence on a few eddy turnover times \citep{Cho2002}.
With $\tau_{\eta} \geq H/c_{\rm S} = \epsilon r / c_{\rm S} = r^{1/2}$ (in code units),
we find a time scale for the decay of MHD turbulence launched from the disk into the
outflow of e.g. $\tau_{\eta}(r=5) \geq 3$, thus in the range of half an orbital period. 
This must be compared to the kinematic time scale for the jet launching.
If we consider the propagation of the initially slow disk wind over a few disk pressure
scale heights $\tau_{\rm wind}(r=5) \simeq H / v_{\rm wind} \simeq 0.5/0.05 = 10$,
we find that this time is of the order of a few eddy turnover times.
Also, we will later see that jet launching is a rapid process with an outflow 
being established already after a few dynamical time steps.

Given the essential role of the magnetic diffusivity for wind/jet launching, we decided to 
investigate
the launching process for different strength and different scale height of diffusivity.
We have run simulations for a variety of combinations of parameter values, 
but for sake of comparison, in this paper, we show results for diffusive scale height values,
$\epsilon_{\rm \eta} = 0.1,0.2,0.3, 0.4$.

Formally, we apply an $\alpha$ prescription \citep{1973shakuraetal}, similar also to previous works 
(see e.g. \citet{Zanni2007}).
We assume the diffusivity tensor to be diagonal with the non-zero components
\begin{equation}
 \eta_{\rm \phi\phi} \equiv \eta_{\rm p} \;\;, \;\; \eta_{\rm rr}=\eta_{\rm zz} \equiv \eta_{\phi}.
\end{equation}
Here, we have defined a poloidal magnetic diffusivity 
$\eta_{\rm p} \equiv \eta_{\rm \phi\phi} = \eta_{\rm p,i} f(r,z)$, and
the toroidal magnetic diffusivity 
$\eta_{\phi} \equiv \eta_{rr} =\eta_{zz} = \eta_{\rm \phi,i} f(r,z)$,
respectively\footnote{The $\eta_{\rm \phi\phi}$ effectively governs the diffusion of the poloidal 
magnetic field,while the $\eta_{rr} $ and $\eta_{zz} $ govern the diffusion of the toroidal magnetic
field}, with a function $f(r,z)$ describing the diffusivity profile.

As it is known from the literature,
an {\em an-isotropic} magnetic diffusivity is required to obtain stationary
solutions \citep{Wardle1993,Ferreira1995, Ferreira1997}.
Most of the numerical simulations followed that approach and did find a stationary-state
from their simulations as well 
(see e.g. \citep{Zanni2007, Tzeferacos2009,Murphy2010}).
Note that simulations of \citet{Casse2002, Casse2004} apply an isotropic
diffusivity and reach a steady state as well.

We define an anisotropy parameter by the ratio of the toroidal to poloidal diffusivity 
components, $\chi = \eta_{\phi,i} / \eta_{\rm p,i}$.
For the diffusivity function $f$ several options can be considered.
For the simulations in this paper (paper I) we apply
\begin{equation}
f = v_{\rm A}(r,z=0) H \exp{\left(-2\frac{z^2}{H^2_{\eta}(r)}\right)},
\label{eq:magdiff}
\end{equation}
where $\eta_{\rm p,i}$ and $\eta_{\phi,i}$ govern the strength of magnetic diffusivity.
The Alfv\'en speed $v_{\rm A}(r,z=0) = {B_{\rm {z}}(r,z=0)}/{\sqrt{\rho(r,z=0)}}$, 
and the disk thermal scale height $H(r) =c_{\rm s}(r,z=0)/\Omega_K(r,z=0)$ are both calculated
along the mid-plane.
The parameter $\epsilon_{\rm \eta}$ measures the scale height of the disk magnetic diffusivity,
similar to the hydrodynamic disk scale height parameter $\epsilon_\eta = H_{\eta}/r$.
For the present paper, we decided not to evolve the diffusivity profile in time.
We find that since both the disk scale height $H$ and the Alfv\'en speed $v_{\rm A}$ do vary as
the disk evolves, the strength of diffusivity may vary substantially from the initially 
prescribed value (up to a factor 10). 
This highly non-linear feedback may disturb the progress of our simulation such that an artificially 
high diffusivity may be derived which will affect the numerical time stepping.
A constant-in-time prescription of diffusivity simplifies our aim of disentangling the governing
physical processes involved in jet launching.

In the follow-up paper II presenting truly bipolar jet launching
we will discuss further models for the magnetic diffusivity.

\begin{figure*}
\begin{center}
\includegraphics[width=3.  cm]{\figurepath/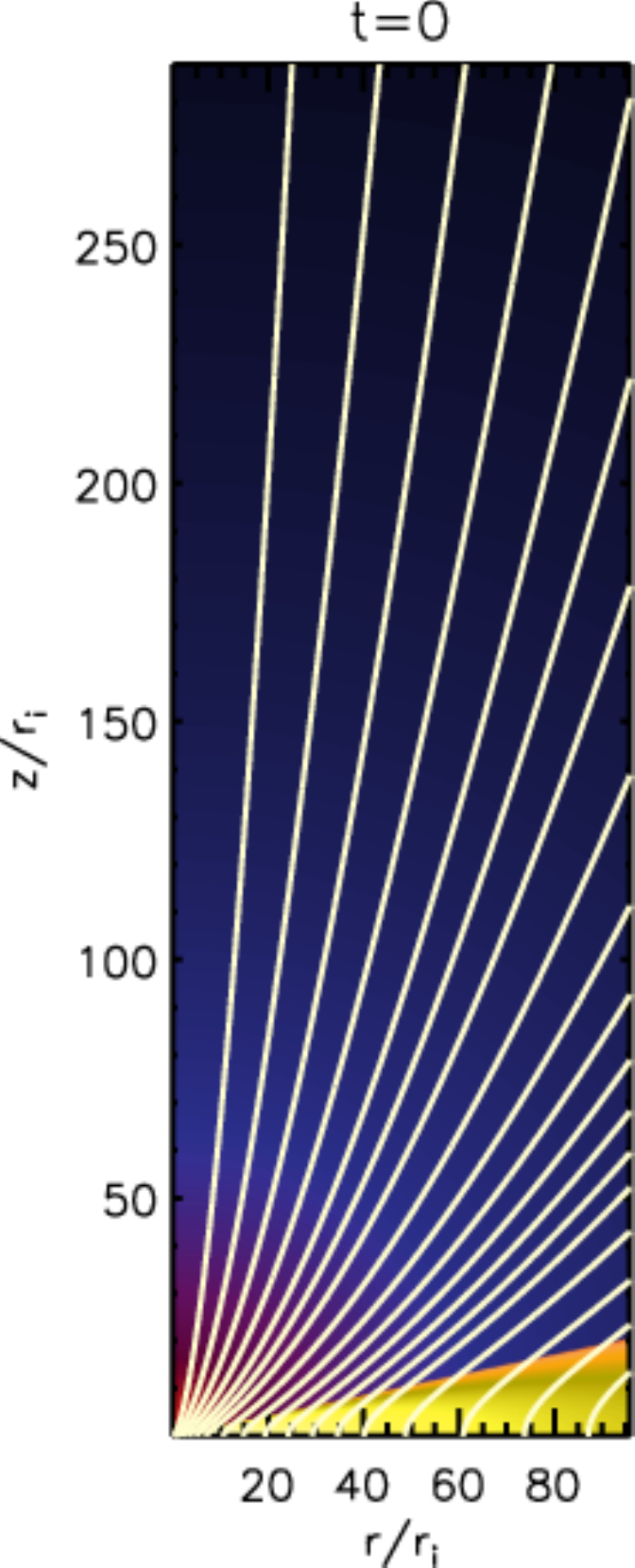}
\includegraphics[width=3.  cm]{\figurepath/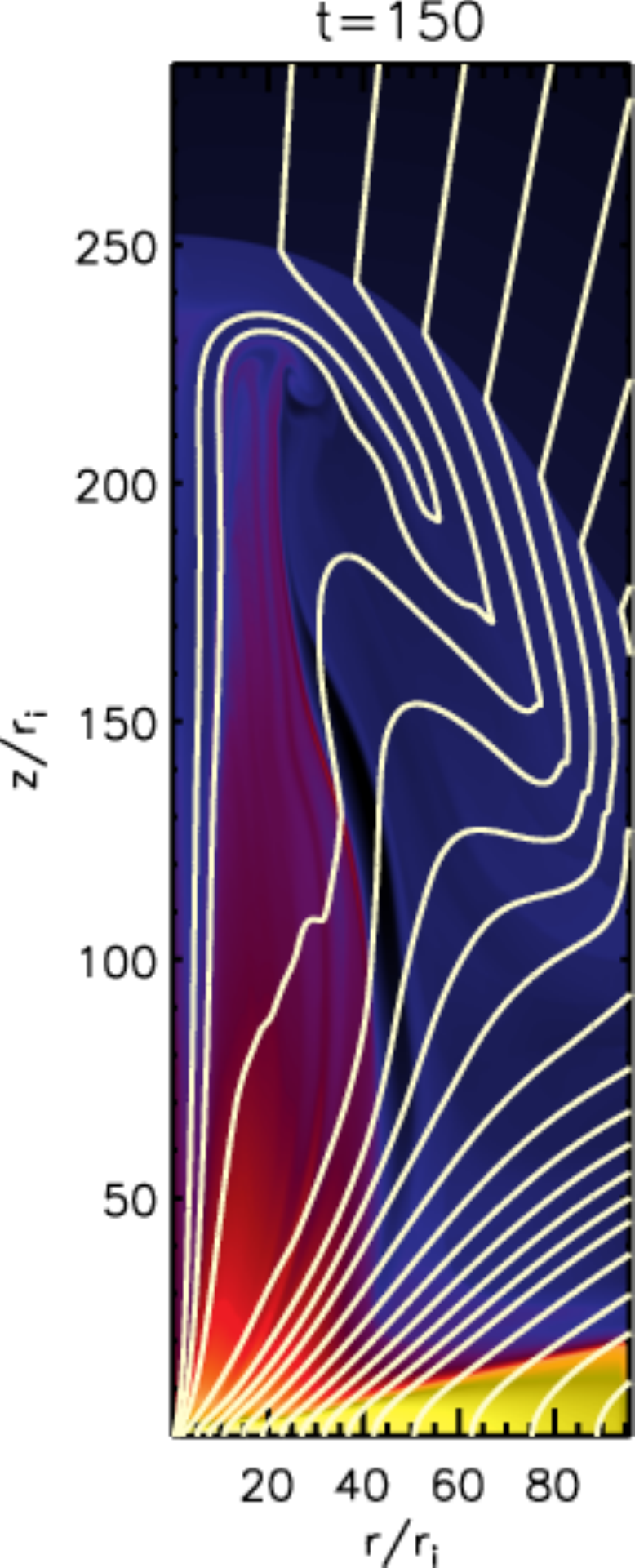}
\includegraphics[width=3. cm]{\figurepath/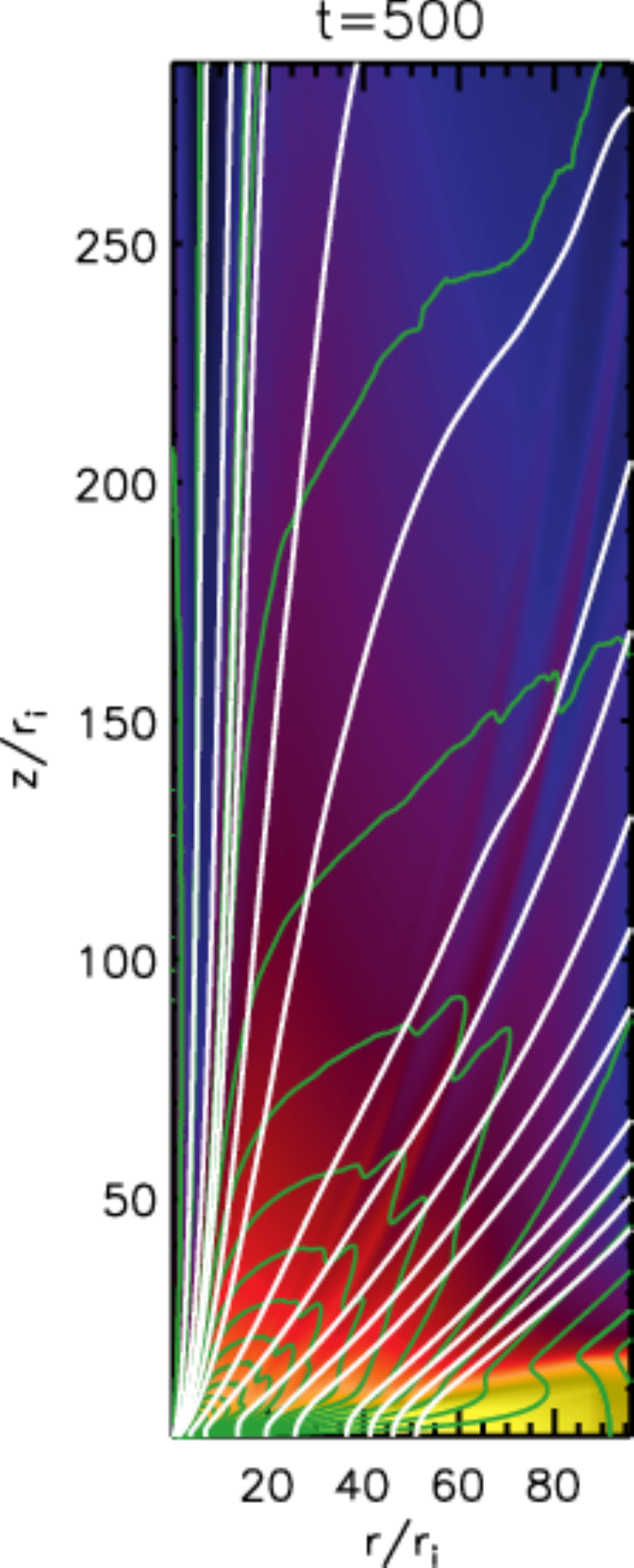}
\includegraphics[width=3. cm]{\figurepath/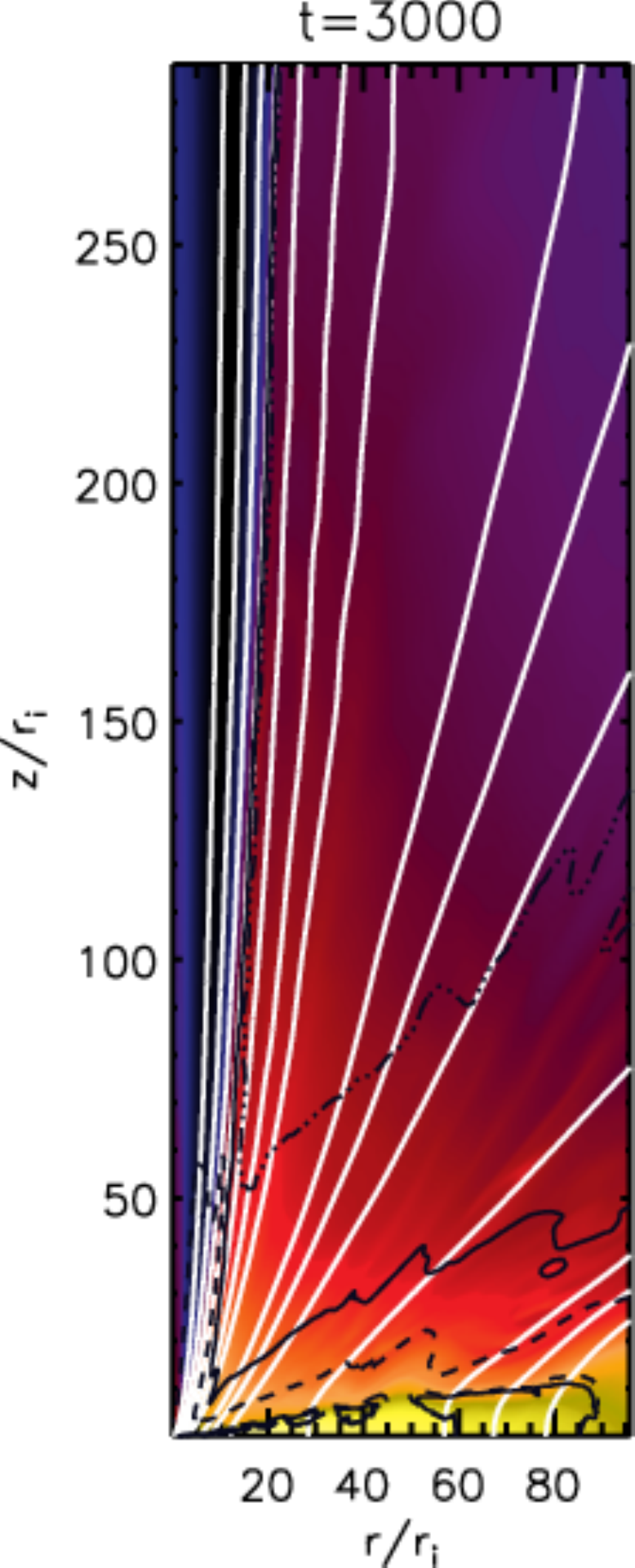}
\includegraphics[width=3. cm]{\figurepath/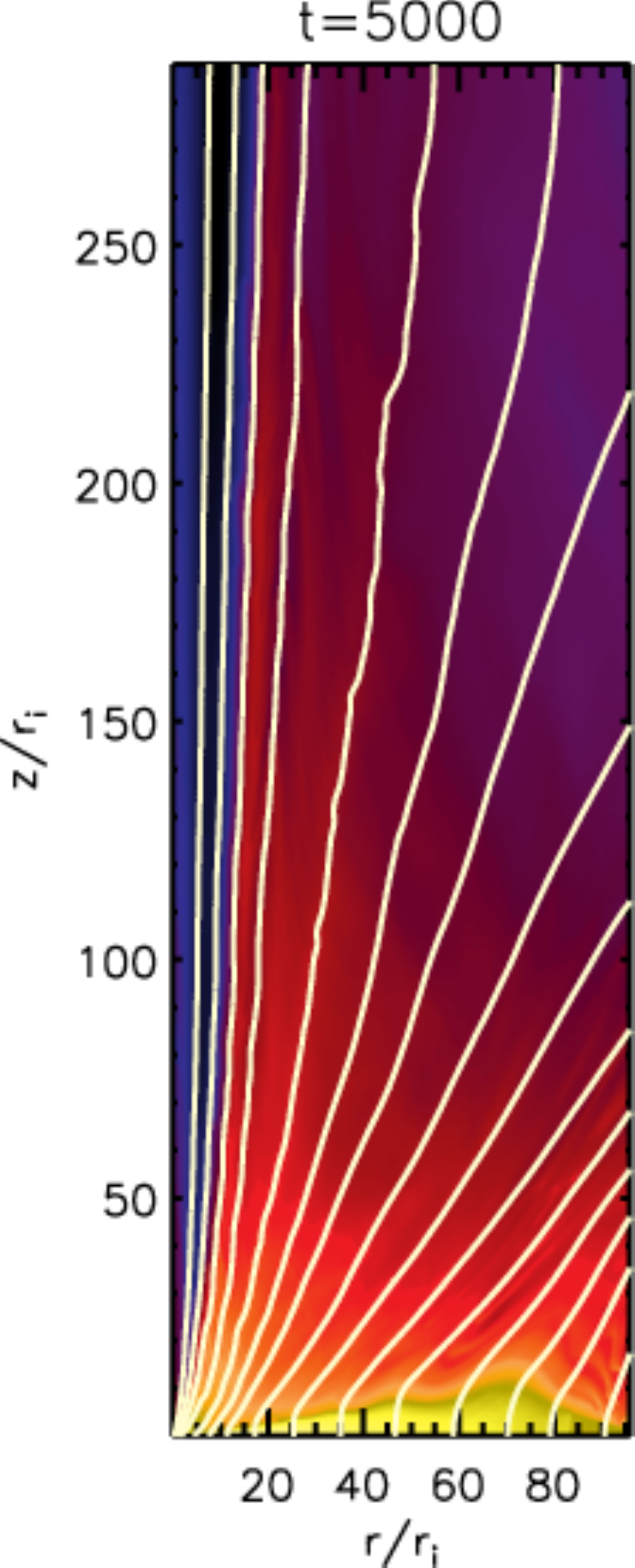}
\includegraphics[width=5cm]{\figurepath/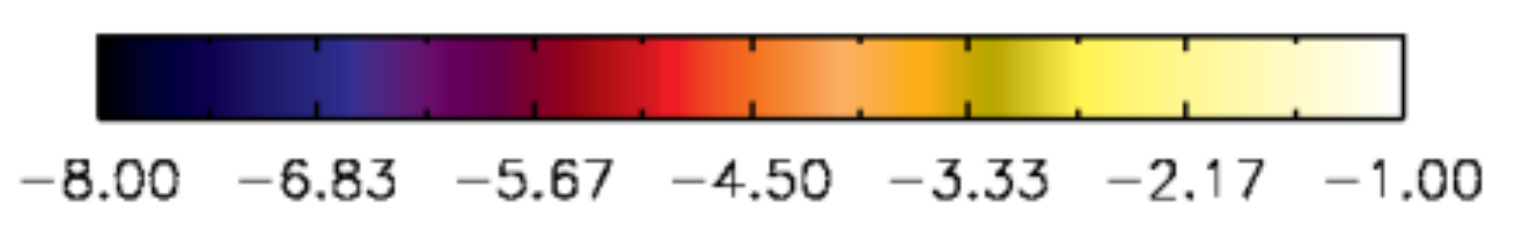}
\caption{Time evolution if the jet-disk structure for reference simulation case 1.
Shown is the evolution of the mass density (color) and the poloidal magnetic field 
(contours of poloidal magnetic flux $\Psi$ for the levels 
0.01, 0.03, 0.06, 0.1, 0.15, 0.2, 0.26, 0.35, 0.45, 0.55,  0.65, 0.75, 0.85, 0.95, 1.1, 1.3, 1.5, 1.7) 
for the dynamical times $t = 0, 500, 1000, 2000, 3000$.
 The dark lines indicate the slow magneto-sonic (dashed), the Alfv\'en (solid), and the fast
magneto-sonic (dotted-dashed) surfaces at $t = 3000$.
The electric current lines are shown for $t=500$ (green solid lines).}
\label{fig:rhocase1}
\end{center}
\end{figure*}

We again stress the importance to provide the reader with the actual number values of the
magnetic diffusivity applied\footnote{We remark that in previous simulations which evolve $\eta$ in 
time, only the initial diffusivity  distribution was provided, butvnot its temporal evolution, and, 
thus, the magnetic diffusivity distribution which is actually involved in producing the outflow
(see e.g. \citet{Zanni2007,  Tzeferacos2009, Murphy2010}).
See also our numerical example below.}.
If we consider a $\eta_{\rm \phi,i} \sim 1$ together with a maximum Alfv\'en speed in the 
disk of about $ 10^{-2}$ (located at the inner disk radius), 
this gives a maximum disk diffusivity of about $\eta(r,z)_{\rm i} \simeq 0.01 - 0.1$, 
through all the simulations.
Figure \ref{fig:diff_prof} shows the distribution of the magnetic diffusivity.
The diffusivity is mainly concentrated in the disk,
increases with disk radius (consistent with a decreasing temperature or ionization degree with radius),
and decreases with height resembling the transition from a cool disk to an ideal MHD outflow.

%-------------------------------------------------------------------------------------------------
\subsection{Main simulation parameters}
The simulations are governed by a set of non-dimensional parameters for the
following physical quantities,
\begin{itemize}
\item
the magnetic field strength defined by plasma-$\beta$, and its initial geometry defined by 
the "bending" parameter $m$,
\item
the magnetic diffusivity, with the three parameters $\eta_{\phi,i}$, $\eta_{p,i}$, $\epsilon_{\eta}$ 
governing its strength and anisotropy, and the diffusivity scale height,
\item
the initial density contrast between disk and corona $\delta$,
\item 
the initial disk thermal scale height parameter $\epsilon = H/r$.
\end{itemize}
An overview over these parameters are shown in Tab.~\ref{tbl:cases} (first half) for the simulations 
presented in this paper. 
The second part of Tab.~\ref{tbl:cases} shows the main dynamical quantities resulting from
our simulations and will be discussed below.
Table \ref{tbl:scale} show the similar numbers for the simulations of different diffusive 
scale height.
%===============================================================================================================
\section{MHD jet launching}
Before presenting detailed results of our parameter study of jet launching conditions, we will first 
discuss the general physical processes involved considering our long-term reference simulation (case 1).
This simulation will then later be compared to simulations applying
 i) different diffusivity profiles,
ii) a higher grid resolution,
iii) and different magnetic field strength %and geometries ; we have same geometry in these parameter runs.....$$
(see Tab.~\ref{tbl:cases}).  
A similar setup will then be used to launch bi-directional outflows, without the constraint on hemispheric symmetry
(see paper II).

%---------------------------------------------------------------------------------------------------------
\subsection{General evolution}
Our reference simulation, case 1 is carried out up to $t = 5000$ dynamical time, corresponding 
to 796 inner disk orbital periods and is thus one of the longest simulations considering the MHD jet launching.
This huge time period corresponds, however, only to 3 rotations at a radius 
$r = 40\,r_{\rm i}$ and correspondingly less at the outer grid radius $r = 96\,r_{\rm i}$.
Consequently, while the inner part of the disk, and thus the outflow evolving from that part, 
has reached a steady state situation, the outer disk corona is still dynamically evolving. 
We will therefore concentrate our discussion mainly to the inner disk areas.

The time evolution of density together with the magnetic field is shown in Fig.~\ref{fig:rhocase1}.
Note that we do {\em not} show field lines integrated from certain foot point radii,
but {\em magnetic flux} contours.
This allows to visualize the diffusion and advection of the magnetic field as a consequence of the disk 
evolution.

The magnetic diffusivity distribution determines the coupling between the plasma and 
magnetic field, and, thus, affects the mass loading into the outflow.
Since the high diffusivity in the outer part of the disk, the coupling is weaker, and,
thus, the mass loading less efficient.
As a general result we observe a continuous and robust outflow launched from the inner part of the disk,
expanding into a collimated jet.

\begin{figure}
\centering 
\includegraphics[width=0.6\columnwidth]{\figurepath/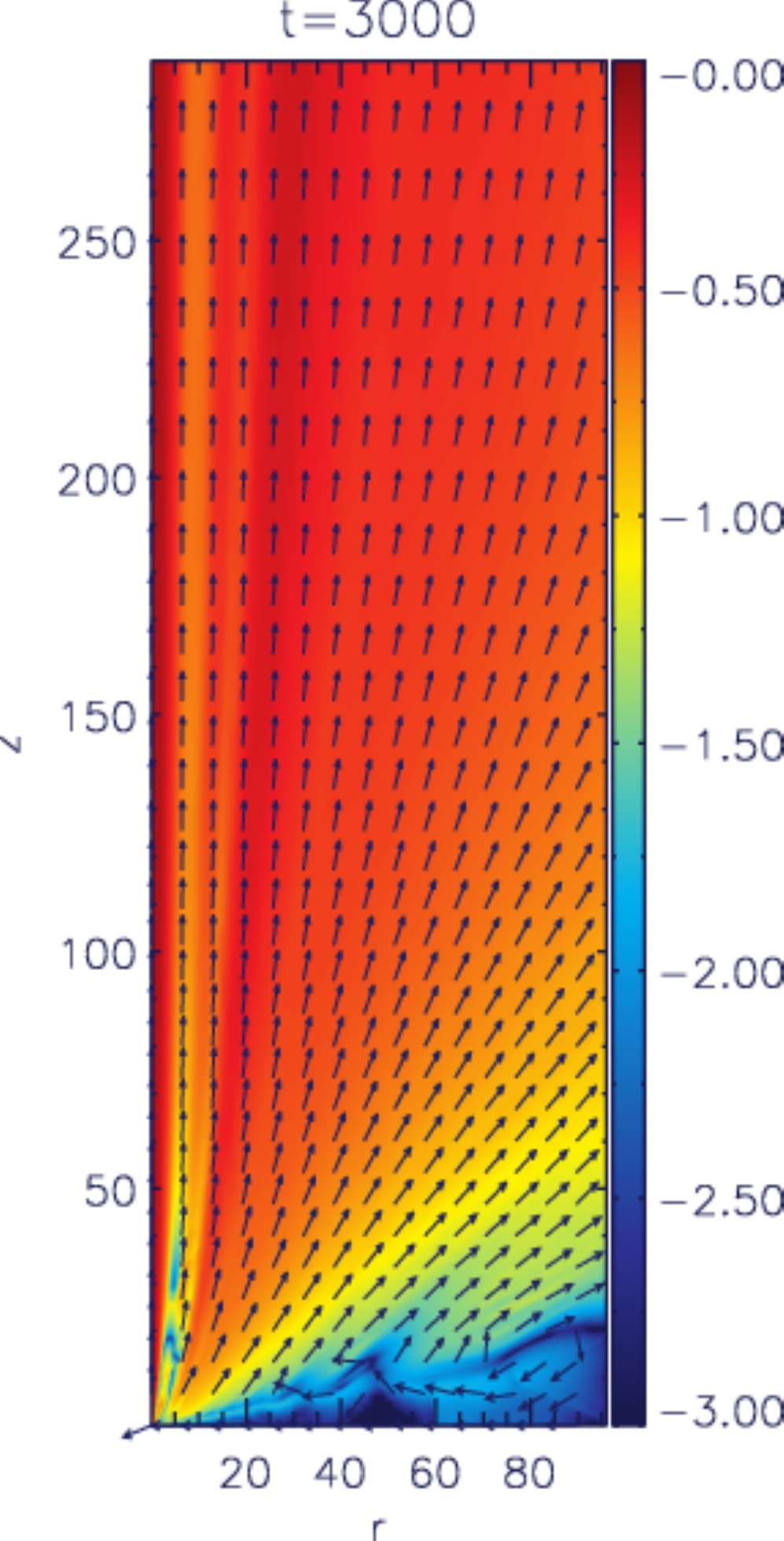}  
\caption{
Accretion-ejection poloidal velocity map at $t=3000$ for reference run case 1.
Shown is $v_{\rm p}$ distribution in logarithmic scale overlaid with which arrows of the 
normalized poloidal velocity vectors indicating the direction of the flow, 
$\vec{v}_{\rm p}/\mid{ \vec{v}_{\rm p}}\mid$.}
\label{fig:poloida-velocity-vecplot}
\end{figure}

Figure \ref{fig:poloida-velocity-vecplot} shows the poloidal velocity distribution in jet and disk
and the normalized poloidal velocity vectors indicating the direction of the mass fluxes. 

The outflow is accelerated to super fast magneto-sonic speed (see the magnetosonic surfaces 
indicated in Fig.\ref{fig:rhocase1})

A bow shock develops at the interface between the outflow and the surrounding hydrostatic corona.
As the outflow propagates, the initial corona is swept out of the computational domain together
with the bow shock.
While the interaction with the ambient gas plays a role within the first evolutionary steps,
the long-term evolution of the outflow - its acceleration and collimation - is purely 
determined by the force balance within the outflow and the physical conditions of the 
launching region.

%==========================================================================================================
\subsection{Disk structure and disk mass evolution}
As jets are launched from disks, the disk evolution is itself an essential part of the jet evolution
and must be carefully considered.
It is expected that the jet mass flux would somewhat correspond to the disk accretion rate 
(which would also depend on actual mass of the disk).  
\begin{figure}
\centering 
\includegraphics[width=0.9\columnwidth]{\figurepath/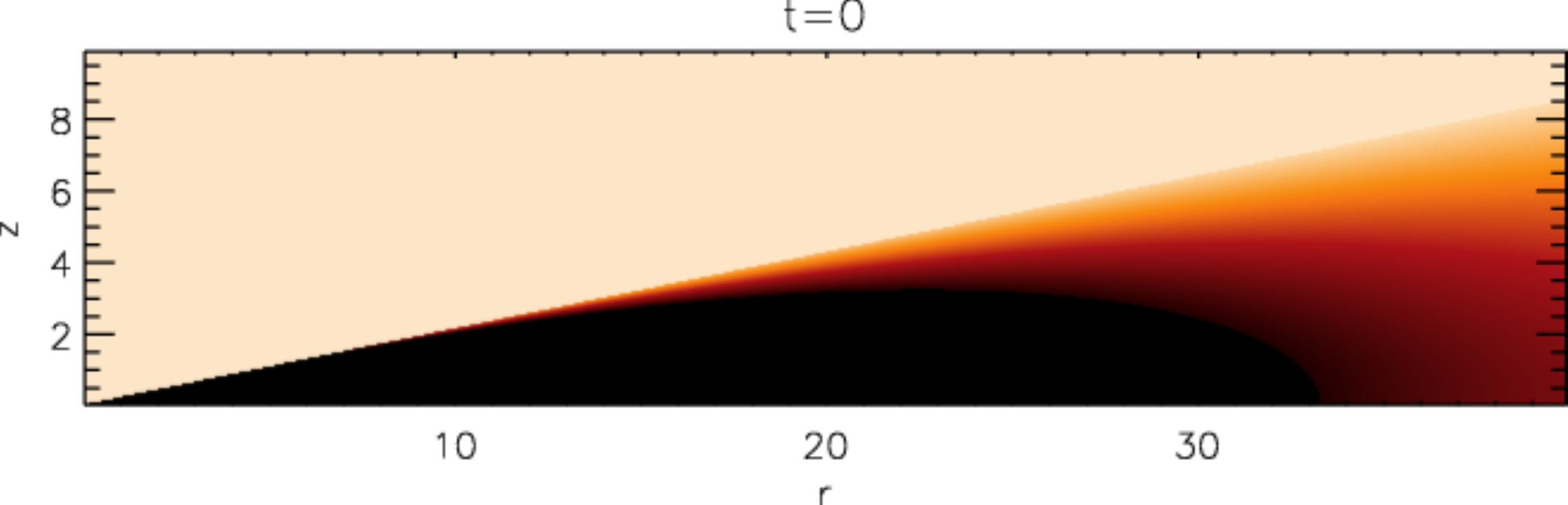}  
\includegraphics[width=0.9\columnwidth]{\figurepath/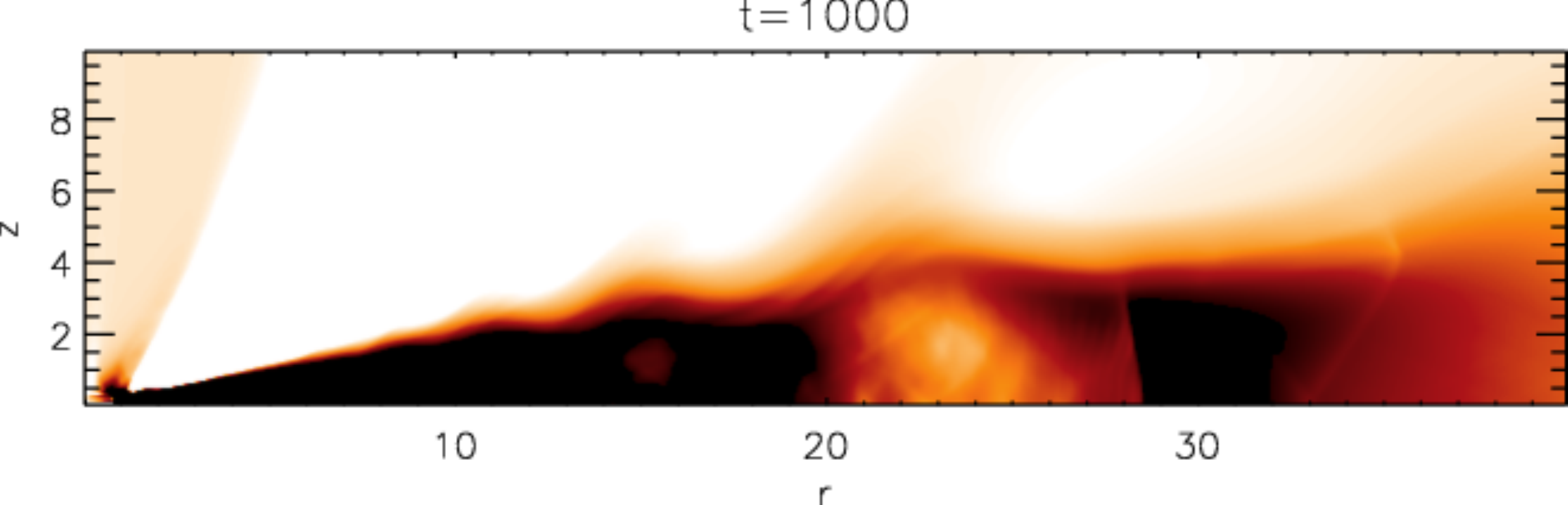}
\includegraphics[width=0.9\columnwidth]{\figurepath/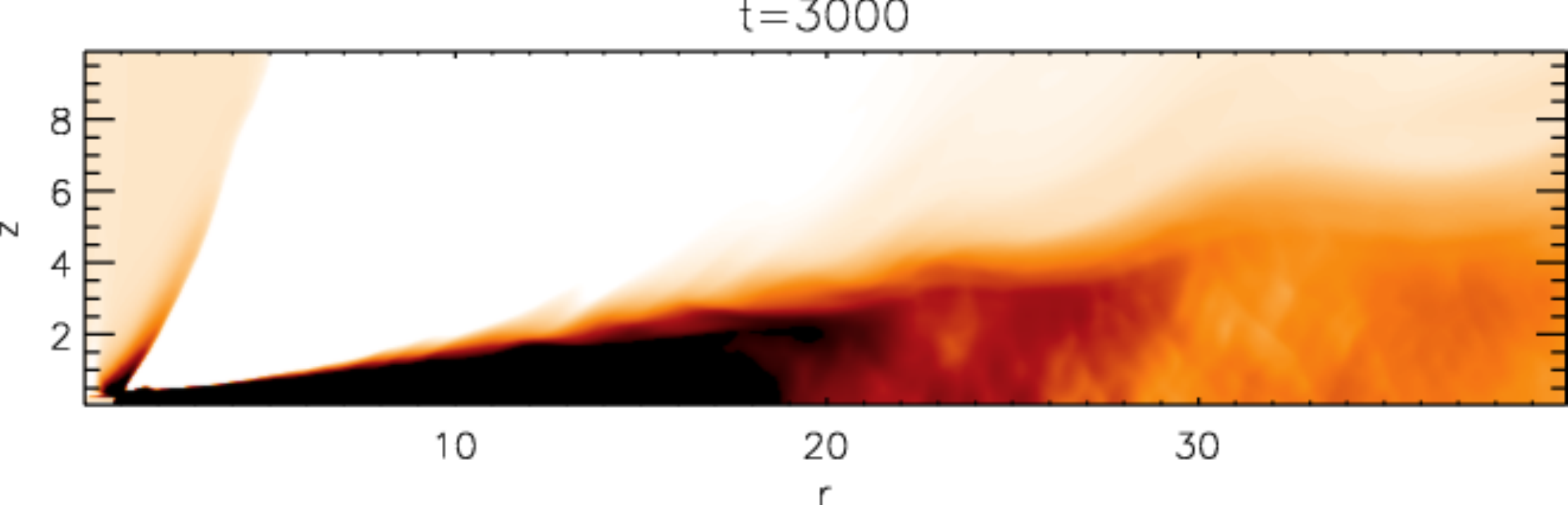}
\includegraphics[width=0.9\columnwidth]{\figurepath/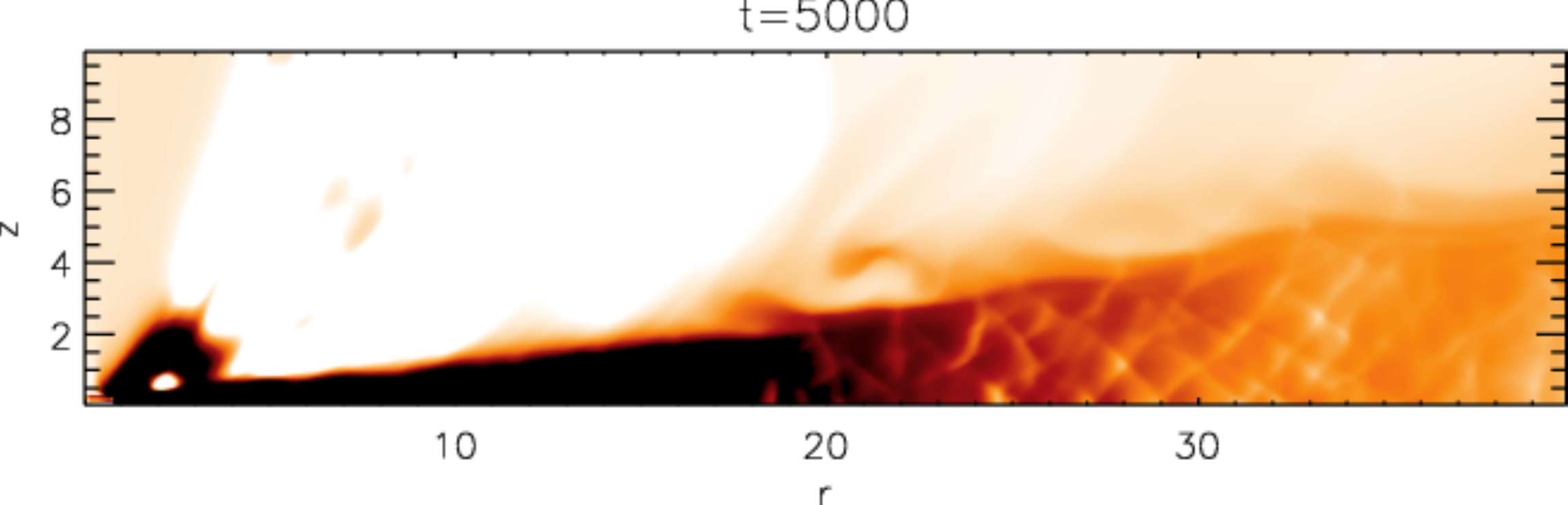}
\includegraphics[width=0.9\columnwidth]{\figurepath/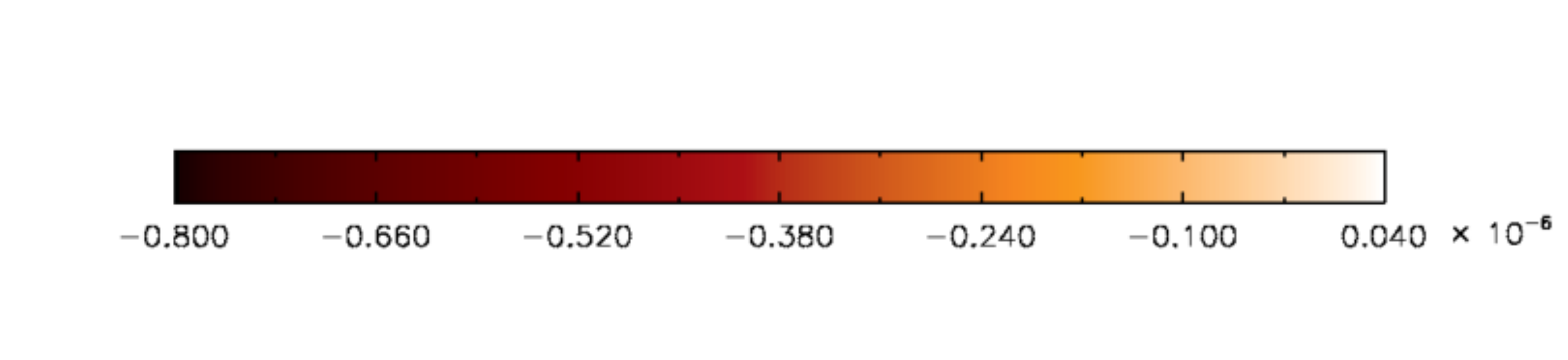}
\caption{Time evolution of the accretion disk. 
Shown is the radial mass flux $(\rho v_r)$ in the disk for reference run (case 1)
for the dynamical times $t = 0, 1000, 3000, 5000$.}
\label{fig:accretion_layer}
\end{figure}

Our simulations show that the disk evolution is complex, with smooth accretion phases are
followed by more turbulent stages.
The time evolution of the accretion stream $\rho v_r$ is shown in Fig.~\ref{fig:accretion_layer}.
We see that accretion starts first at small radii and fully establishes after $t=1000$.
Accretion shocks and vortices may destroy the quasi-steady state situation.
Close to the outer disk radius, mass is lost through the outer boundary.
At late evolutionary stages we see a complex pattern in the $\rho v_r$ distribution
indicating small shock and turbulent motion within the disk.

Nevertheless, the negative mass fluxes in Fig.~\ref{fig:accretion_layer}, together
with the negative velocity vectors in Fig.~\ref{fig:poloida-velocity-vecplot} clearly 
indicate that accretion is established throughout the whole disk.

\begin{figure}
\centering 
\includegraphics[width=0.8\columnwidth]{\figurepath/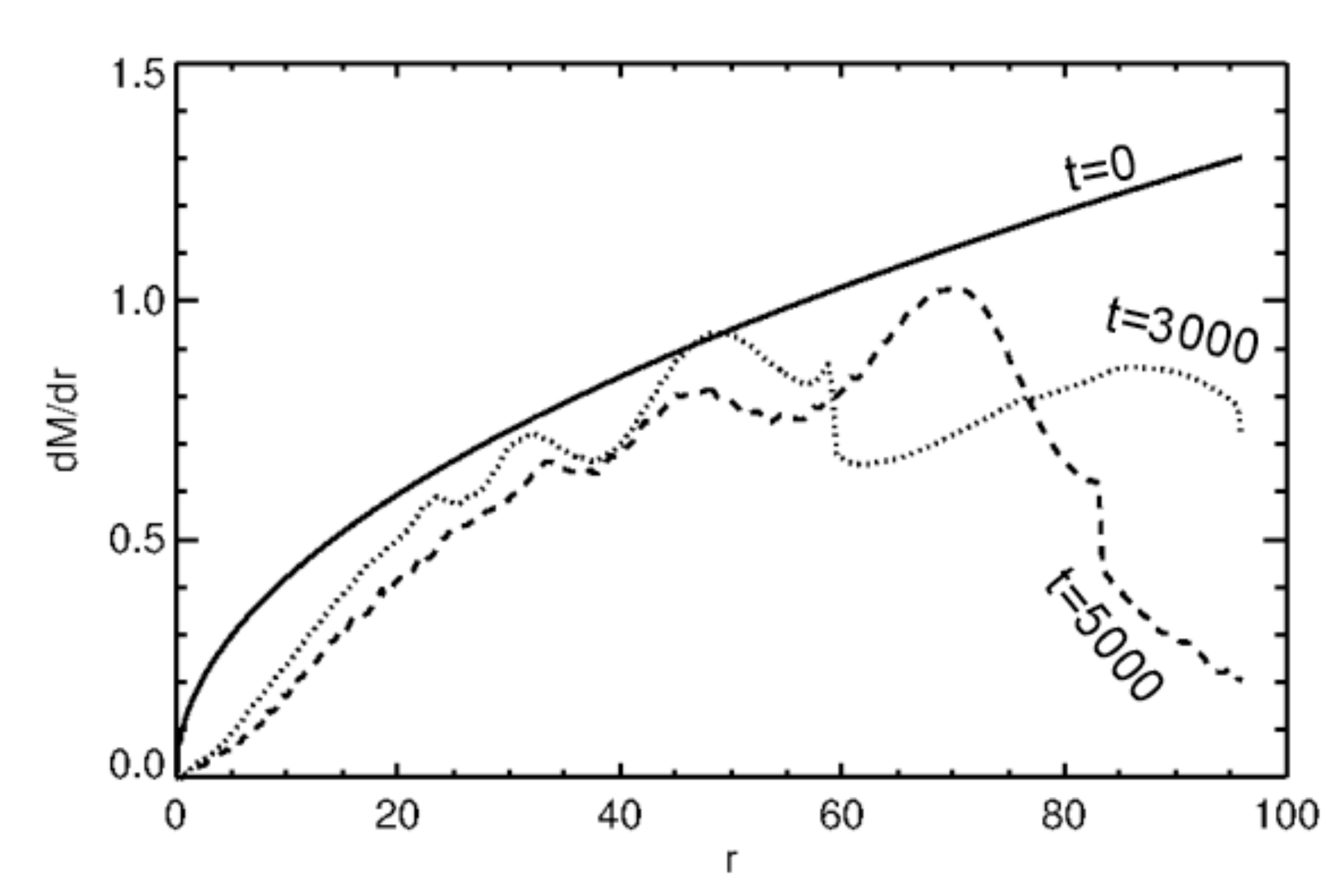}
\includegraphics[width=0.8\columnwidth]{\figurepath/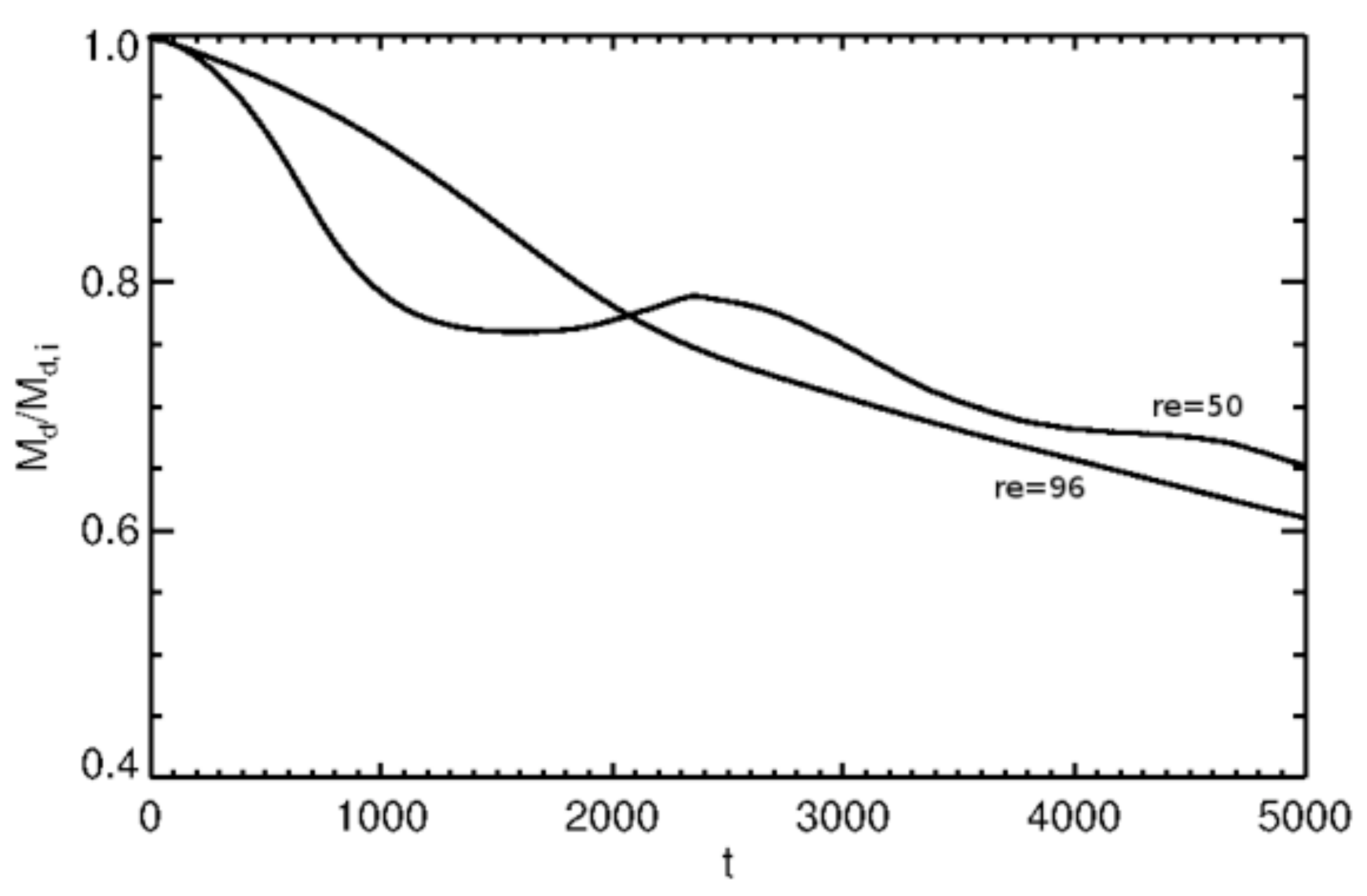}
\caption{Time evolution of the disk mass.
Shown is the radial profile of the mass distribution $dM/dr$ (integrated mass per disk ring, in code units) 
at times $t=0, 3000, 5000$ (top),
and the time evolution of the total disk mass $M_{\rm d}$ normalized to the initial disk mass $ M_{\rm d,i}$ 
for integration radii $r_2 = 50, 96$ (bottom).}
\label{fig:accelement}
\end{figure}

We now discuss the time evolution of the disk mass for our reference simulation.
In order to measure the disk mass, we need to carefully define the control volume defining a 
corresponding {\em "disk surface"}.
Here, we consider as disk surface the location where the mass density at each radius falls bellow $10\%$ 
of the mid-plane density (because of the vortex motion we could not use the negative velocity as
proxy for accretion).
Measuring the mass contained in subsequent disk rings $dM/dr = 2\pi r\rho dz$, we see that most 
of the mass is indeed stored within the outer disk areas (Fig.~\ref{fig:accelement}, top).
We may identify two essential effects affecting the long-term disk evolution.
Firstly, the mass reservoir in the outer disk has decreased - mainly due to outwards mass loss across
the outer disk boundary.
Secondly, the mass of the inner jet-launching disk has also decreased - due to both outflow  activity
and accretion into the sink.

Comparing the mass content of the disk at the initial and final stage, we find that the
total disk mass decreases from $M \simeq 520$ at $t=0$ to $M \simeq 320$ at $t=5000$
(in code units). Integrating the mass loss by accretion and ejection into the jet
(see Fig.~\ref{fig:accelement})
over 5000 dynamical time steps, we find a total mass loss of $M\simeq 75$. 
The difference, i.e. $M \simeq 125$, 
is the mass which is lost from the outermost disk in vertical and radial direction.

In other words,
a substantial amount of the disk mass loss is lost from the very outer part of the disk
and does not directly influences the jet formation.
The disk mass which is lost from the inner part is partly lost as disk wind/jet, partly
accreted into the sink and partly replenished by accretion from outer radii.

The time evolution of the disk mass gives a similar picture (Fig.~\ref{fig:accelement}, bottom).
Integrated over the whole domain, the disk loses about $38\%$ of its mass until $t = 5000$  
in the reference simulation case 1.
If we decrease the integration domain (outer radius $r=50$) the inner disk looses less 
mass while part of its mass is being accreted from outer disk radii. 
In general this implies that up to time $t=5000$, there is 
 i) still sufficient mass to support the accretion process, but also that 
ii) the disk mass shows a considerable decrease which may have changed the internal disk properties.
Thus, for an even longer-lasting simulation one would have to invent another mass source for the disk
accretion
(e.g. by a physical mass inflow boundary condition at the outer disk radius 
properly taking into account angular momentum transfer, or a floor
 density distribution in the disk).

\begin{figure*}
\begin{center}
\includegraphics[width=10cm]{\figurepath/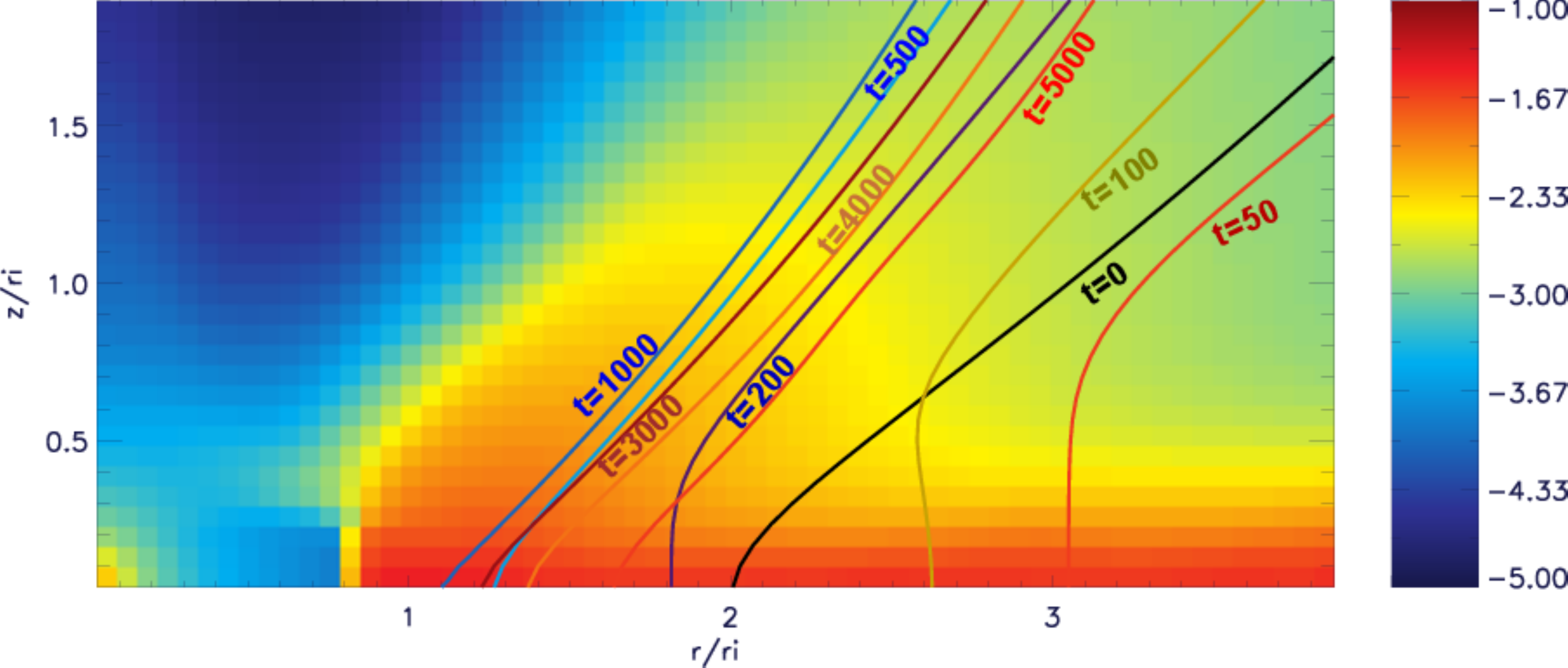}
\caption{Diffusion and advection of magnetic flux.
Shown is the evolution of the magnetic flux surface $\psi = 0.1$ for times
$t = 0,  50, 100, 200, 500, 1000, 3000, 4000, 5000$ (colored lines)for reference simulation case 1.
This flux surface is initially rooted at $(2,0)$. 
Superimposed is the density distribution at $t=5000$.}
\label{fig:adve_diff}
\end{center}
\end{figure*}

Along with the hydrodynamic disk evolution, the magnetic field distribution evolves as well, 
subject to the competing processes of advection and diffusion.
This also changes the launching conditions, as the overall profiles of plasma-$\beta$, 
resp. magnetization, are affected.
Figure~\ref{fig:adve_diff} displays the time evolution of the magnetic flux surface $\Psi = 0.1$ for reference
simulation case 1, initially rooted at the radius $r = 2.0$.
We see that this flux surface is first moving (diffusing) outwards {"}driven{"} by magnetic pressure gradient and tension, 
and then, once disk accretion has established again moves inwards being advected with the mass flux.
After about $t=3000$ both processes balance and a quasi steady-state situation in the system evolution
is reached.
Considering simply flux conservation, we may estimate the change in poloidal magnetic field strength and
the corresponding change in magnetic diffusivity and plasma-$\beta$.

Estimating an average field strength ${\bar B}_{\rm p}$ and a corresponding magnetic flux
$\Psi \simeq {\bar B}_{\rm p} r^2$, 
flux conservation tells us that $\Psi_{\rm t=1000} \approx \psi_{\rm t=50}$.
Therefore, ${\bar B}_{\rm p,t=1000} \approx 10 \;B_{\rm p,t=50}$, 
since this flux surface (e.g. $\Psi = 0.1$) is now rooted at a different radius.
With the increased poloidal magnetic field strength the Alfv\'en speed $v_{\rm A} \sim B_{\rm p} $ 
increases, and, thus, also the magnetic diffusivity parameterized as $\eta \sim v_{\rm A}$.
Thus, we expect that in previous work (e.g. \citet{Zanni2007, Tzeferacos2009}) 
the actual (time-dependent) value of the magnetic diffusivity may differ substantially from its 
initial value.
We believe that this will impact the derived mass fluxes.
A similar estimate can be made for the plasma-$\beta$ or magnetization. 
Since $\beta \sim B^{-2}$, the increase in plasma-$\beta$ is by a factor 100, which also is expected 
to affect the jet formation severely.
We will come back to this point later when we compare different parameter runs.

%==========================================================================================================
\subsection{Accretion rate and ejection efficiency}
The main goal of this paper is to investigate what fraction of the mass accretion
is loaded into the outflow,
and how the mass loading is affected by various disk parameters.
The derived mass ejection-to-accretion ratio can be an important ingredient for studies of
AGN or YSO feedback, or another similar self-regulating outflow scenario.

Efficient accretion is feasible only if angular momentum is sufficiently removed. 
Since we do not consider a physical viscosity in our simulations, angular momentum removal from the disk is 
mostly accomplished by the torque of the magnetized disk wind. 
Therefore, in our simulations disk accretion can only work if a disk wind has been established.
In the following we discuss the inflow and outflow rates in our disk-jet simulations, 
concentrating first on reference run (case 1).

In order to calculate the integrated properties of inflow and outflow, such as mass flux
or angular momentum flux, we need to carefully select the integration domain\footnote{Averaging 
in time is applied for all the fluxes evolutions.}.
The derived fluxes depend strongly on the choice of the integration boundaries - thus, on 
how we distinguish between material belonging to {\em accretion} or {\em ejection}.
In reality there is no {\em disk surface}, but a smooth transition between accretion and 
ejection.
The initial setup of a thin disk with aspect ratio $\epsilon = 0.1$ dynamically evolves
in time, and so does the disk surface.
\begin{figure}
\centering
\includegraphics[width=0.7\columnwidth]{\figurepath/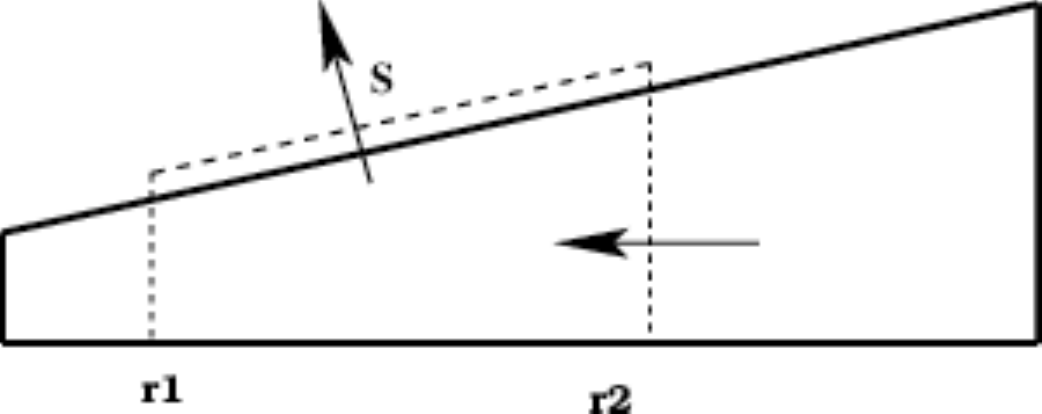}
\caption{Control volume to measure the accretion and ejection rates in the jet-disk structure.
Integration is at/between the radii $r_1$ and $r_2$ along the disk.
The integration surface denoted by $\rm{S}$ is inclined and is parallel to the initial disk 
surface.}
\label{fig:cv}
\end{figure}

Our control volume to measure the disk accretion and ejection rates is defined by 
an axisymmetric sector enclosed by two surfaces perpendicular to the equatorial plane 
located at $r=r_1$ and $r = r_2$, and two other surfaces being the disk mid plane,
and a surface $\rm{s_1}$ which is close by and parallel to the initial disk surface
(see Fig.~\ref{fig:cv}). We usually adopt $r_1 = 1.0$.

The accretion rate is then calculated as
\begin{equation}
 \dot M_{\rm acc}(r) = - 2 \pi r \int_0^{a H} \rho v_r dz,
\label{eq:acc}
\end{equation}
where the parameter $a$ controls the scale height of the integration, and $H = H(r,t=0) = \epsilon r$
is the initial thermal scale height of the disk.
Considering the evolution of the large-scale disk velocity (see Fig.~\ref{fig:accretion_layer}), 
we have chosen $a = 1.6$ for our reference simulation.
Similarly, we calculate the ejected mass flux as
\begin{equation}
 \dot M_{\rm ejec}(r)\arrowvert_{\rm S} =  \int_{r_1}^{r} \rho \vec{v}_{\rm p}\cdot d\vec{A}_{\rm S},
 \label{eq:ejec}
\end{equation}
where the integration is done along the inclined surface area S from $r_1$ to $r_2$ considering 
the area element $d\vec{A}_{\rm s}$.

\begin{figure}
\centering
\includegraphics[width=0.9\columnwidth]{\figurepath/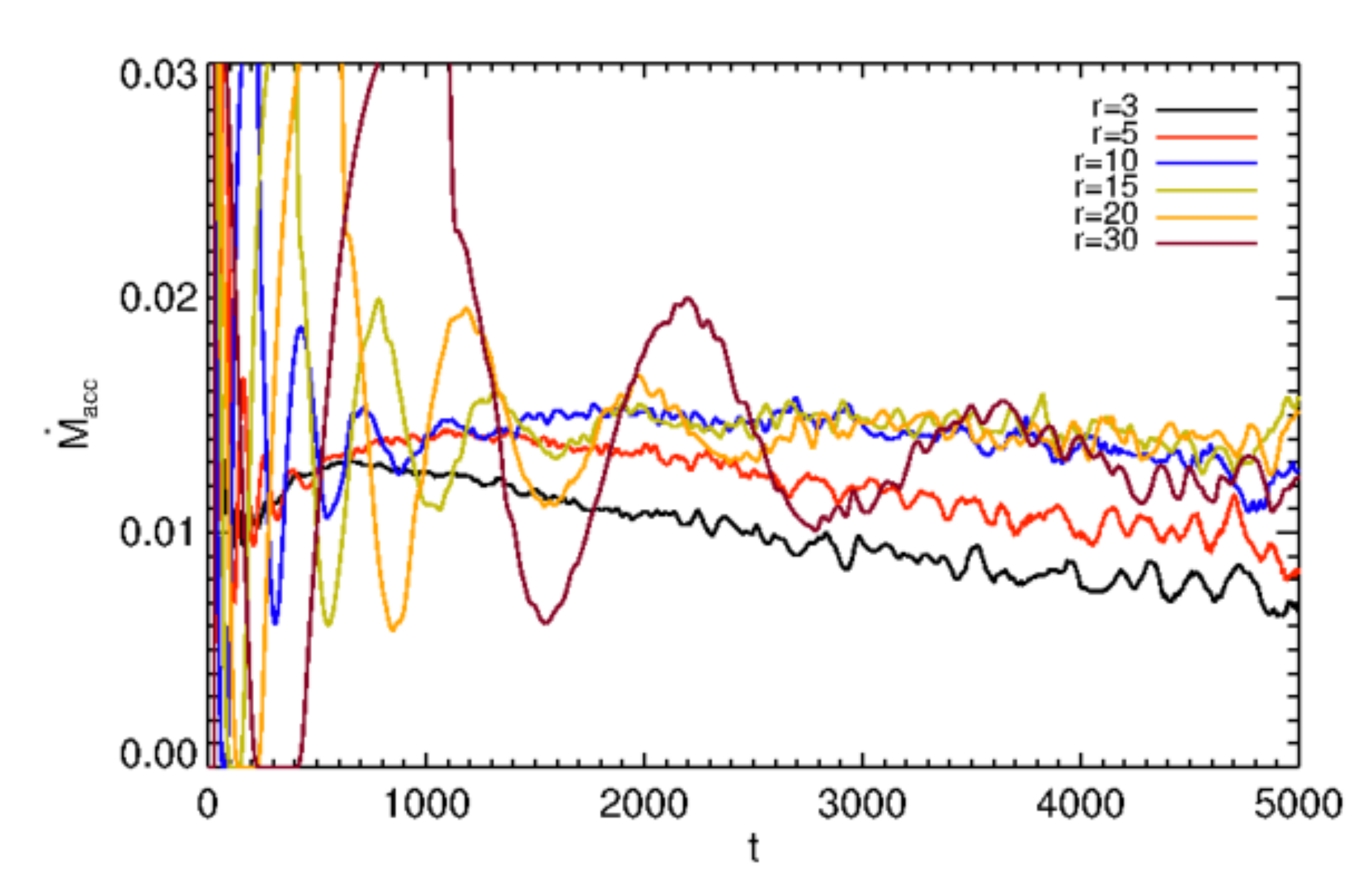}
\includegraphics[width=0.9\columnwidth]{\figurepath/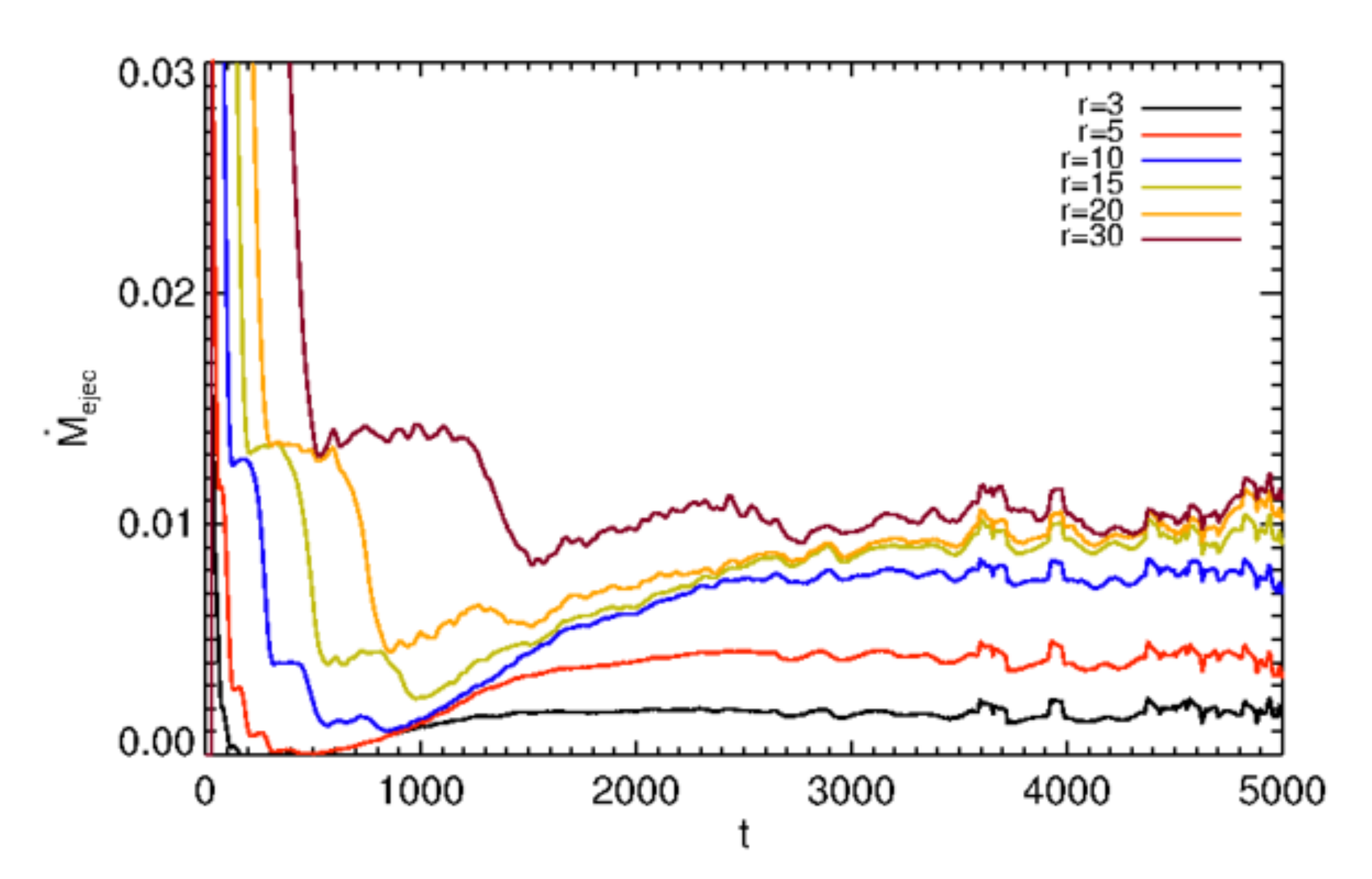}
\includegraphics[width=0.9\columnwidth]{\figurepath/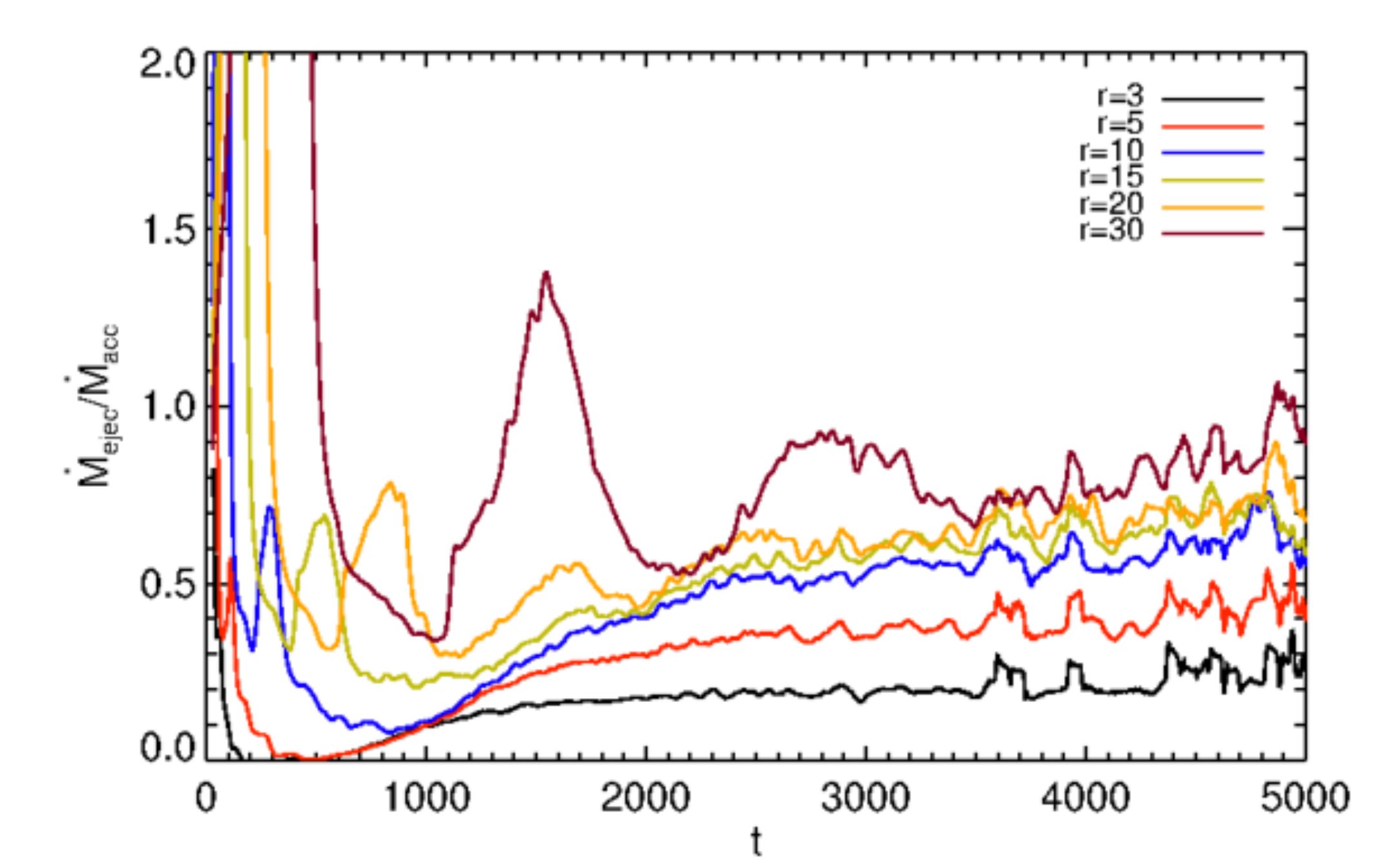}
\caption{Time-evolution of the mass fluxes. 
Shown is evolution of accretion rate (top), the ejection rate (middle), 
and the ejection-to-accretion ratio (bottom) for the reference simulation case 1,
all in code units.
Colors indicate mass fluxes calculated for increasing outer radii of the control volume,
$r_2 = 3, 5, 10, 15, 20, 30$. }
\label{fig:accr_plot}
\end{figure}

The time evolution of the accretion and ejection rate for the reference simulation is shown in 
Fig.~\ref{fig:accr_plot}.
We have calculated the accretion and ejection rates for different sizes of the integration domain 
(thus with different outer radius $r_2$).
We observe that for small radii the accretion rate saturates at an approximately constant value.
For the outer parts of the disk, however, a longer dynamical time is required for the disk to 
evolve into a new dynamical steady state.
This is visible in the time evolution of the accretion rates - for increasing radii $r_2 = 3, 5, 10$, 
the time when the plateau phase is reached, is increasing to $t= 300, 800, 1500$.
Beyond the plateau phase, the mass fluxes slightly decrease, most probably due to the overall decrease 
of the disk mass itself, and the subsequent change in the internal disk dynamics.
The accretion rates at large radii are larger than those at smaller radii - the mass
difference is ejected as outflow.
For the control volume with larger $r_2$ the ejection rate increases, which is simply a consequence of 
the increased integration area. 

The last panel in Fig.~\ref{fig:accr_plot} shows the ejection-to-accretion rate 
$\dot{M}_{\rm ejec}/\dot{M}_{\rm acc}$.
Again, depending on the size of the control volume this fraction changes.
Although the outflow quickly evolves from the disk surface, and extends soon to large radii, 
several orbital periods are required to establish full accretion.
At earlier times and for large radii, the accretion process is not fully established, 
resulting in a somewhat arbitrarily high ejection-to-accretion ratios even above unity.
For the inner part of the disk within radii $< 10 $, about 60\% of the accreting material 
is diverted into the outflow for our reference simulation.
This is a large fraction and similar to simulations in the literature, but definitely more than
derived in steady state models \citep{Pelletier1992, Ferreira1995}.
For the other cases we have investigated, also smaller rates were obtained (see below).

Applying a radially self-similar approximation of the MHD equations,
\citet{Ferreira1995} have introduced a so-called {\em ejection index} $\xi$,
\begin{equation}
  \frac{\dot M_{\rm ejec}}{\dot M_{\rm acc}} = 1 - \left( \frac{r_{\rm i}}{r_{\rm e}} \right)^\xi,
\end{equation}
essentially resulting from local mass conservation (where $r_{\rm i}= r_1$ and $r_{\rm e} = r_2$ in our notation).
With self-similar solutions \citet{Ferreira1997} constrained the ejection index to $ 0.04 < \xi < 0.08$.
In comparison, for our reference run we find both larger numerical values and also a wider range for 
the ejection index, $0.1 < \xi < 0.5 $ . 
Table \ref{tbl:cases} provides an overview over the ejection indexes we measure.
We can apply our reference run to different astrophysical sources (see \ref{app:units-normalization}), 
thus for a  stellar jet with the central
mass $\approx 1 M _\odot$ and $\rho_{\rm i} \approx 10^{-10}$, the  accretion rate is 
$\approx  10^{-6} M_\odot {{\rm yr} ^{-1}}$. 
This value for AGN with the central mass $\approx 10^8 M_{\odot}$ and $\rho_{\rm i} \approx 10^{-12}$ is about $ \approx 7  M_\odot {{\rm yr} ^{-1}}$.
%
%=================================================================================================================
\subsection{Launching forces and outflow formation}
In order to investigate the launching, the acceleration and the collimation processes of the outflow, we now
consider the forces acting in the disk-outflow system.%

\begin{figure}
\centering
\includegraphics[width=0.9\columnwidth]{\figurepath/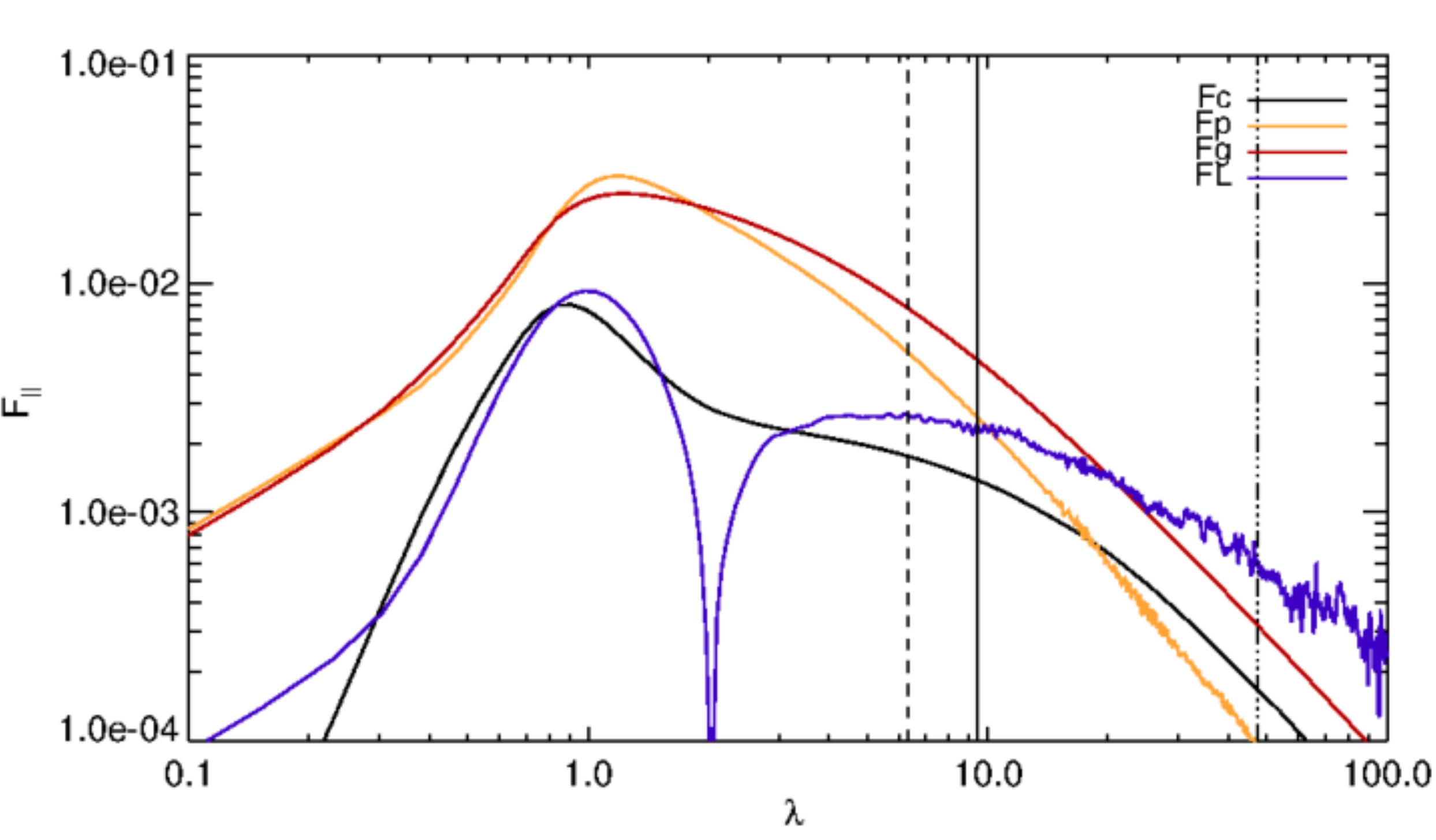}
\includegraphics[width=0.9\columnwidth]{\figurepath/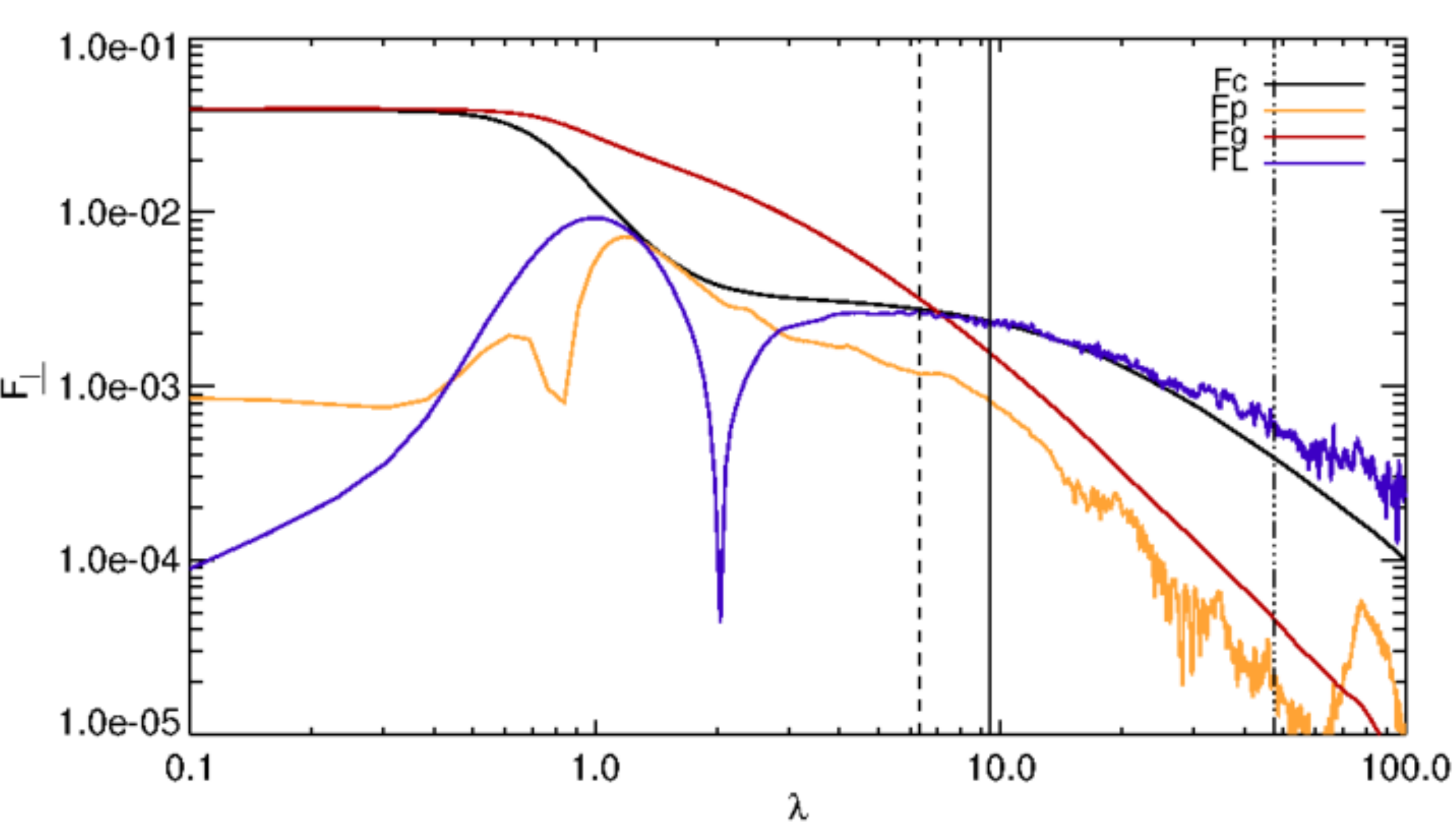}
\caption{Accelerating and collimating forces. 
Shown are the profiles of the parallel (top) and  perpendicular (bottom) specific forces
in logarithmic scale for simulation case 1 at time $t=1000$ along the distance $\lambda $ 
of the field line rooted at (5,0).
Here $F_{\rm C}, F_{\rm G}, F_{\rm L}, F_{\rm P} $ 
denote the centrifugal force, gravity, the Lorentz and the gas pressure gradient forces ( all 
in absolute value), respectively. 
The vertical lines present the slow-magnetosonic (dashed line), the Alfv\'en (solid), and the 
fast-magnetosonic (dot-dashed) point, respectively.}
\label{fig:forces_case1}
\end{figure}

To identify the forces for {\em acceleration and collimation} explicitly, it is preferable to project
them along or perpendicular to a certain flux surface, respectively.
By comparing these projected force components, we may disentangle the dominant driving and collimation 
mechanism.
Figure \ref{fig:forces_case1} shows the force components along a flux surface (field line) rooted at $(5,0)$,
and also the critical MHD surfaces - the slow-magneto-sonic, the Alfv\'en, and the fast-magneto-sonic surface.
The forces involved in {\em driving the outflow} are the centrifugal force $F_{\rm C}$, the Lorentz force
$F_{\rm L}$, the gas pressure gradient $F_{\rm P}$, and gravity $F_{\rm G}$. 
Among them, the pressure gradient and the centrifugal force are de-collimating while gravity and Lorentz force 
have collimating components. 
The pressure gradient, centrifugal force and Lorentz force also contribute to accelerate the outflow.
In agreement with the literature we see that upstream the slow magneto-sonic point, the acceleration is 
mainly by the gas pressure gradient.
For the super-slow outflow the Lorentz force and the centrifugal force dominate.
For the super-fast flow the Lorentz force plays a major role.
In case of the collimating forces the situation is different.
Before the Alfv\'en point, the de-collimating centrifugal force dominates.
Beyond the Alfv\'en point the Lorentz force has a comparable contribution, 
and finally after the fast-magneto-sonic point, the Lorentz force dominates
and collimates the outflow.
The Lorentz force behavior is a sign of the electric current distribution in the disk/jet system. 
It compresses the disk while is collimating the corona.

The forces involved in {\em launching the outflow} are mainly the vertical gas pressure gradient
which counteracts the tension forces of the poloidal field and gravity
(see Fig.~\ref{fig:forces_case1}).
The thermal pressure gradient is always positive, and thus supports launching.
It increases from the disk mid plane to the surface and then decreases in the corona.
The tension term is mostly negative, thus compressing the disk, and does not support launching.
The magnetic pressure (toroidal component and total) are both positive above the disk, and, thus,
may support the launching process.
The toroidal component of the Lorentz force $F_{\phi,\rm L}$ provides the magnetic torque braking the 
disk, and loading the outflow with both angular momentum and energy (not shown).
In order to brake the disk, the torque must be negative in the disk and change sign at the disk
surface \citep{Ferreira1997}. This is confirmed by our simulation.

\begin{figure}
\centering
\includegraphics[width=0.9\columnwidth]{\figurepath/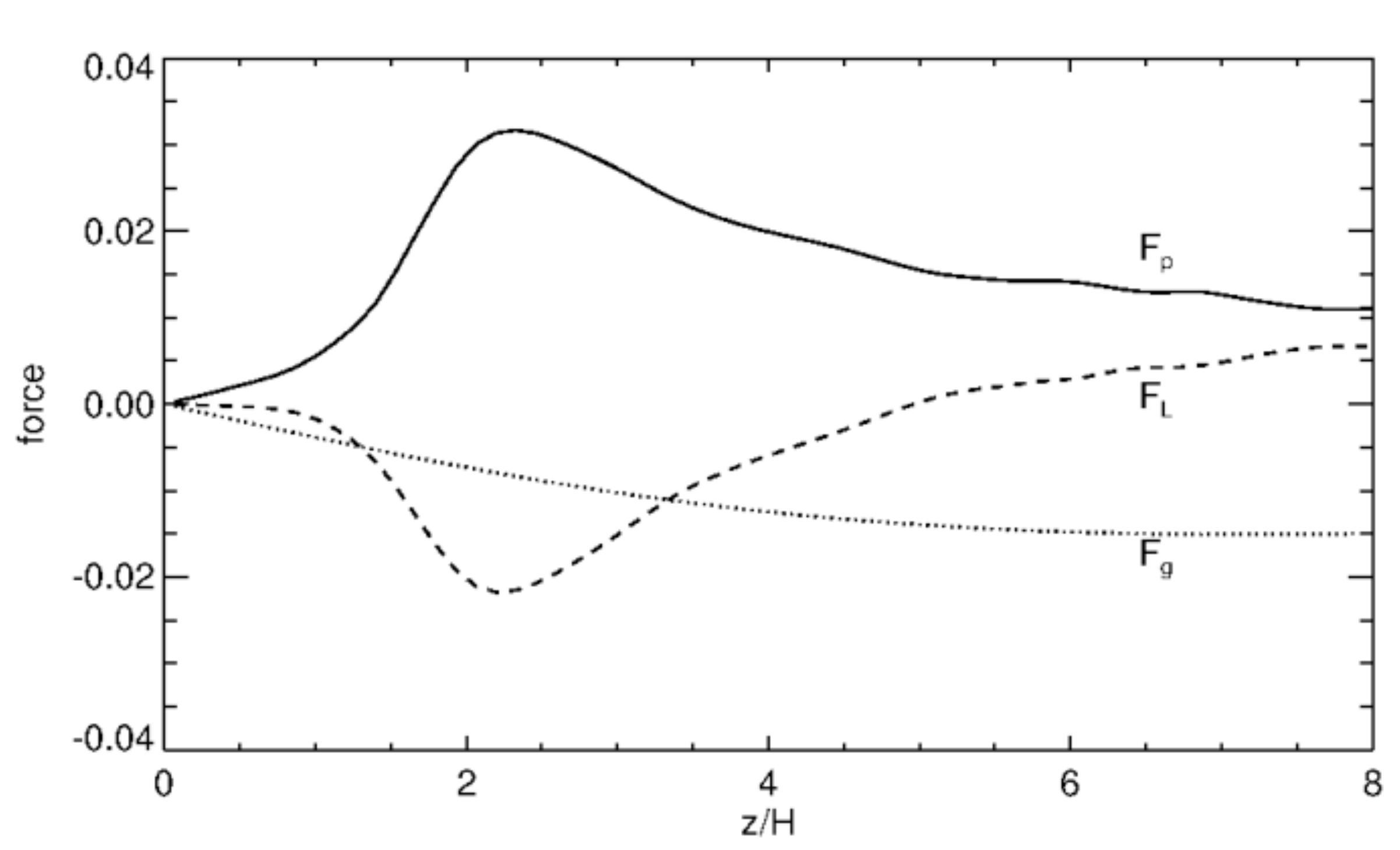}
\hspace{0.5 cm}
\includegraphics[width=0.9\columnwidth]{\figurepath/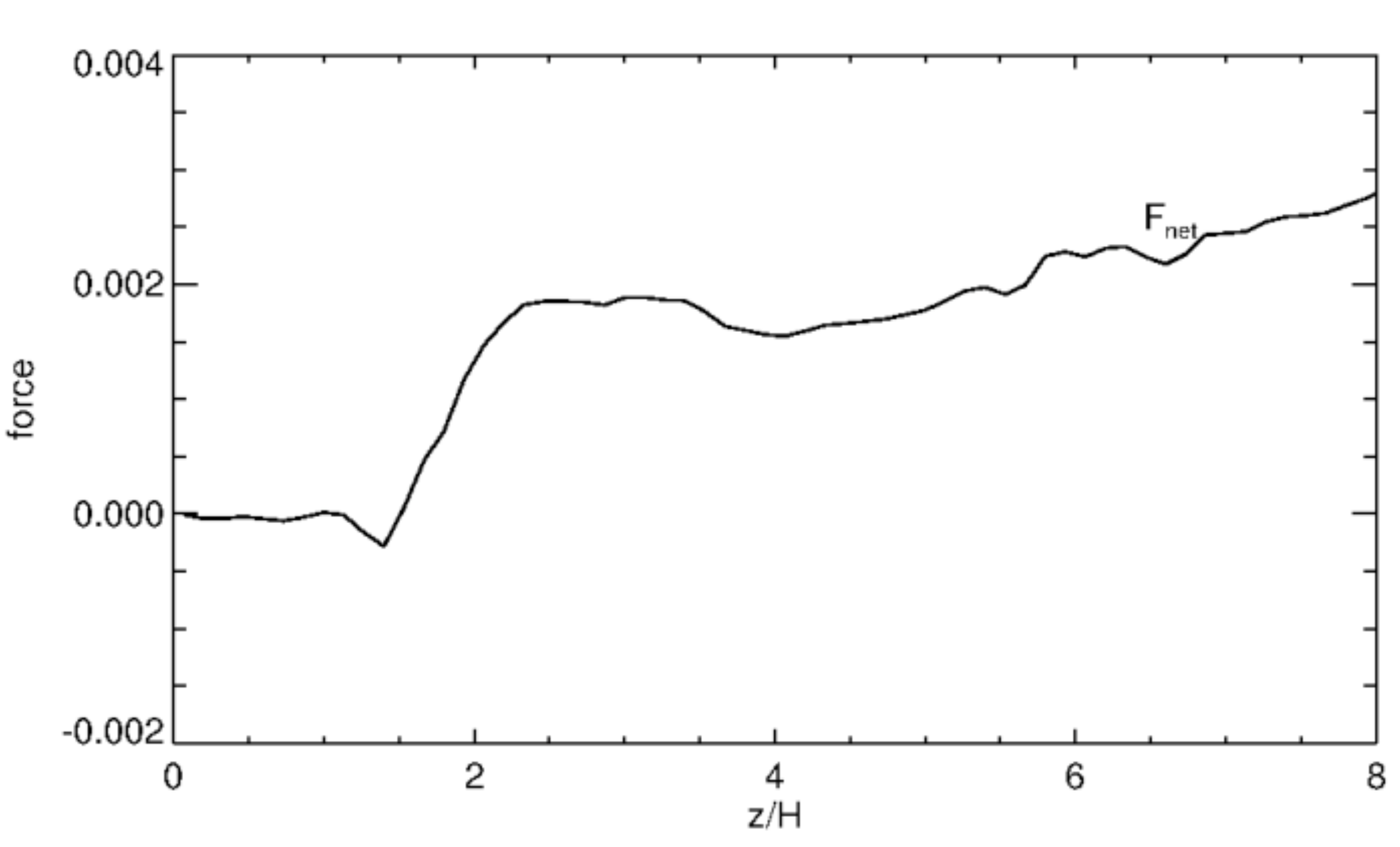}
\caption{Launching forces of the outflow. 
Shown is the vertical profile of the vertical components of gravity, 
the pressure gradient force, and the Lorentz force (top), as well as the vertical profile of 
the net force (bottom) at time $t=1000$ at a radius $r=5$ for in case 1.}
\label{fig:launching-forces2}
\end{figure}

Figure \ref{fig:launching-forces2} show for comparison the three main force components (top) and 
the net force (bottom). 
We see that in the disk the {\em vertical force components} almost balance, while above the disk
a considerable
net force remains which launches and accelerates the outflow.
In the disk the (positive) gas pressure is almost balanced by the (negative) Lorentz force.
However, the gas pressure slightly wins and counteracts gravity.
Essentially the figure shows that launching is a process which happens at the disk surface.
The outflow material is not lifted from the mid plane into the disk wind.
It is the disk material which is accreting along the disk surface which is loaded into the 
outflow.
The net (specific) force which is responsible for launching and initial acceleration is about 10\% of 
the value of the single force components.

\begin{table*}
\caption {Input parameters and derived dynamical parameters of our simulation runs following
a magnetic diffusivity profile Eq.~\ref{eq:magdiff} with a scale height
of the disk diffusivity $\epsilon_\eta =0.4$.
Displayed are the input parameters of grid resolution $\Delta r, \Delta z$, 
magnetic poloidal and toroidal diffusivity $\eta_{\rm p,i}$ and $\eta_{\phi,\rm i}$, 
the diffusivity anisotropy $\chi$, 
and the plasma-beta parameter $\beta$.
The following outflow parameters are derived from our numerical simulations:
the asymptotic jet radius $r_{\rm jet}$, calculated as mass flux weighted (see Eq.~\ref{eq:rjet}), 
the launching radius $r_{\rm l}$,
the typical speed of the outflow, $v_{\rm jet}$, 
the average collimation degree $\zeta_{\rm jet}$ (see Eq.~\ref{eq:angle}),
the mass accretion rate $\dot{M}_{\rm acc}$,
and mass ejection rate $\dot{M}_{\rm ejec}$, 
the ejection index $\xi$,
the diffusive scale height $\epsilon_\eta$
the kinetic and magnetic angular momentum losses by the outflow, $\dot{J}_{\rm kin}$ and $\dot{J}_{\rm mag}$,
respectively, as well as the total angular momentum loss $\dot{J}_{\rm tot}$,
and its asymptotic kinetic luminosity $L_{\rm kin} = 0.5 \dot{M}_z v_{\rm jet,z}^2$.
All parameters are given in code units.}
\begin{center}
 \begin{tabular}{cccccccccccccccccc}
\hline
\hline
\noalign{\smallskip}
                       & case 1  & case 2  & case 3  & case 4  & case 5   & case 6    & case 7    & case 8    & case 9 & case 10 \\
$\Delta r$             & 0.064   & 0.064   & 0.064   & 0.064   & 0.064    & 0.025     & 0.025     & 0.064     & 0.064  & 0.064  \\
$\Delta z$             & 0.066   & 0.066   & 0.066   & 0.066   & 0.066    & 0.025     & 0.025     & 0.066     & 0.066  & 0.066    \\
$\epsilon_\eta$        & 0.4     & 0.4     & 0.4     & 0.4     &  0.4     & 0.4       & 0.4       & 0.4       & 0.4    & 0.4      \\
$\eta_{\rm p,\rm i}$ & 0.03    & 0.01    & 0.15    & 0.09    & 0.03     & 0.03      & 0.03      & 0.03      & 0.03   &0.03   \\
$\eta_{\phi,\rm i} $ & 0.09    & 0.03    & 0.09    & 0.27    & 0.01     & 0.09      & 0.09      & 0.09      & 0.09   & 0.09   \\
$\chi$                 & 3       & 3       & 3/5     & 3       & 1/3      & 3         & 3         & 3         & 3      & 3     \\
$\beta$                & 10      & 10      & 10      & 10      & 10       & 5000      & 10        & 50.0      & 250    & 500    \\
\noalign{\smallskip}
\hline
\noalign{\smallskip}  
$r_{\rm jet,z=180}$   & 43.8   & 73.8    & 24.46  & 28.57    & 33.19   & --         & --          & 47.8     & 50.0    &  34.13    \\
$r_{\rm jet,z=60}$    & 25.01  & 38.41   &15.37     & 17.72    & 25.02   & 21.69     & 19.6       & 27.54    & 24.18   & 17.22      \\
$r_{\rm l}$           & 3.8    & 8.2     & 3.0      & 4.2      & 3.0     &  --       & 3.5        & 2.8      & 0.8     & 0.7        \\
\noalign{\smallskip} 
$v_{\rm jet,z=170}$           & 0.4     & 0.23   & 0.85    & 0.56      & 0.8      & 0.47   & 0.42       & 0.56    & 0.49 & 0.27       \\
$v_{\rm jet,z=280}$           & 0.45    & 0.3    & 0.92    & 0.61      & 0.9      & --     & --         & 0.5     & 0.89  & 0.22  \\
\hline
\noalign{\smallskip} 
$\zeta_{\rm jet,1}$  \footnotemark   & 2.37  & 0.013 & 5.28     & 5.1   & 2.58  &  0.89  &  4.6    & 1.56   & 0.32  & 2.08   \\
\noalign{\smallskip} 
$\dot M_z$         & 0.0005 & $6\times 10^{-6}$ & 0.0002 &  0.0002  & 0.0007 & $5\times 10^{-5}$ & 0.0008 & 0.0001   &0.0004  & 0.0002  \\
\noalign{\smallskip} 
\noalign{\smallskip}
$\zeta_{\rm jet,2}$  & 2.90   & 0.3                 & 4.45     & 6.7   & 2.28    &  --       &  --    & 2.46    & 0.42  & 1.84   \\
\noalign{\smallskip} 
$\dot M_z$        & 0.001   & 0.0002              & 0.0003   & 0.0004 & 0.001   & --       & --     &  0.0003 & 0.0006 & 0.0004  \\     
\noalign{\smallskip}   
\hline
\noalign{\smallskip}
$\dot{M}_{\rm acc}$ \footnotemark          &   0.015 & 0.022 &  0.0038 & 0.0035   & 0.022    & 0.001 & 0.013    & 0.005   & --  & --\\
$\dot{M}_{\rm ejec}$                       &  0.008  & 0.016 &  0.001  &  0.001   & 0.007    & 0.005 & 0.009    & 0.004   & --  & --  \\
\noalign{\smallskip}
${\dot{M}_{\rm ejec}}/{\dot{M}_{\rm acc}}$ &  0.5    & 0.7   &  0.2    &  0.2     & 0.31     & --    & 0.6      & --     & --  & --\\
\noalign{\smallskip} 
$\xi $                                     &  0.3    & 0.5   &  0.09  &  0.09   & 0.1      &  --   & 0.39     & --      & --  & -- \\
$\dot{J}_{\rm kin}$                        & 0.008   & 0.005 & 0.002   & 0.002    & 0.027    & 0.005 & 0.005    & 0.008 & --  & --   \\
$\dot{J}_{\rm mag}$                        & 0.03    & 0.01  & 0.0025  & 0.005    & 0.034    & 0.0   & 0.016    & 0.008   & 0.0 & 0.0\\
$\dot{J}_{\rm tot}$                        & 0.038   & 0.015 & 0.004   & 0.007    & 0.06     & 0.005 & 0.021    & 0.017   &--   & --\\
\noalign{\smallskip} 
\hline
\noalign{\smallskip} 
${L}_{\rm kin}$    & $1\times 10^{-4}$& $9\times 10^{-6}$& $1.3\times 10^{-4}$& $7.4\times 10^{-5}$& $4\times 10^{-4}$& $5.5\times 10^{-6}$& $7.1\times 10^{-5}$& $3.7\times 10^{-5}$& $1.9\times 10^{-4}$& $10^{-6}$\\
\hline 
\end{tabular}
 \end{center}
\label{tbl:cases}
\tnote[a)]{The collimation degree is measured from two different areas, $\zeta_{\rm jet,1}$ enclosing 
$0\leq r \leq 40, 0\leq z \leq 160$,and $\zeta_{\rm jet,2}$ with the area $0\leq r \leq80, 0\leq z \leq 250$.}\\
\tnote[b)]{The mass flux values for the two simulations case 9 and 10 with high $\beta_{\rm i}$ are omitted 
due to their highly perturbed behavior.}
\end{table*}

\begin{figure*}
\centering
\includegraphics[width=3cm]{\figurepath/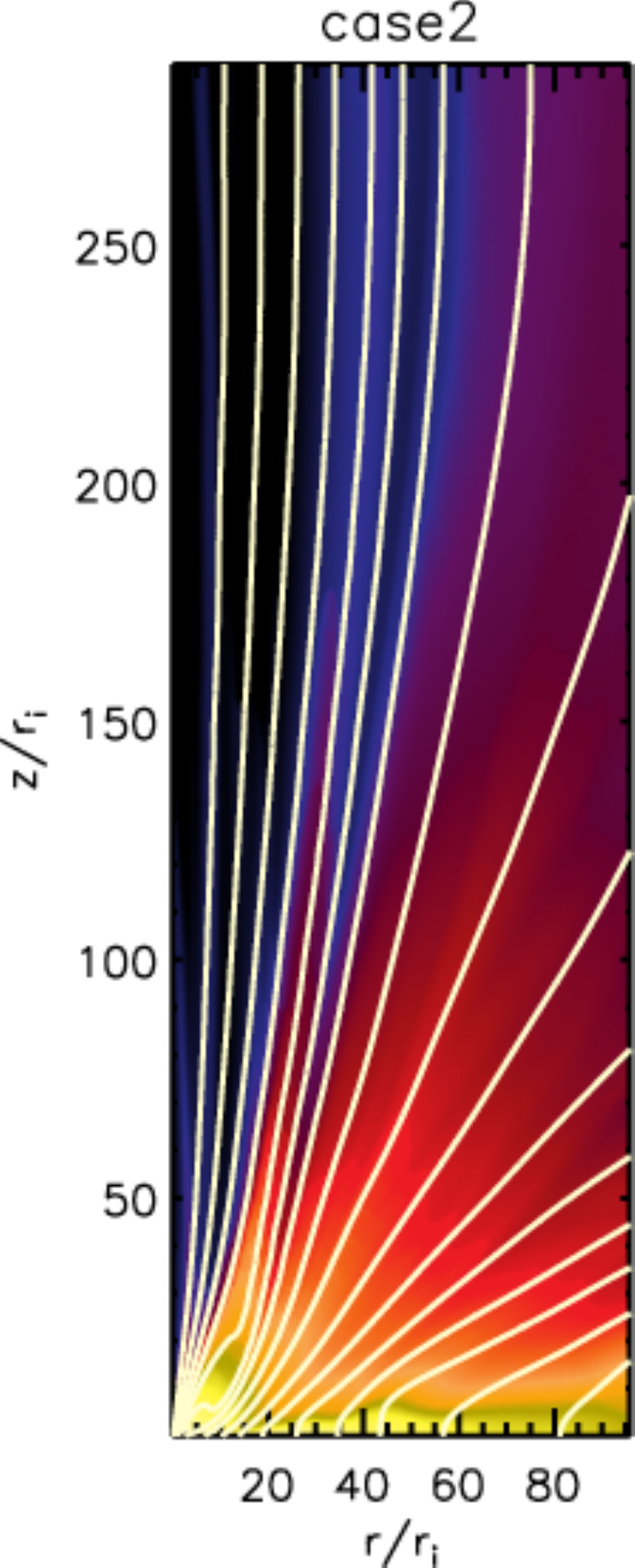}
\includegraphics[width=3cm]{\figurepath/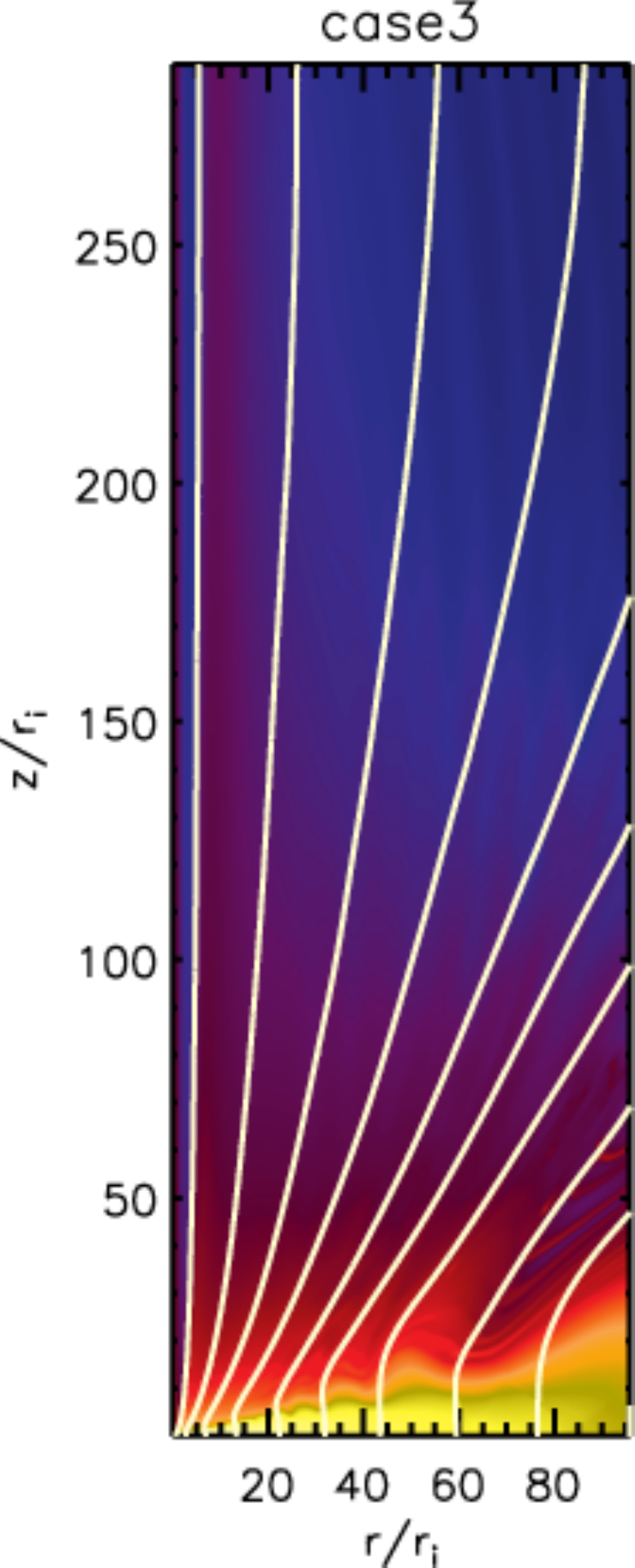}
\includegraphics[width=3cm]{\figurepath/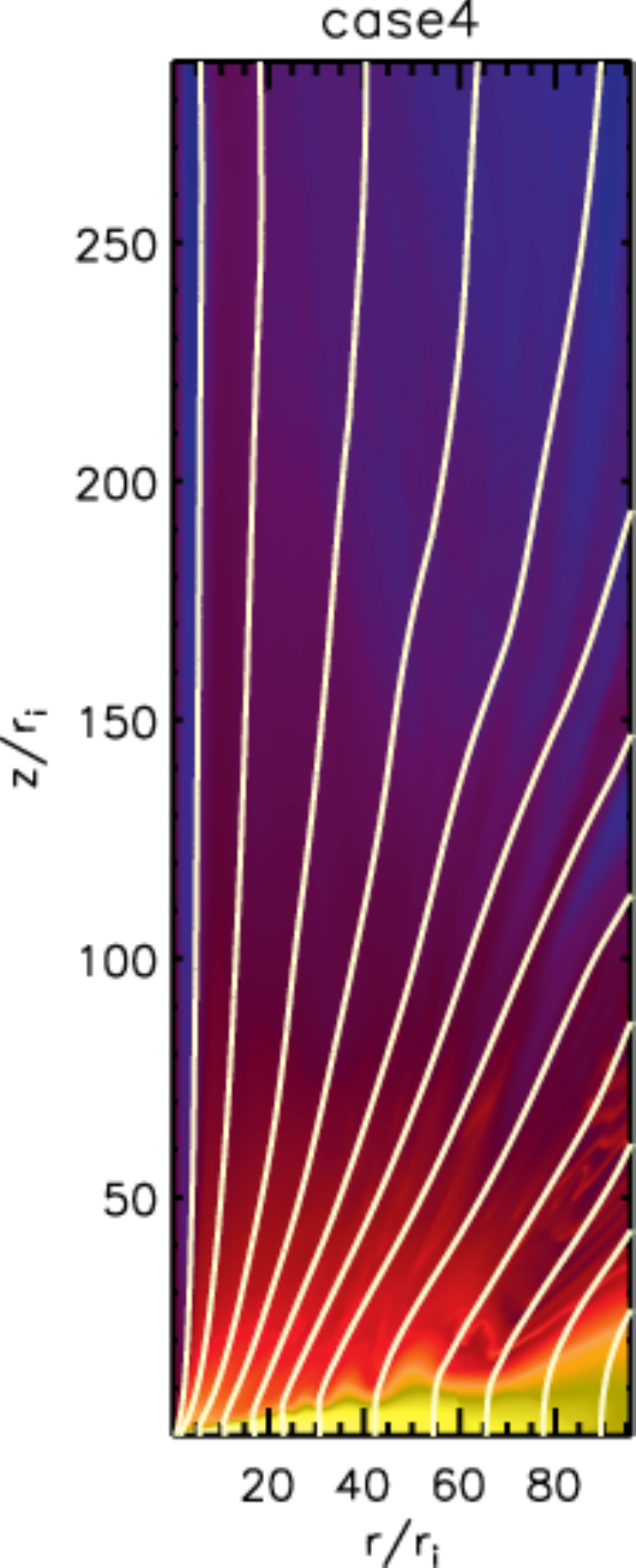}
\includegraphics[width=3cm]{\figurepath/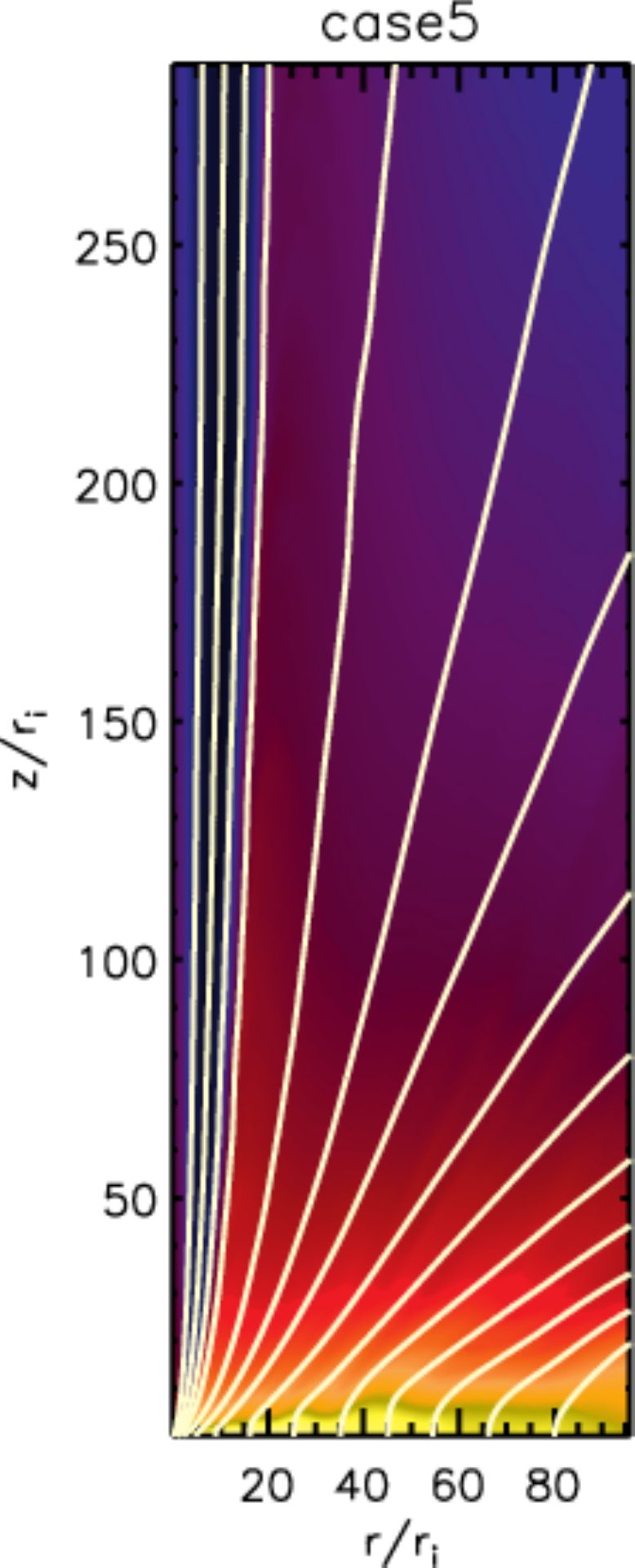}\\
\includegraphics[width=3cm]{\figurepath/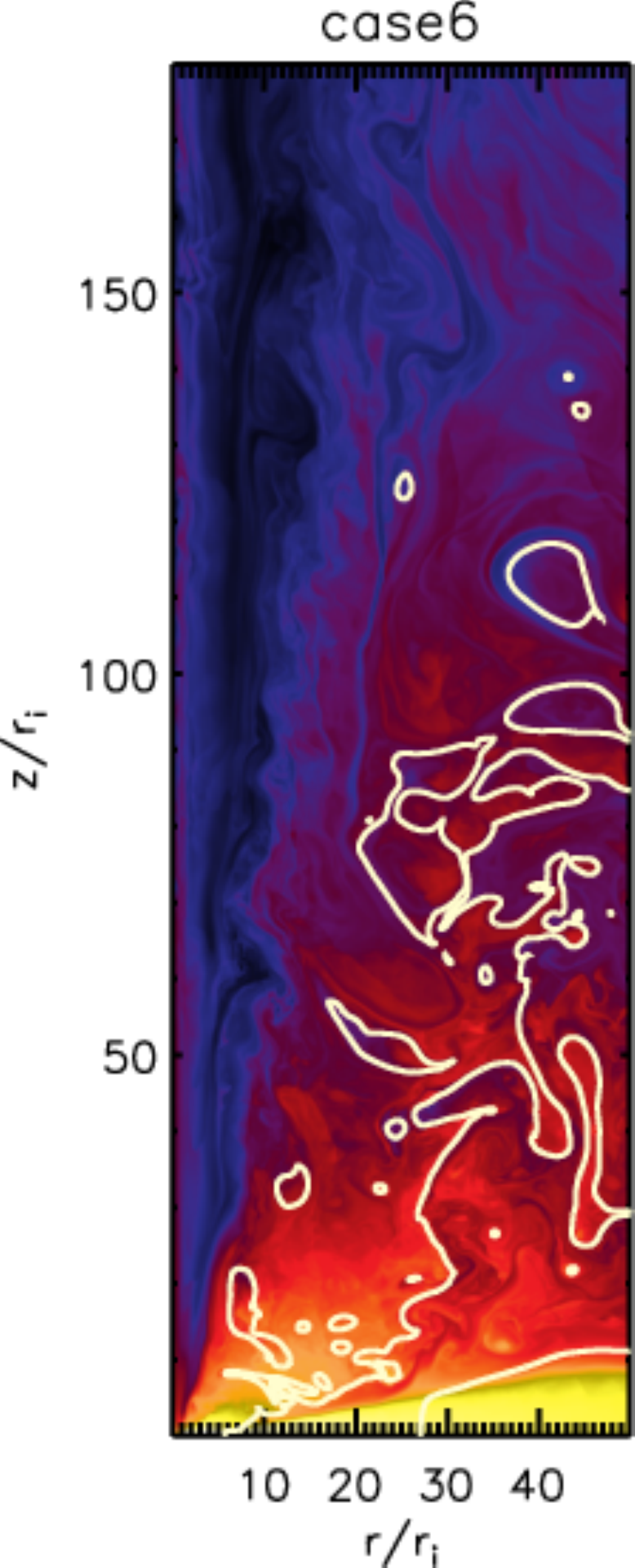}
\includegraphics[width=3cm]{\figurepath/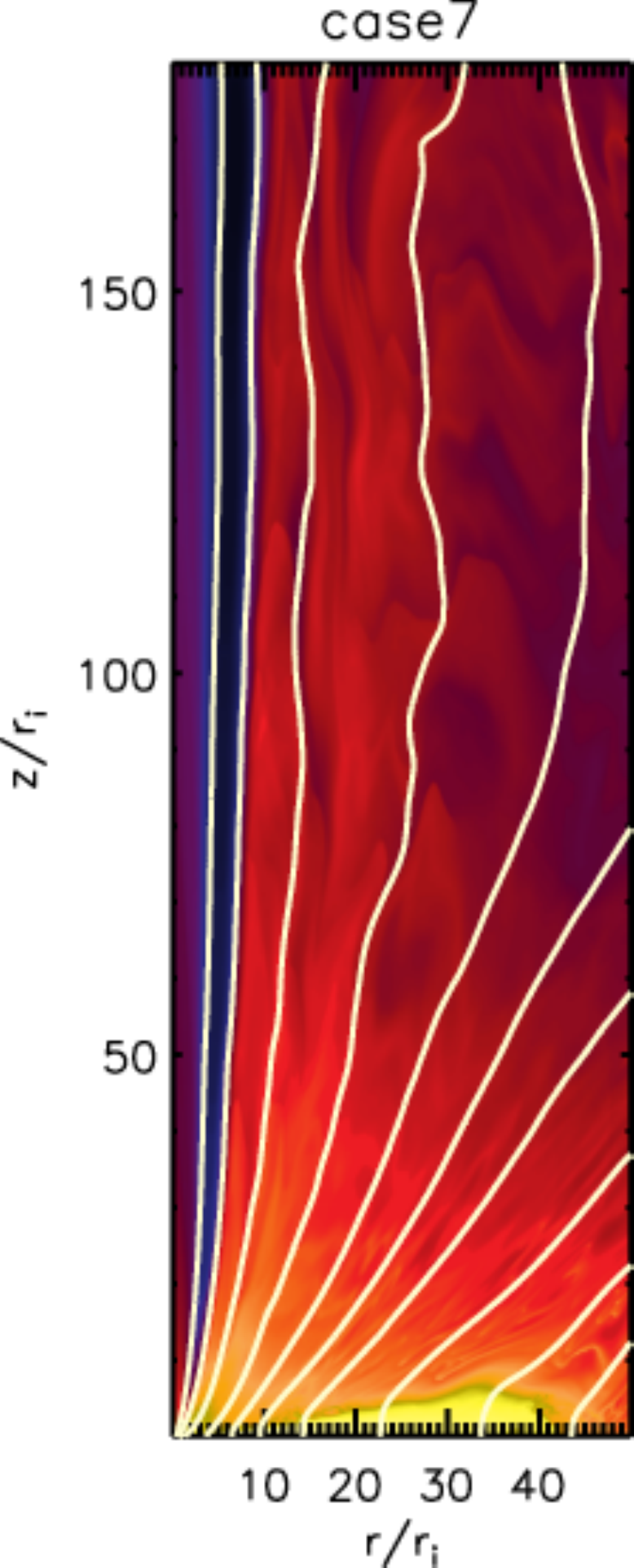}
\includegraphics[width=3cm]{\figurepath/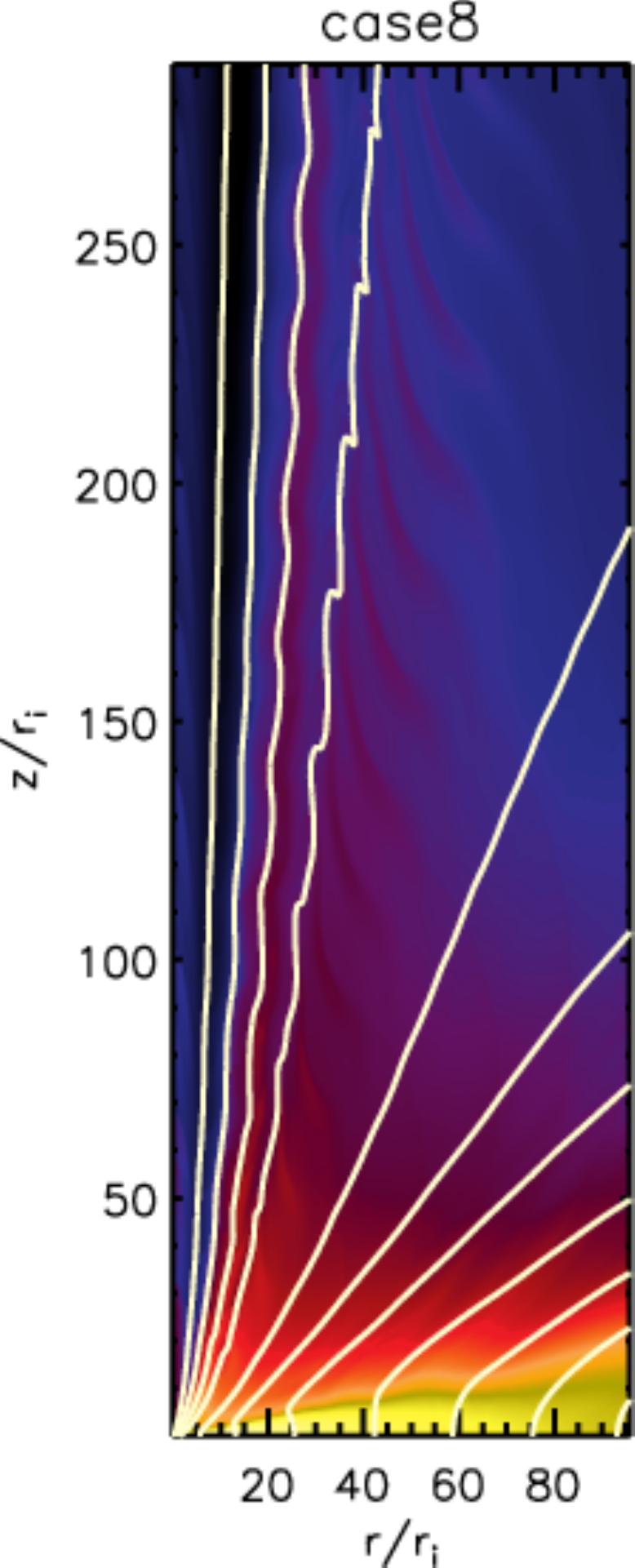}
\includegraphics[width=3cm]{\figurepath/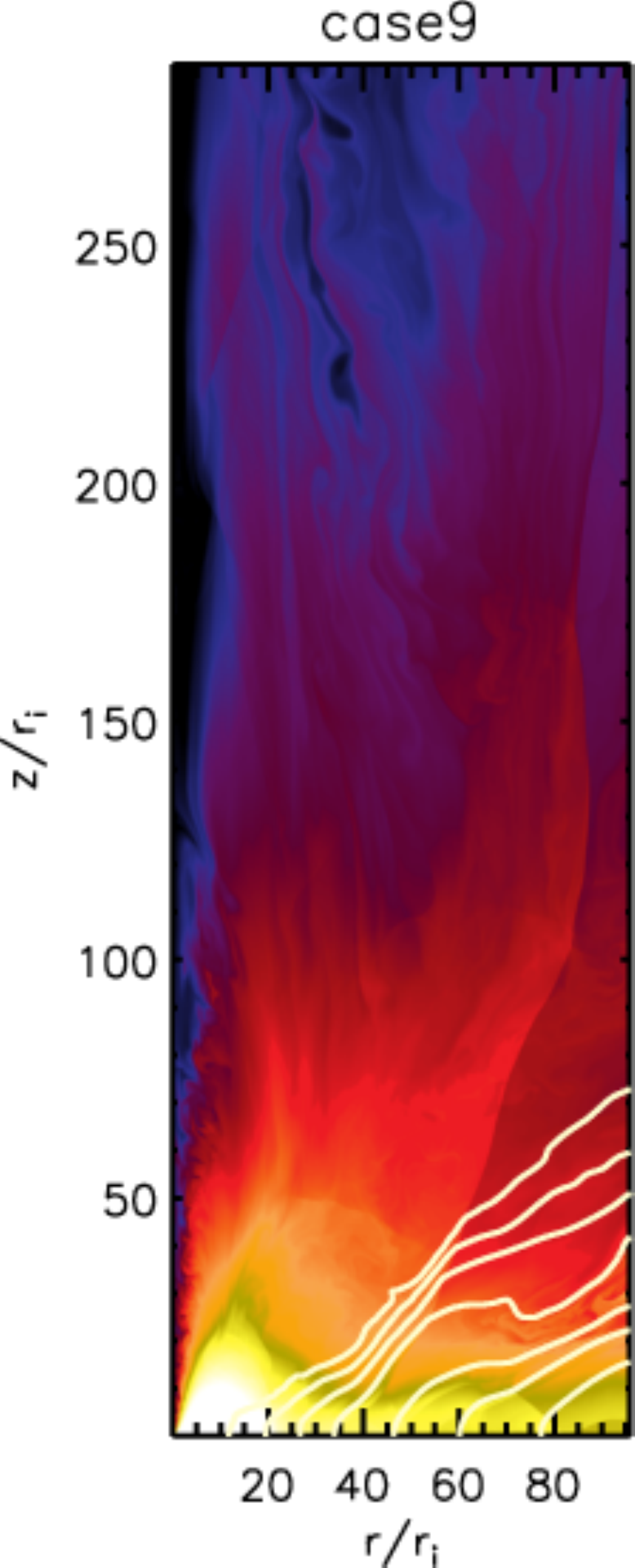} \\
\includegraphics[width=5cm]{\figurepath/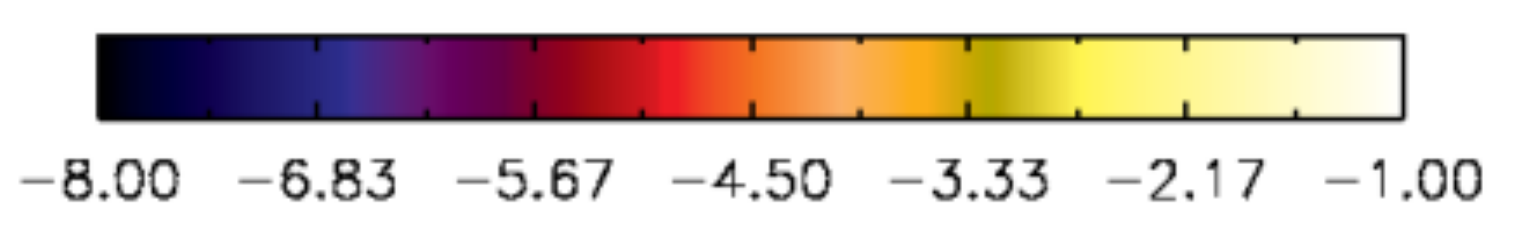}
\caption{Comparison of  different parameters runs (see Tab.~\ref{tbl:cases})at $t=3000$.
Shown is the density distribution  nd the poloidal magnetic field (contours of magnetic flux for levels
$\Psi =  0.01, 0.03, 0.06, 0.1, 0.15, 0.2, 0.26, 0.35, 0.45, 0.55, 0.65, 0.75, 0.85,  0.95, 1.1, 1.3, 1.5, 1.7$)
for simulation runs with 
rather low diffusivity (case 2), 
rather  high poloidal diffusivity with anisotropy $\chi=3/5$ (case  3), 
rather high diffusivity  strength (case 4), 
rather low toroidal diffusivity with anisotropy $\chi=1/3$ (case  5), 
simulations with a rather weak magnetic field with plasma-$\beta  = 5000$ (case 6), plasma-$\beta=100$ (case  8), 
plasma-$\beta=250$ (case  9), and also a simulation with three times higher resolution (case 7),
all to be compared to our reference simulation case 1, see Fig.~\ref{fig:rhocase1}.}
\label{fig:rho_parameterrun}
\end{figure*}
%----------------------------------------------------------------------------------------------------------------------
\subsection{Jet radius and opening angle}
The radius of the asymptotic jet and its opening angle can be measured by the observations
rather easily.
Therefore it is interesting to provide a comparison from the simulations.
Simple energy conservation arguments tell us that jets with the observed kinetic energy must
originate from a disk area very close  to the central object. 
This is a fact which seem to be hold for all astrophysical jet sources.
We will later consider the question of how these asymptotic properties are related to the
conditions of the jet launching process.
Here we first provide a clear numerical definition of these properties and apply them to
our reference simulation.
In order to do a quantitative comparison, we need to define a {\em "radius of the outflow"}. 
We suggest a definition of an {\em axial mass flux weighted} jet radius, measured at a certain  
distance $z = z_{\rm m}$ from the source (see also \citep{Porth2010}),
\begin{equation}
r_{\rm jet}\arrowvert_{z_{\rm m}} = 
\frac{\left[\int_0^{r_{\rm m}} r \rho v_{\rm z} dr\right]_{z_{\rm m}}}
     {\left[\int_0^{r_{\rm m}}   \rho v_{\rm z} dr\right]_{z_{\rm m}}}.
\label{eq:rjet}
\end{equation}
This radius measures the bulk of the mass flux contained in the jet at a specific distance 
$z_{\rm m}$ from the mid plane.
In order to derive the jet launching area, we now adopt that flux surface which follows
the bulk of the mass flux, thus passing through the point 
$(r_{\rm jet}(z_{\rm m}),z_{\rm m})$, and trace the same flux surface back to the disk 
surface where it is rooted.
This foot point radius defines the launching area of the outflow.
For reference run case 1 we find a mass flux weighted asymptotic jet radius for the high-velocity component of
$r_{\rm  jet} \simeq 44$ at $z_{\rm m} = 180$, corresponding to $\simeq 4$\,AU 
for a protostellar scaling applying $r_{\rm i} = 0.1 $\,AU.
For the jet launching radius we measure $r_{\rm l} \sim 0.4$ AU in our reference run.
Similarly, the launching area for the extended low velocity component (LVC) of the DG Tauri wind measured
by \cite{Anderson2003} extending from $ \sim 0.3 \;{\rm to} \; 4$ AU from the star.
They have invented a method relying on the MHD energy and angular momentum conservation along the jet.
By comparing the observed kinetic energy and jet rotation they infer the necessary disk rotation and 
launching area in the wind-launching region.

The asymptotic jet opening angle is the other jet characteristics. 
One way to measure the jet opening angle is by the inclination angle of the flux surface defined by the 
asymptotic jet radius as discussed above. 
Another measure of the collimation has been suggested by \citet{Fendt2006} who assigned 
an {\em average collimation degree} $\zeta$ of the outflow by comparing the vertical and radial mass fluxes 
in the outflow
(applying a proper normalization per surface area of a cylinder of height $z = z_{\rm m}$ and 
radius $r = r_{\rm m}$),
\begin{equation}
 \zeta_{\rm jet} = 2 \frac{z_{\rm m}}{r_{\rm m}}
                   \frac{ [\int_0^{r_{\rm m}} r         \rho v_{\rm z} dr]_{z_{\rm m}} }
                        { [\int_0^{z_{\rm m}} r_{\rm m} \rho v_{\rm r} dz]_{r_{\rm m}} },
\label{eq:angle}
\end{equation}
 where only positive poloidal velocities are considered.
Outflow collimation would simply imply that $\zeta_{\rm jet} > 1$.
For our reference run, we find $\zeta_{\rm jet} = 2.37$ implying that about two times more mass is propagating 
along the jet that away from the jet axis.
For comparison, the flux surface which encloses the bulk of the jet mass flux, has an (half) opening
angle of about $10\degr$ at $(r,z) = (44,180)$, but collimates slightly more further downstream.
We note that this definition is related also to the {\em concentration of mass flux} across the jet, and, thus, 
provides a somewhat different information than the opening angle of the field lines.
For example, a cylindrical jets with zero degree opening angle may have a narrow or a broad
radial density or mass flux profile. 
With our definition the {\em narrow mass flux} would be interpreted as {\em more collimated}.

%%%================================================================================================================
\section{Comparison of parameter runs}   
In this section we compare simulations governed by different input parameters which rely on a
 magnetic diffusivity profile following Eq.~\ref{eq:magdiff} with a  scale height
of the disk diffusivity $H_{\eta} = \epsilon_\eta r$, grid resolution $\Delta r, \Delta z$, 
maximum magnetic poloidal and toroidal diffusivity $\eta_{\rm p,i}$ and $\eta_{\phi,\rm i}$, 
the diffusivity anisotropy $\chi$, and the plasma-beta parameter $\beta$ (see Tab.~\ref{tbl:cases}).
We understand these parameters as the governing parameters of jet launching from accretion disks.
All simulations have been carried out up to $t = 3000$ (reference run up to $t = 5000$).
We found that this time is sufficient to reach a quasi steady state of accretion-ejection 
(roughly after $t = 1000$).

We first show a comparison of the global mass density distribution at $t=3000$ for simulation setups 
resulting from a different strength $\eta$ or different anisotropy $\chi$ of diffusivity, 
a different plasma-$\beta$, or a different grid resolution (Fig.~\ref{fig:rho_parameterrun}).
The immediate result is that both the disk structure and the dynamical evolution of the outflow, 
change substantially compared to the reference run (case 1).
Jet-like outflows have been formed in all cases, although in some cases like case 6 with $\beta = 5000$, 
or case 9 with $\beta=250$ the outflow appears highly filamentary.
Denser outflows are observed as in case 7 with higher resolution, or case 2 lower diffusivity, or
case 5 with lower toroidal diffusivity, while in other cases the outflow is more tenuous.
Outflow and disk evolution is interrelated -
a denser outflow thus implies to a geometrically thinner  accretion disk, as more of the accreting matter 
has been diverted into the outflow.
The smoothness of the outflow varies for the different setups.
We obtain a much more filamentary and perturbed structure for outflows for which
the (physical or numerical) diffusivity in the disk is lower. 
A similar statement can be made for a low magnetic field strength.

\begin{figure}
\centering
\includegraphics[width=0.9\columnwidth]{\figurepath/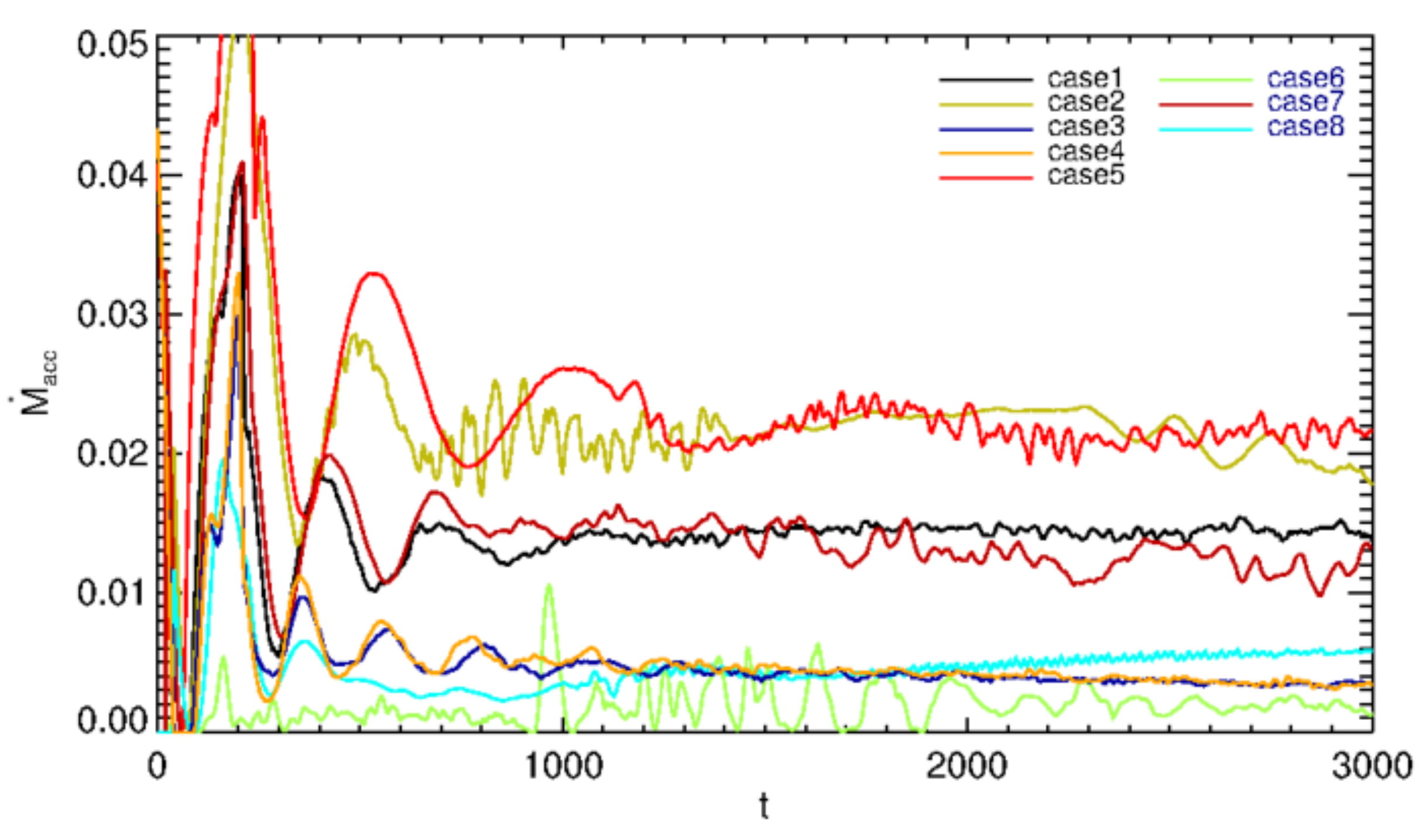}
\includegraphics[width=0.9\columnwidth]{\figurepath/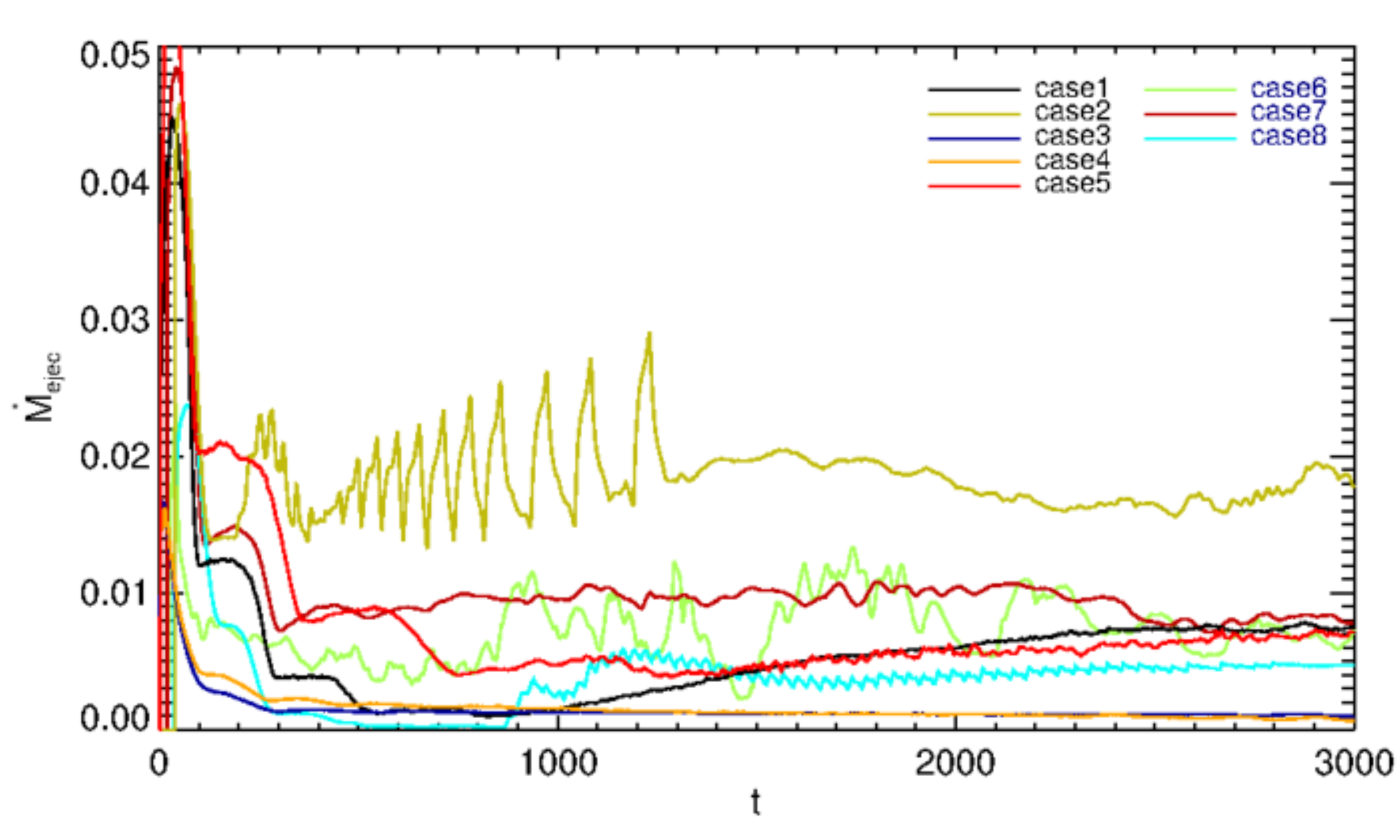}
\caption{Comparison of mass fluxes for the different cases. 
Shown is the time evolution of the accretion ejection rate (top) 
and the ejection rate (bottom) for simulation runs applying a different
parameter setup (see Tab.~\ref{tbl:cases}).}
\label{fig:flux_parameterrun}
\end{figure}

\begin{figure}
\centering
\includegraphics[width=0.9\columnwidth]{\figurepath/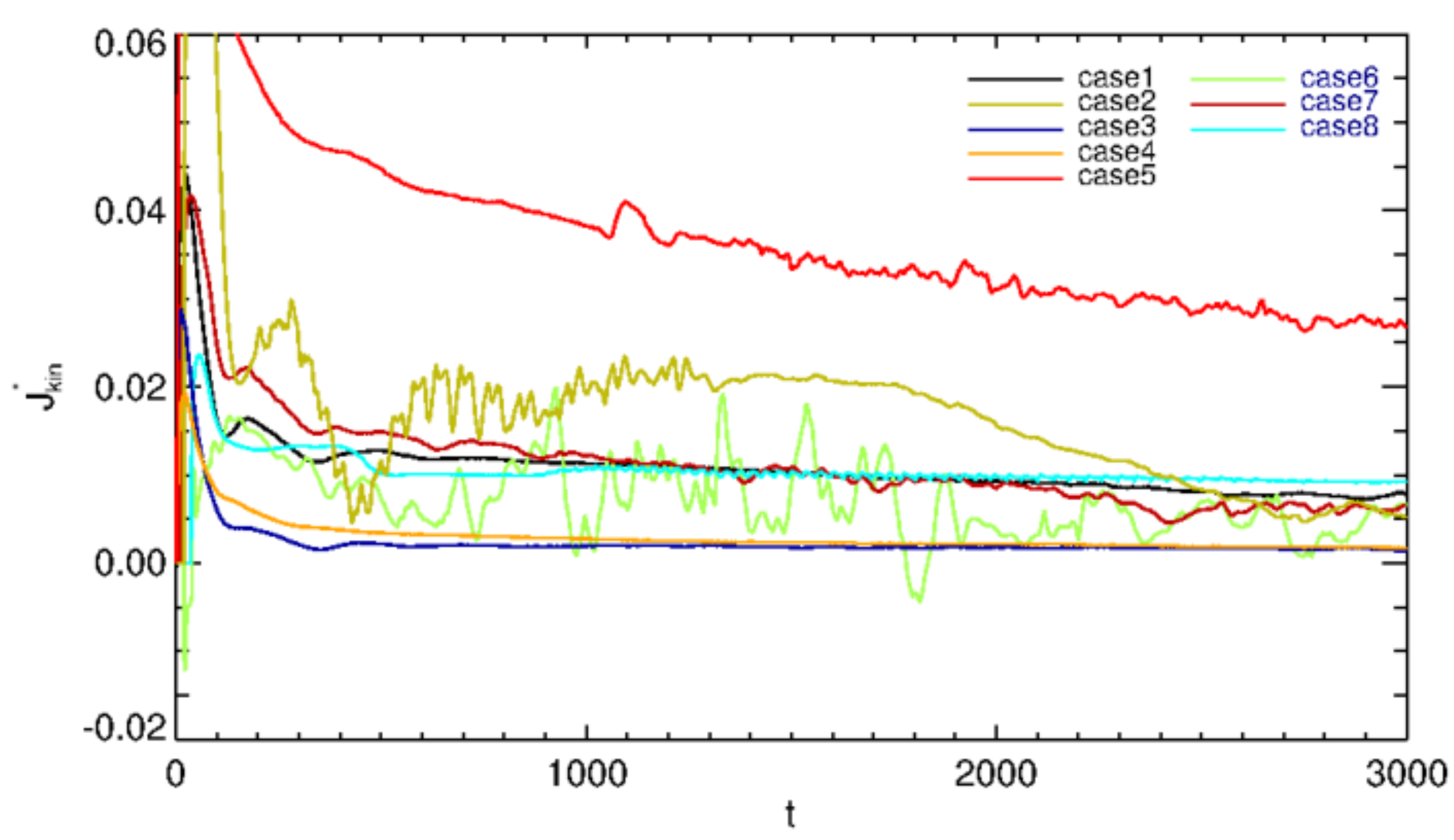}
\includegraphics[width=0.9\columnwidth]{\figurepath/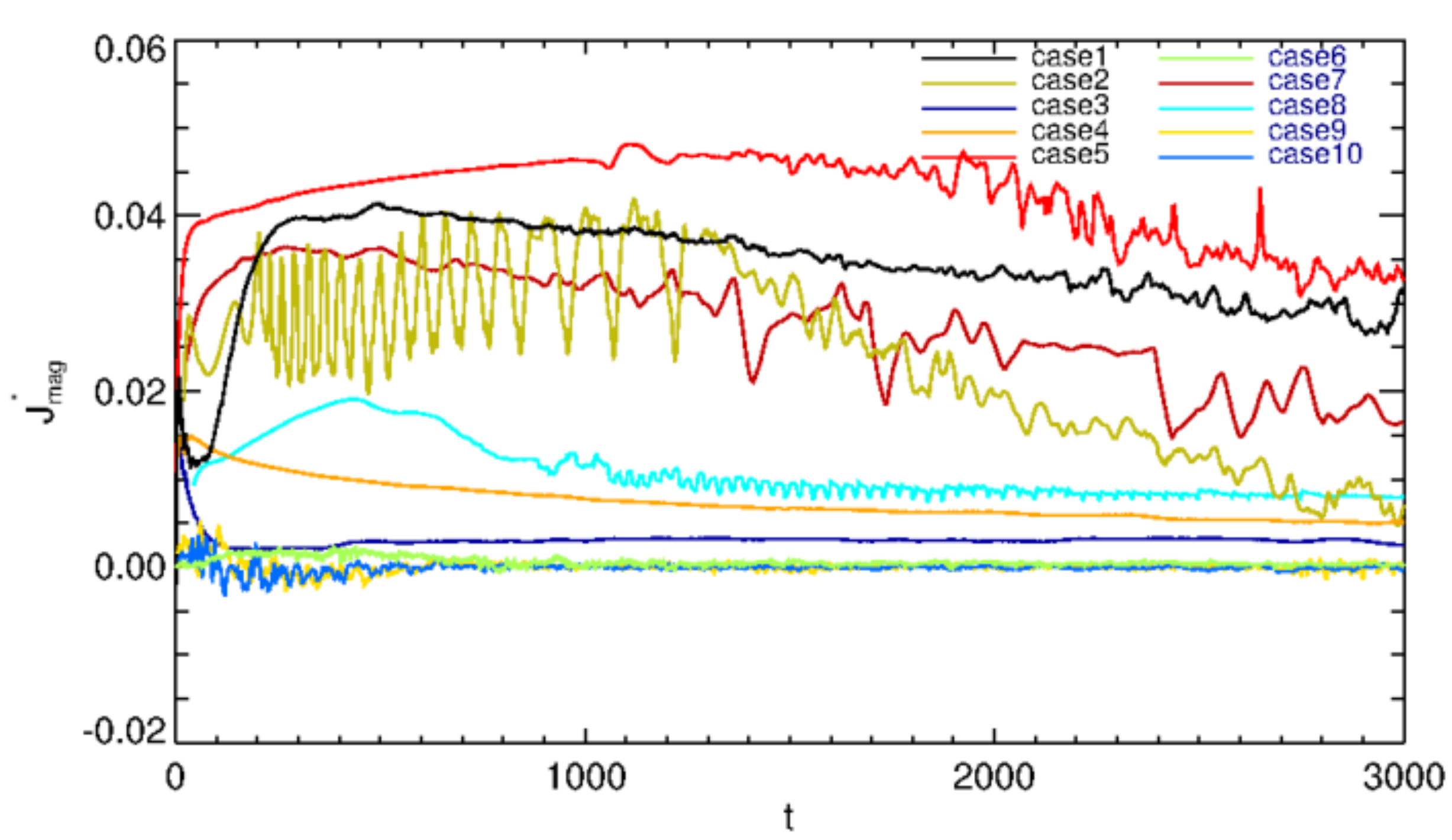}
\includegraphics[width=0.9\columnwidth]{\figurepath/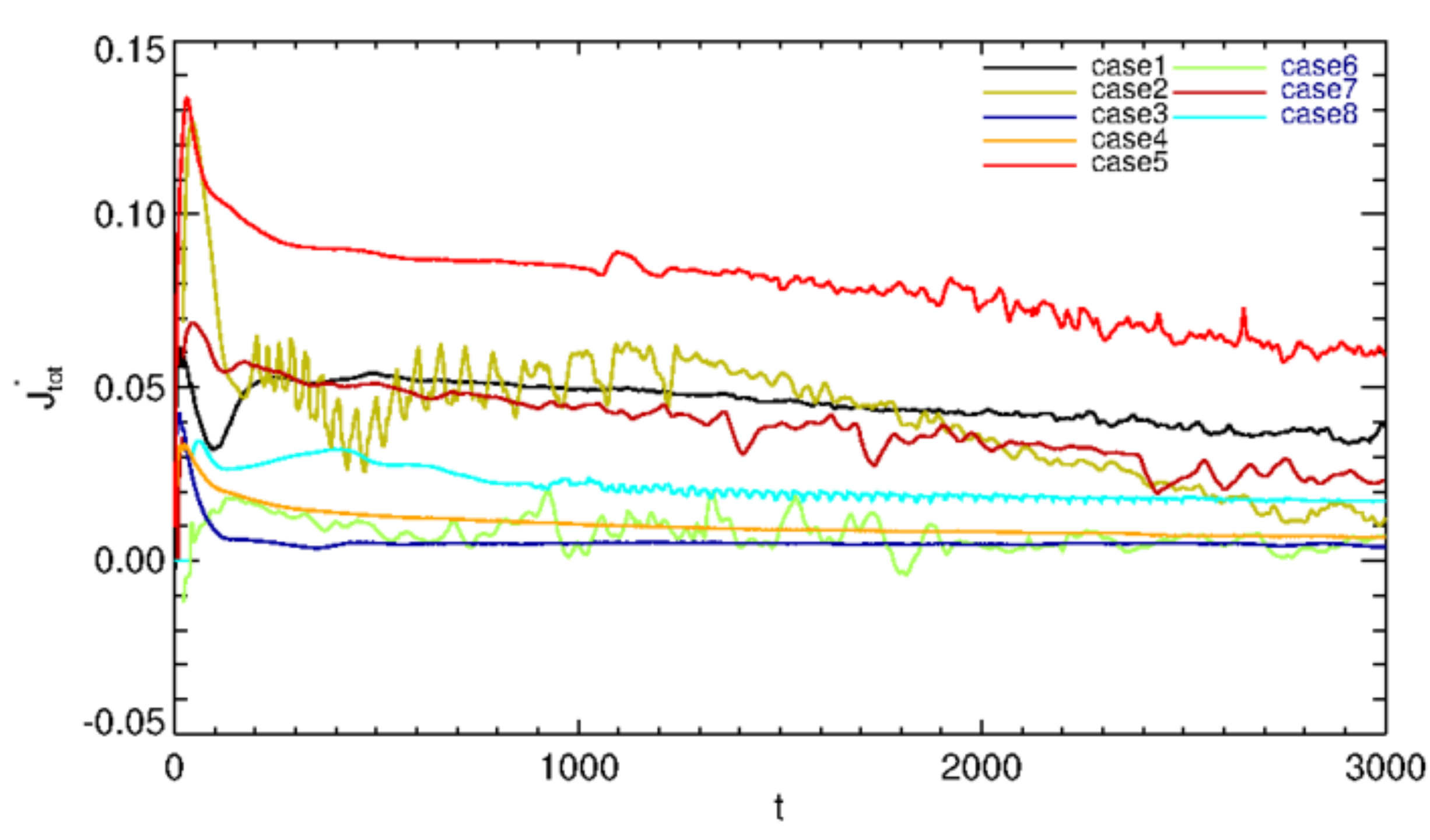}
\caption{Comparison of angular momentum fluxes for the different cases. 
Shown is the time evolution of the vertical angular momentum flux from the disk, 
calculated for a control volume with $r_1 = 1.0$, $r_2 = 10.0$ (see Tab.~\ref{tbl:cases}).}
\label{fig:ang-flux}
\end{figure}
We now consider the main properties of the accretion-ejection system in quasi steady state for all simulation runs.
The two most prominent physical quantities are the mass and the angular momentum fluxes.
Similar to the mass fluxes defined in Eqs. \ref{eq:acc} and \ref{eq:ejec} we integrate the ejection torque 
that is the torque exerted on 
the disk by the outflow applying the same control volume (see \S 3.3), 
\begin{equation}
 \dot J_{\rm kin}=   \int_{r_1}^{r} \!\!r \rho v_\phi \vec{v}_{\rm p}\cdot d\vec{A}_{\rm s}, \,\,
 \dot J_{\rm mag}= - \int_{r_1}^{r} \frac{r B_\phi}{4\pi} {\vec B}_{\rm p}\cdot d\vec{A}_{\rm s},
\end{equation}
with the kinetic and magnetic angular momentum flux, $\dot J_{\rm kin}$ and $\dot J_{\rm  mag}$, carried by 
the outflow.
Figure~\ref{fig:flux_parameterrun} shows the time evolution of the 
accretion rate (top) and the ejection rate (bottom) for the different cases.
The accretion rate is calculated for radius $r = 10$ for all cases\footnote{For high plasma-$\beta$ (case 9, 10), 
the mass fluxes evolution looks strongly fluctuating and are therefore omitted.}.
For comparison, we show the corresponding vertical angular momentum fluxes evolution (Fig.~\ref{fig:ang-flux}). 
For all cases investigated, accretion sets in after several 100s of rotations and
is fully established within $t \simeq 1000$.
After some initial fluctuations, the accretion rate levels off into a steady state,
depending on the physical parameters prescribed in the simulation.
 
Depending on the efficiency of the angular momentum transfer from the disk, the disk establishes a different 
accretion rate.
We find that the simulation runs with the highest accretion rates,  
also have the highest angular momentum flux from the disk (Figs.~\ref{fig:flux_parameterrun},\ref{fig:ang-flux}). 
These are the the runs case 2 with rather low diffusivity and case 5 with rather low poloidal diffusivity,
strongly indicating that with a low diffusivity and, thus, a strong coupling between field and matter,
the magneto-hydrodynamic torque of the jet on the disk is most efficient, and subsequently leads to a
more efficient disk accretion.
Similarly, the high plasma-$\beta$ (thus with a low field strength, simulations runs (case 6, 8, 9 and 10)
exhibit a weak magnetic angular momentum removal and, consequently, are inefficient accretors 
(for a summary see Tab.~\ref{tbl:cases}).
In summary, we find that the extraction of angular momentum from the disk by the outflow 
and accretion are clearly interrelated. 

In the following sections we discuss how the governing system parameters such as 
diffusivity, plasma-$\beta$, and numerical resolution affect the mass flux evolution.

%----------------------------------------------------------------------------------------------------------
\subsection{Possible impact of numerical diffusivity}
Before we further investigate the physical effects, we discuss the results of our resolution study.
Numerical diffusivity will add to the physical diffusivity, as it is a
natural consequence of the finite  difference scheme applied in the PLUTO code.
This effect could be in particular important in the jet launching regime close to the disk
surface where strong gradients in density, pressure, or magnetic diffusivity exist, and which
may not be resolved.
\citet{Murphy2010} have claimed that jet launching could be in fact present just due to 
numerical diffusivity.

In order to estimate the impact of numerical diffusivity we have repeated our reference simulation case 1 
with a three times higher resolution (case 7), but with the same physical diffusivity and anisotropy parameter 
as for the reference simulation.
We find that the leading disk and outflow properties are similar to the reference case.
The accretion rate is slightly decreased from $\dot M_{\rm acc} = 0.015$ (case 1) to $\dot M_{\rm acc} = 0.013$ (case 7), 
while the ejection rate is slightly increased from $\dot M_{\rm ejec} = 0.008$ to $\dot M_{\rm ejec} = 0.009$ 
(see Tab.~\ref{tbl:cases}).
Due to the smaller computational domain for the high resolution simulation we cannot compare the
asymptotic jet radii (at $z=180$), however, we can compare the radius of the bulk mass flux similar to
Eq.~\ref{eq:rjet} for lower altitudes (at $z=60$).
The simulation with higher resolution seems to result in a slightly more collimated jet, with a jet radius
of $r_{\rm jet} = 19.6$ compared to $r_{\rm jet} = 25$ for the reference simulation.
This results in a similarly smaller jet launching radius.
The maximum jet velocities are just the same $v_{\rm jet} = 0.5$.

Figure \ref{fig:flux_parameterrun} shows the time evolution of the mass fluxes.
We see that the accretion rates and ejection rates for case 1 and case 7 saturate
at the same level.
It looks that the high resolution simulation needs more time to establish an outflow,
while the accretion evolves similarly in both runs.
One may see indication for a slightly larger accretion rate for the low resolution
run, which would fit into the picture that the disk material can more easily diffuse
across the field lines due to the numerical diffusivity.

The same picture holds for the angular momentum fluxes (see Fig.~\ref{fig:ang-flux}), which are very similar
for the kinematic part, and slightly offset for the magnetic part.
%-------------------------------------------------------------------------------------------------------------
\subsection{Impact of the magnetic field strength}
There is a common agreement in the literature that jet formation requires
a certain amount of magnetic flux to be present in the jet launching regime.
On the other hand, the maximum magnetic flux which can be supported by the
accretion disk is limited by the disk equipartition field strength (we will
neglect the question of the origin of the magnetic field in this paper).

We have also studied the impact of the magnetic field strength on jet launching, governed by the 
plasma-$\beta$ parameter.
We note that we start all simulations with the same magnetic field {\em profile}, however, due to 
diffusion and advection in the disk, the field distribution may change substantially.

The magnetic field strength determines 

the amount of magnetic energy which is available for jet acceleration,
and is directly interrelated with the length of the lever arm of the magnetic torque 
on the disk (e.g. \citet{Pelletier1992}). 

Due to the larger magnetic torque in case of a strong field, the disk angular momentum  
could be removed more efficiently.
In order to investigate these effects quantitatively, we will compare simulations with 
different plasma-$\beta$ such as case 1 with initial  $\beta=10$, case 8 with $\beta=50$, 
case 9 with $\beta=250$, and the weak field case 6 with $\beta=5000$.
Note that the plasma-$\beta$ is, however, a space and time dependent function of the simulation,
$\beta = \beta(r,z,t)$ and not only one single parameter we prescribe initially at the inner
disk radius.
In the high plasma-$\beta$ regime, the flow of matter controls the dynamical structure of the 
system.
For all cases of a weak magnetic field, the disk-jet system evolution seems highly perturbed - visible
in the global structure of the outflow 
(Fig.~\ref{fig:rho_parameterrun}, case 6,8,9).

Note that the regimes with high plasma-$\beta$ are known to be dominated by the internal, turbulent
torque, which is not taken into account in our simulations by an $\alpha$-viscosity.

In particular, for case 9 and 10, the mass flux evolution is exceptionally perturbed (so we decided 
not to include them in all plots and just show the magnetic angular momentum flux).
Figure \ref{fig:ang-flux} proves that with increasing plasma-$\beta$, less angular momentum is removed 
from the disk, and, subsequently, the accretion rate is reduced. 
This is in particular visible when we compare the simulation with $\beta=50$ with the reference run with 
$\beta=10$.
In case of the weakest magnetic field, thus for the simulations with $\beta=250 $ and $\beta=5000$, the 
magnetic angular momentum removal is approximately negligible.
For the weak field cases, the accretion rate decreases, and in case 9, 10 and 6 no efficient accretion is 
observed.
In summary, we confirm the hypothesis that efficient magneto-centrifugal jet driving requires a strong 
magnetic flux (i.e. a low plasma beta), together with a large enough magnetic torque in order to produce a 
powerful jet.

 \begin{figure}
  \centering
  \includegraphics[width=0.4\columnwidth]{\figurepath/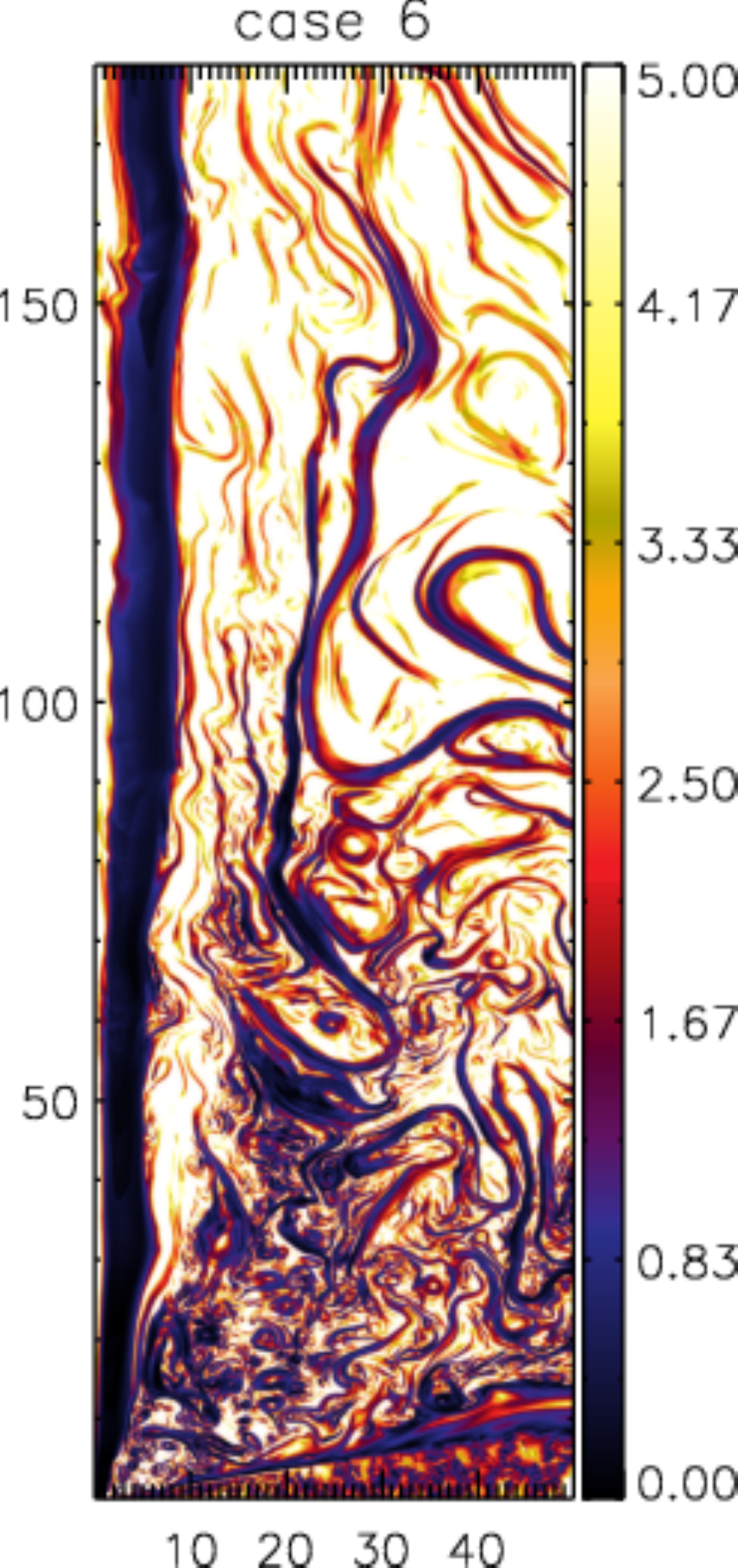}
 \hspace{0.5cm}
  \includegraphics[width=0.4\columnwidth]{\figurepath/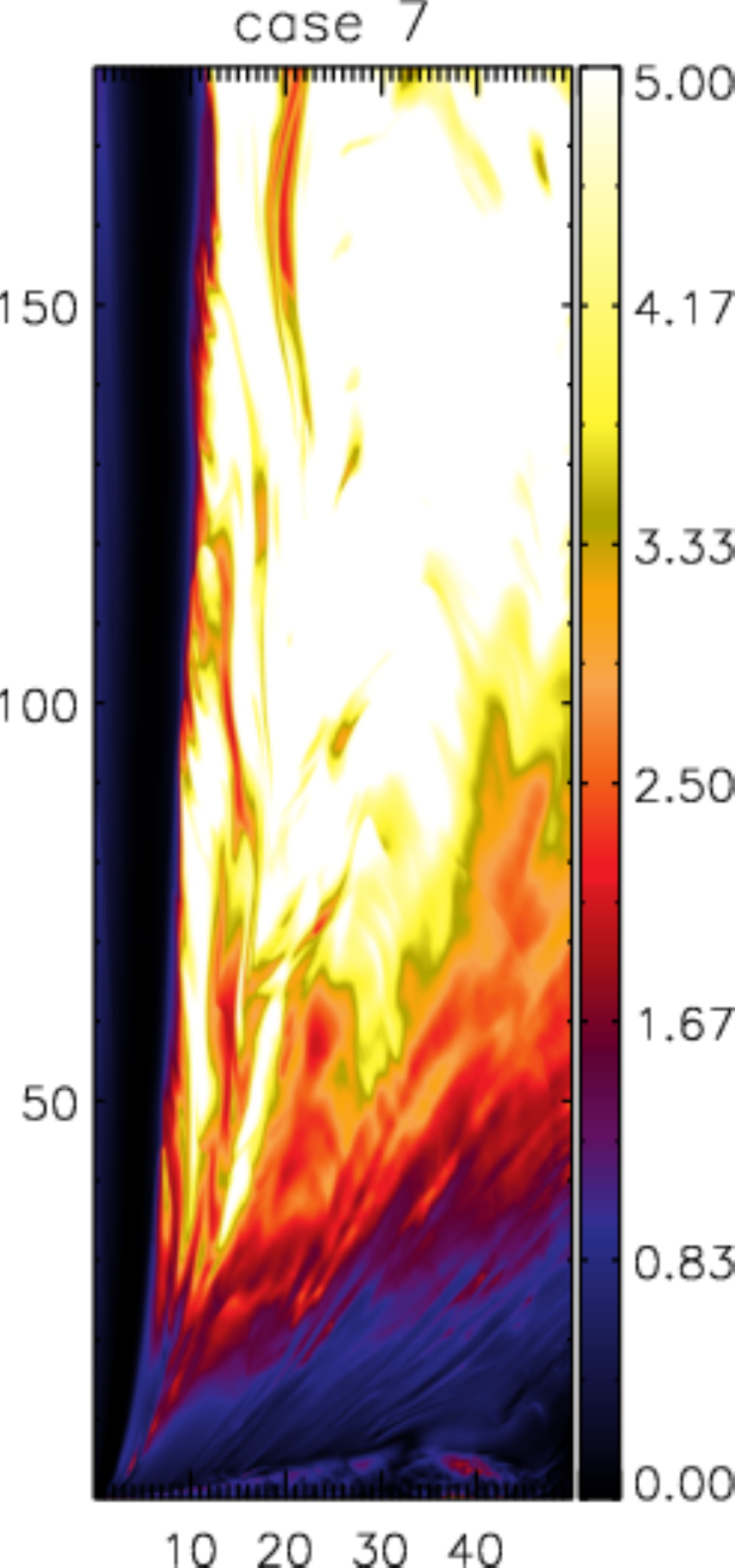}\\
   \centering
   \caption{Comparison of outflow launching for low and high plasma-$\beta$ cases.  
  Display of the  Alfv\'enic  Mach  number  for  two simulations with three times 
  higher  resolution. 
  The left  panel  shows the weak field case $\beta=5000$  (case  6).
  The right  panel  shows the strong  field case $\beta  = 10 $ (case  7).  }
  \label{fig:high_res1}
  \end{figure}

Simulation case 6 applies a very weak magnetic field with $\beta=5000$. 
Figure \ref{fig:high_res1} compares the two simulations case 6 with the standard setup case 7 in high 
resolution.
No smoothly structured outflow is obtained for case 6, but a highly turbulent outflow with non-negligible 
outflow speed and mass flux.
Due to the weak poloidal field, the outflow in case 6 is super-Alfv\'enic right from the launching point, 
thus magneto-centrifugal acceleration cannot play a role.
Interestingly, the size of the turbulent features increases with distance from
the origin. Also the poloidal magnetic field is highly tangled (Fig.~\ref{fig:rho_parameterrun}).
The mass flux we measure for case 6 is about $\dot M_{\rm ejec} \simeq 0.001$ with outflow velocities of 
$v_{\rm jet} \simeq 0.4$. While the velocities are comparable to case 7 (or case 1 with lower resolution),
the ejected mass flux is substantially lower (factor 2).
Accretion in case 6 is weak (the smallest of all simulations), consistent with the low angular 
momentum losses (the smallest of all simulations).
The accretion rate is even smaller than the ejection rate (factor 5),
and we may call such a disk an ejection disk instead of an accretion disk.

The question is what is launching and accelerating such a turbulent, high plasma-$\beta$ outflow?
Our interpretation is that the {\em initial} acceleration to super-escape speed is by the toroidal
magnetic pressure gradient (induced by the differential rotation between the rotating disk and the 
static corona).
When the bow shock has left the domain, this acceleration process decays,
and the remaining outflow acceleration is due to weak Lorentz force (as we are in the super Alfv\'enic 
regime).
The vertical mass flux for case 6 (very weak field) is one order of magnitude below the mass flux
in case 7 (with strong field $\beta=10$), measured at the same altitude (see Tab.~\ref{tbl:cases}).
In summary, launching conditions as in case 6 with $\beta=5000$ do not produce a strong jet flow,
but a light disk wind of super-escape speed.

We point that the physical regime of acceleration does not only depend on the field strength, 
but also on the mass flux. 
Magnetocentrifugal effects are more evident for outflows with low mass flux,
while in heavy jets the magnetic pressure gradient may play a substantial role
(see e.g. \citet{Andersonli2005}).

\begin{figure}
\centering
\includegraphics[width=0.9\columnwidth]{\figurepath/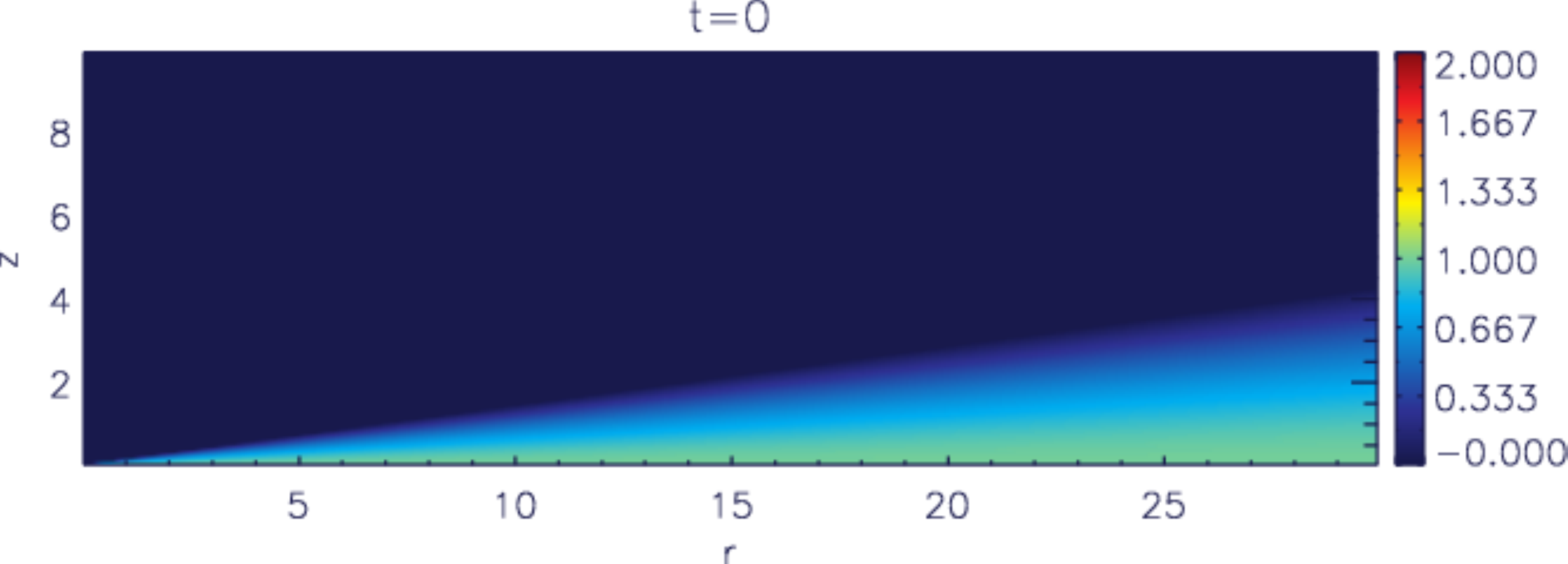}
\includegraphics[width=0.9\columnwidth]{\figurepath/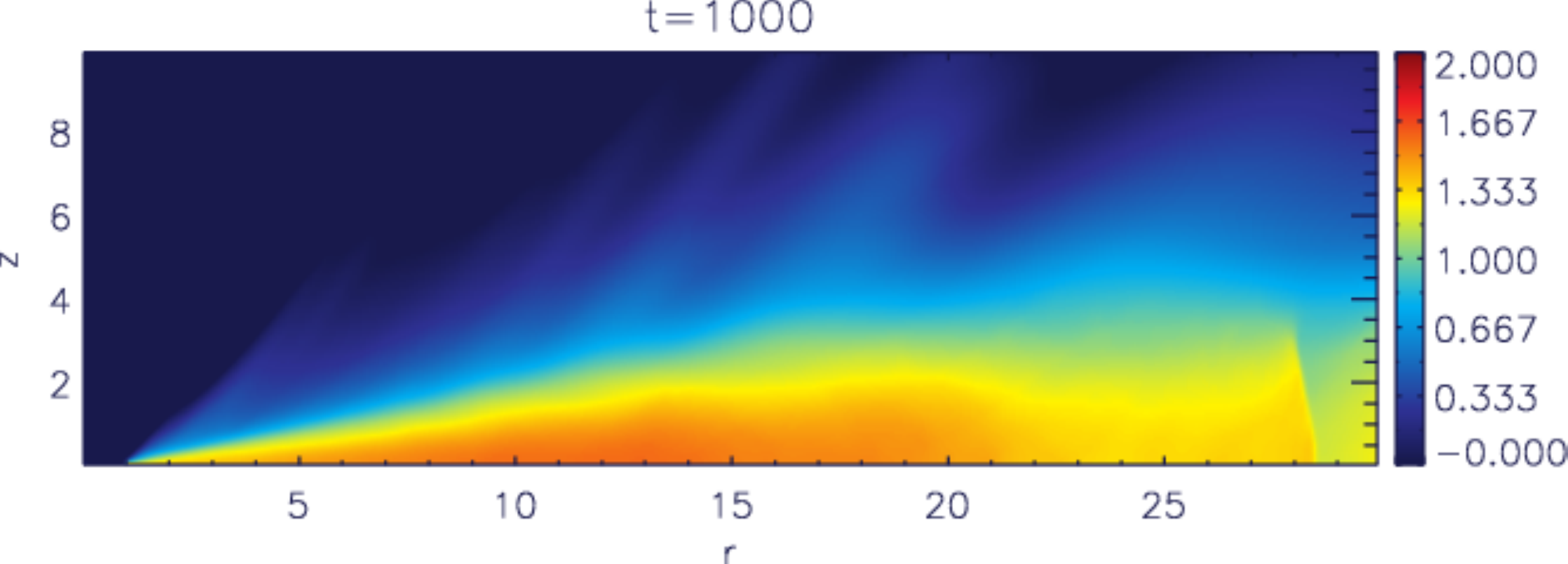}
\includegraphics[width=0.9\columnwidth]{\figurepath/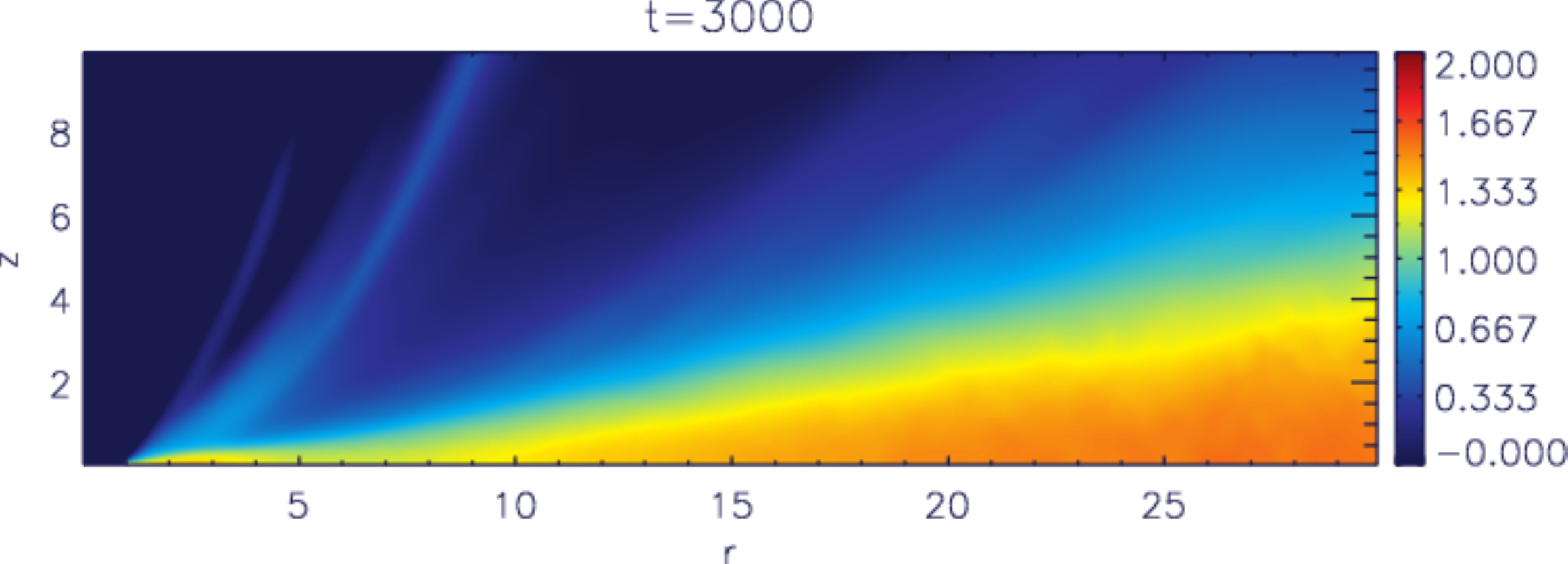}
\caption{Time evolution of the plasma-$\beta$ in the inner disk-jet system. 
Snapshots of the plasma-$\beta$ distribution (in logarithmic scale) for case1 at 
time $= 0, 1000, 3000$. }
\label{fig:log_beta_case1}
\end{figure}

\begin{figure}
\centering
\includegraphics[width=1.\columnwidth]{\figurepath/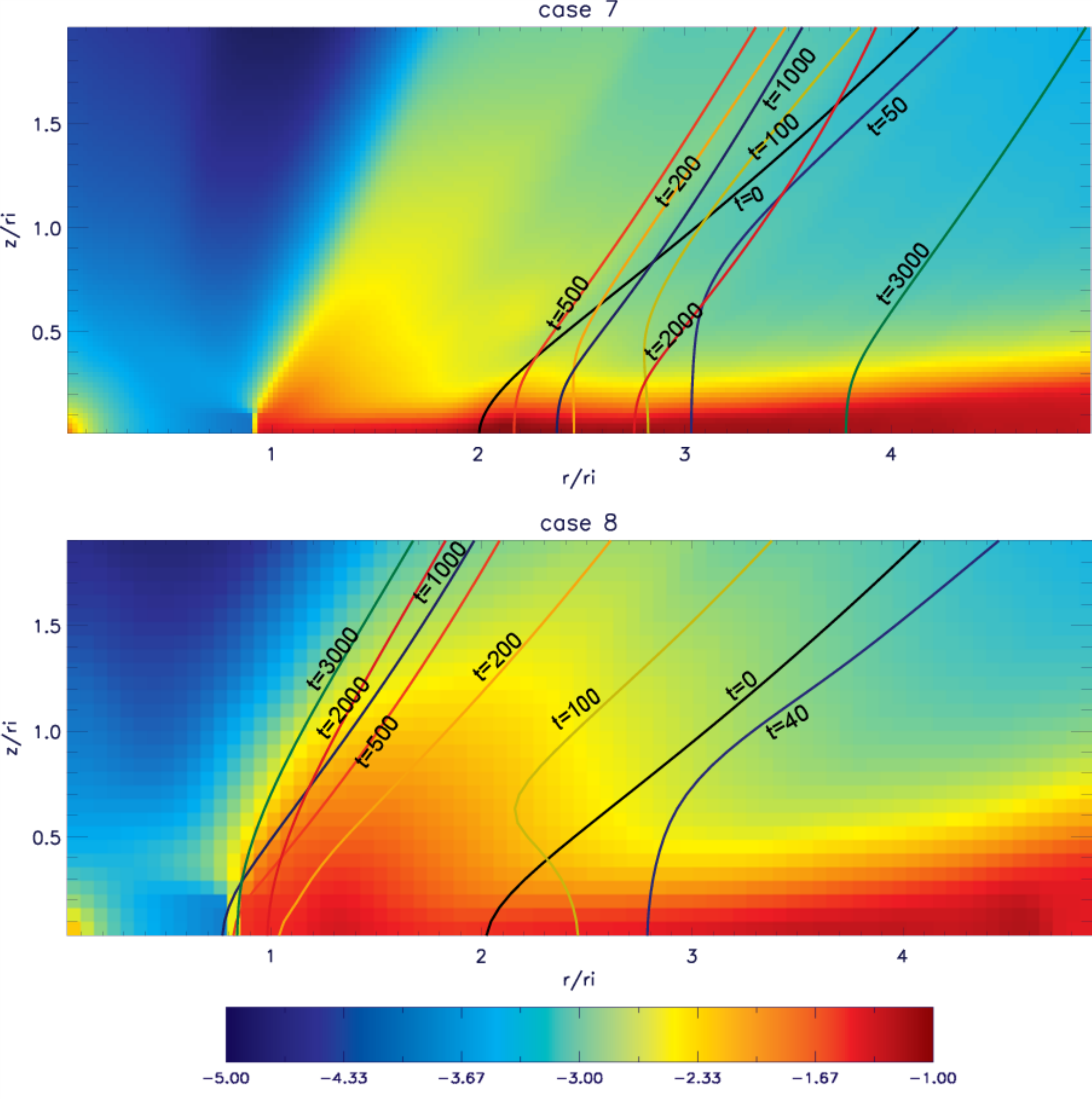}
\caption{Advection and diffusion of magnetic flux over time.
Show is the evolution of the magnetic flux surface $\Psi = 0.1$ rooted initially at $(r,z) = (2,0)$.
for case 7 (high resolution) and case 8($\beta=50$).
Colors show the density distribution at t=3000 (logarithmic scale).}
\label{fig:case8and7_advecdiff}
\end{figure}

The relative field strength (measured by the plasma-$\beta$) is a function of 
space and evolves in time.
The initial field and gas pressure distribution, denoted by the $\beta_{\rm i}$, is changed by the
dynamical evolution including advection and diffusion  of the disk magnetic field.
 Figure \ref{fig:log_beta_case1} shows the evolution of the plasma-$\beta$ for case 1.
It is clearly visible that the plasma-$\beta$ along the disk surface changes substantially.
The inner disk (for $r<15$) reaches a steady state with a $\beta$ somewhat higher than the
initial value.
Later, around $t=3000$ when the disk mass decreases the gas pressure decreases as well,
resulting in a lower plasma-$\beta$ in the disk corona.
The disk corona therefore remains sufficiently magnetized and can support a fast jet.
In case 1 the magnetic field structure reaches a steady state balanced by field diffusion
advection.

In simulation case 7 with the same parameter setup but a higher resolution a steady state
is reached as well.
However, even for late time steps the magnetic flux continues to diffuse outwards.
This is somewhat surprising since the numerical diffusion is smaller and cannot be the 
reason (Fig.~\ref{fig:case8and7_advecdiff}, top).
For case 8 with the higher initial plasma-$\beta$, advection seems to dominate outward
diffusion of flux. The poloidal field distribution reaches a steady state with a magnetic 
flux concentration close to the inner edge of the disk (Fig.~\ref{fig:case8and7_advecdiff}, bottom). 
Thus, this setup seems favorable for jet launching as well, although the initial plasma-$\beta$ 
was high.

As an extreme example, for simulation case 9 with the initial $\beta = 250$ the plasma-beta is 
considerably changed during the disk evolution.
Case 9 shows very weak accretion (see discussion above), so outward diffusion of magnetic flux
dominates advection.
Outward diffusion of flux implies a decrease of magnetic field strength. 
As a result the disk magnetization decreases, and the plasma-$\beta$ reaches number values above $10^5$.
Therefore this disk is not able to launch strong jets. So far no steady state is achieved. 

The interrelation between the magnetic field strength and outflow launching has been discussed by 
other authors \citep{Tzeferacos2009, Murphy2010}, 
indicating that regions with weak field are not able to generate an outflow, and that both
the collimation degree and the ejection rate increase with stronger field.
\citet{Murphy2010} have concluded that jet launching for cases of weak magnetization 
may be artificially supported by numerical diffusivity within the disk surface layer, which should
heat the gas, producing additional gas pressure.
They suggest that ejection is possible as the magnetization reaches unity at the disk surface 
due to the steep density decrease.
No ejection was reported when the mid plane magnetization becomes too small. 
Nevertheless, the asymptotic jet velocity remained too low to explain the observed jets speed. 
From our simulations, we find that even for weak magnetization in the disk the disk corona is 
sufficiently magnetized for jet launching, and, depending on the efficiency of mass loading, 
fast jets could be driven (for comparison, see Figure \ref{fig:log_beta_case1} for case 1).
The mass loading depends on resistivity, and we will discuss this aspect in the next section.

We finally consider the possibility to observe the MRI in our simulations.
This is interesting since we are dealing with magnetic fields of various strength in combination
with a differentially rotating system.
To generate the MRI in numerical simulations, two conditions are essential - a large enough (but not too large) 
plasma-$\beta$ in the disk as to initiate instability, and also a grid of high enough resolution in order
to avoid damping of small wavelengths.
The wavelength of most unstable MRI mode is given by $\lambda_{\rm MRI} = 2\pi v_{{\rm A},z}/\Omega_{\rm K}$ 
for a Keplerian rotation $\Omega_{\rm K}$ (see e.g. \citep{Romanova2011MNRA}).
When  we calculate this wavelength for a simulation with $\beta = 10$, we obtain 
$\lambda_{\rm MRI} \simeq 0.057$,
implying that we cannot resolve the MRI with our setup as the grid size is about half of the MRI wavelength.

We point out that the magnetic diffusivity applied in our simulations also contributes to suppress 
the generation of MRI turbulence.

%------------------------------------------------------------------------------------------------------
\subsection{Magnetic diffusivity strength and anisotropy} 
The strength and anisotropy of the magnetic diffusivity - controlled by the coefficients 
$\eta_{\phi,\rm i},\eta_{\rm p,i}$ and $\chi$ (see Tab.~\ref{tbl:cases}) - are essential parameters
governing the coupling between the magnetic field and the plasma.
Efficient magneto-centrifugal driving requires strong coupling between the magnetic field and the rotating 
disk material, thus a rather low diffusivity.
A similar argument holds for the launching by magnetic pressure gradient.
On the other hand, mass loading from accretion to ejection requires a certain degree of 
diffusivity in order to transport mass across the field lines. 

So far, no general model for the magnetic diffusivity distribution in disk-jet structures
is available.
In many simulations considering magneto-centrifugal driven outflows, the outflow is 
governed by ideal MHD, while the diffusivity is confined to the disk.
In order to investigate the impact of diffusivity on jet launching, we have therefore 
performed simulations with varying strength of the diffusivity components 
$\eta_{\phi,\rm i}$ and $\eta_ {\rm p,i}$.

\begin{figure}
\centering
\includegraphics[width=0.99\columnwidth]{\figurepath/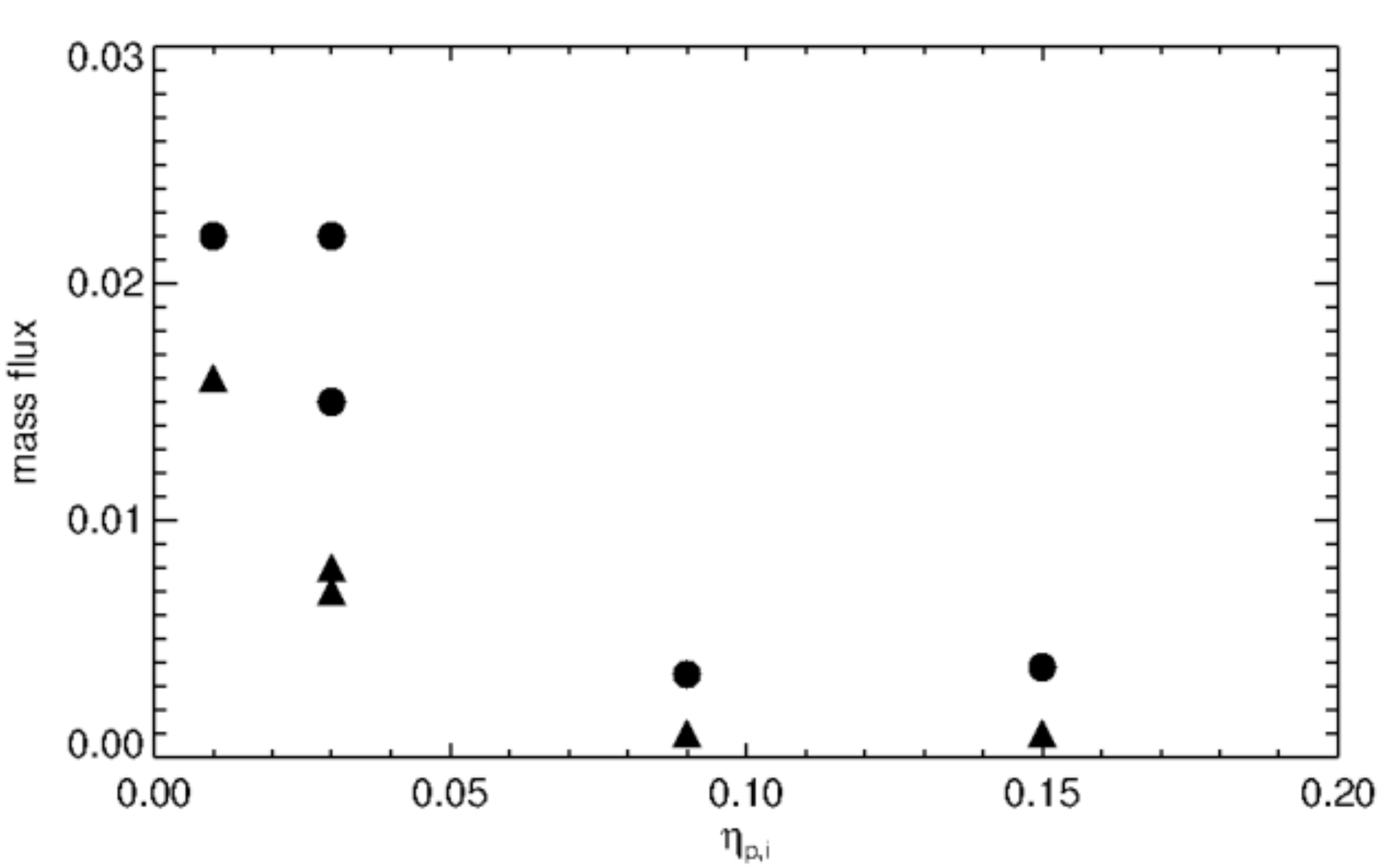}
\caption{Mass flux $\dot{M}_{\rm ejec}$ vs poloidal diffusivity. 
Shown is the ejected jet mass flux (triangles) and the accretion rate (circles) as function of the 
poloidal diffusivity for simulations with the same magnetic field strength, 
i.e. case 1 ($\eta_{\rm p,i}=0.03$), case 2 ($\eta_{\rm p,i}=0.01$), case 3 ($\eta_{\rm p,i}=0.15$), 
case 4 ($\eta_{\rm p,i}=0.09$), case 5 ($\eta_{\rm p,i}=0.03$).} 
\label{fig:eta-massflux} 
\end{figure}

We first compare simulations with the same anisotropy parameter $\chi = 3$ as reference run case 1.
In general, the strength of magnetic diffusivity governs the disk accretion rate.
Our reference run  case 1 reaches quasi steady state, establishing a balance between inward advection 
and outward  diffusion of magnetic flux. 
This  happens after $t=1000$ and is disturbed again in the late stages of the  dynamical evolution, 
most probably due to the change in the overall disk dynamics due to the decreasing disk mass. 
A change in diffusivity will also affect this balance. 
We now compare case 4 with higher and case 2 with lower diffusivity.
In the less diffusive case, advection dominates and we obtain a higher accretion rate.
We find that not only the accretion rate, but also the ejection rates are higher for disks with lower 
diffusivity.
This trend is shown in Fig.~\ref{fig:eta-massflux}, where we display the accretion rate and also the 
ejection rate as function of the the poloidal diffusivity.

Two physical processes  affect the disk dynamics - {\it advection} and {\it diffusion}. 
In a less diffusive disk, the plasma is stronger coupled to the field, and advection of magnetic
flux dominates.
Consequently, the magnetic flux surfaces are located further in the inner part of the disk, 
{"}rotate{"} faster, and a stronger $B_{\phi}$ is induced.
The stronger toroidal field may lead to a stronger acceleration of the outflow, 
by stronger Lorentz forces (either magnetic tension of pressure forces).
In addition, by comparing the vertical profiles of net launching forces at $t=1000$ for less diffusive 
disk (case 2), and reference run (case 1), we find a larger net force in case 2 by which implies that 
more disk material can be lifted into the outflow.

This is exacly what we observe in the mass fluxes - the ejected mass flux in case 2 is two times the
ejected mass flux in case 1.
Note that a stronger $B_{\phi}$ will impose stronger pinching forces
$\sim (B_{\phi} \cdot \nabla) B_{\phi}$ on the disk and the ejected material, thus opposing the loading process.
However, the magnetic pressure gradient of the toroidal field $\sim \nabla B_{\phi}^2$ has a
radially outward direction, thus supporting ejection.
By comparing the vertical profiles of both terms for the two cases with $\beta=10$ and $\beta=50$,
we find that i) the toroidal magnetic field pressure gradient component in vertical
direction $(\nabla B_{\phi}^2)_z$ is always dominating the pinching force of the
toroidal field, and that ii) for higher plasma-$\beta$ the $ (\nabla B_{\phi}^2)_z$ is larger
than for the lower plasma-$\beta$ case.
Thus, in our simulations the toroidal field is supporting ejection.

In summary, although a certain magnitude of diffusivity is required for mass loading,
the lower diffusivity allows for enhanced mass loading of accreting material into the outflow.
Consequently, less diffusive disks tend to have higher ejection rates.
This result is in agreement with the previous  literature \cite{Zanni2007}.

\begin{figure}
\centering
\includegraphics[width=0.99\columnwidth]{\figurepath/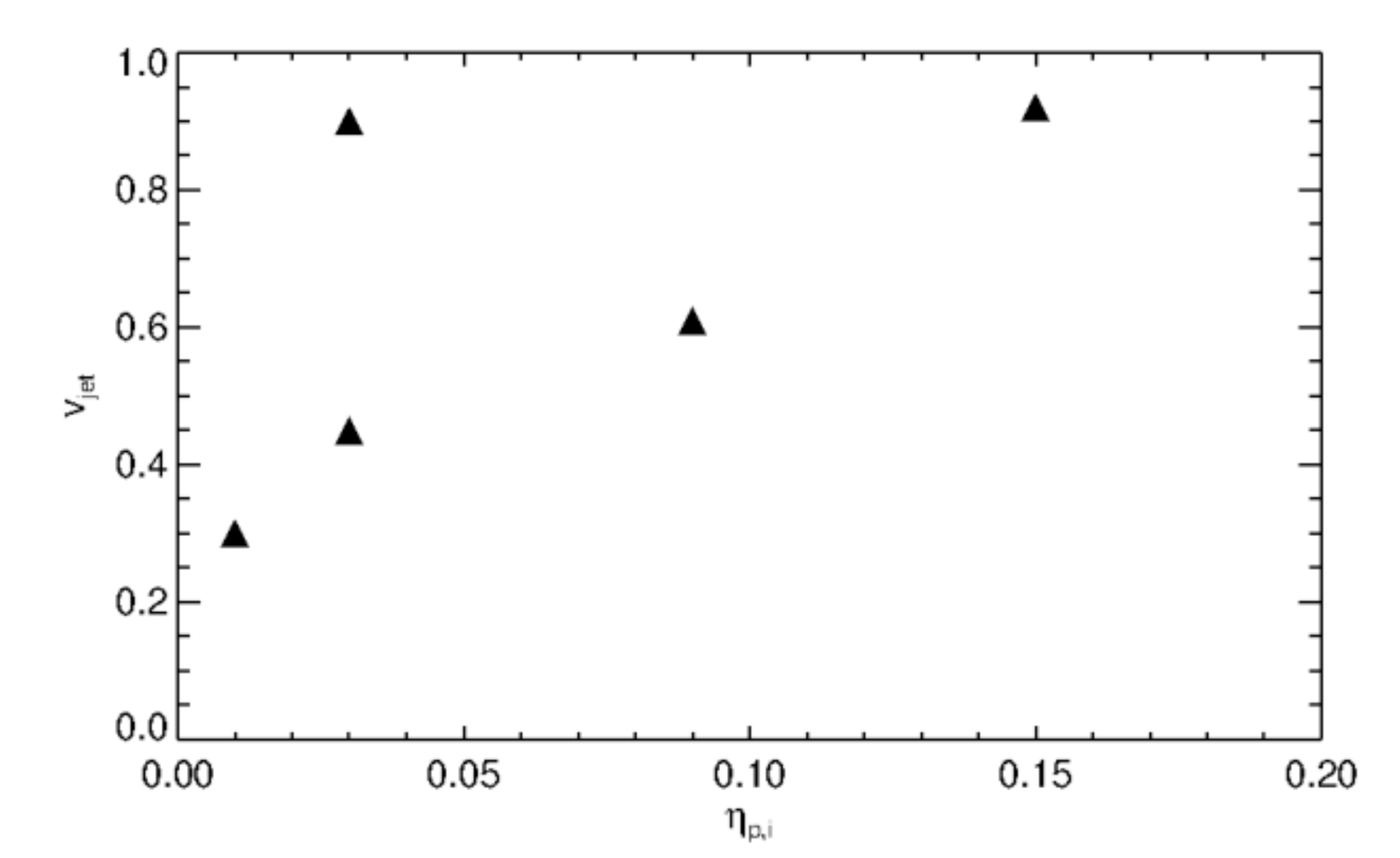}
\caption{Jet velocity vs poloidal diffusivity at $t=3000$. Shown is the typical jet velocity as a 
function of but different diffusivity,
i.e.  case 1 ($\eta_{\rm p,i}=0.03$), case 2 ($\eta_{\rm p,i}=0.01$), case 3 ($\eta_{\rm p,i}=0.15$), 
case 4 ($\eta_{\rm p,i}=0.09$), case 5 ($\eta_{\rm p,i}=0.03$).} 
\label{fig:eta-vjet}
\end{figure}

Our simulations confirm that the strength of magnetic diffusivity does affect the asymptotic speed 
of the outflow.
Figure \ref{fig:eta-vjet} shows the  typical jet velocity versus the poloidal
diffusivity.
Comparing different simulation runs (see Tab.~\ref{tbl:cases}), 
we find that the outflow from the less diffusive disk (case 2) remains slower, while the outflows 
formed from more diffusive disks (case 3 and 4) are accelerated to higher speed.
For case 2 with three times lower magnetic diffusivity the asymptotic speed is reduced by $\approx 30\%$,
while for case 4 with three times higher magnetic diffusivity the increase is about $\approx 30\%$
(see Tab.~\ref{tbl:cases}).
This is reasonable from a physical point of  view, since in a weakly diffusive disk more material is 
diverted into the outflow.
Thus, with the same amount of magnetic flux  available, correspondingly less magnetic energy could be 
transferred per outflow mass unit.
The more massive outflows are only accelerated to lower speed if launched with similar 
magnetization.
Alternatively, a higher magnetic diffusivity resulting in a weaker mass load, leads to a faster outflow.

When comparing our results to the previous literature, one should keep in mind that we have applied
a mass flux weigthed velocity which we consider as the typical velocity of the bulk mass flux. 
The velocity value is generally lower as the maximum speed we measure in the outflow
and which is mostly given in the literature.
While the typical speed ranges from about 0.5 to 0.9 (in code units), the maximum speed in
the outflow ranges from 1.2 to 1.8 inner disk Keplerian velocities and is comparable to
other paper in the literature.
Furthermore, we have detected a slight time evolution in the velocity (see also the mass flux evolution
shown in Fig.~\ref{fig:flux_parameterrun}).
In our reference run case1, the maximum asymptotic speed varies from $v_{\rm p,max} = 1.8$ at $t=500$ 
to $v_{\rm  p,max} = 1.2$ for $t > 1000$.
For comparison, \citet{Zanni2007, Tzeferacos2009} give a maximum speed 
of $v_{\rm p,max} \simeq 1.5 ... 4.5$ at $t=400$.

\begin{figure*}
\centering
\includegraphics[width=4cm]{\figurepath/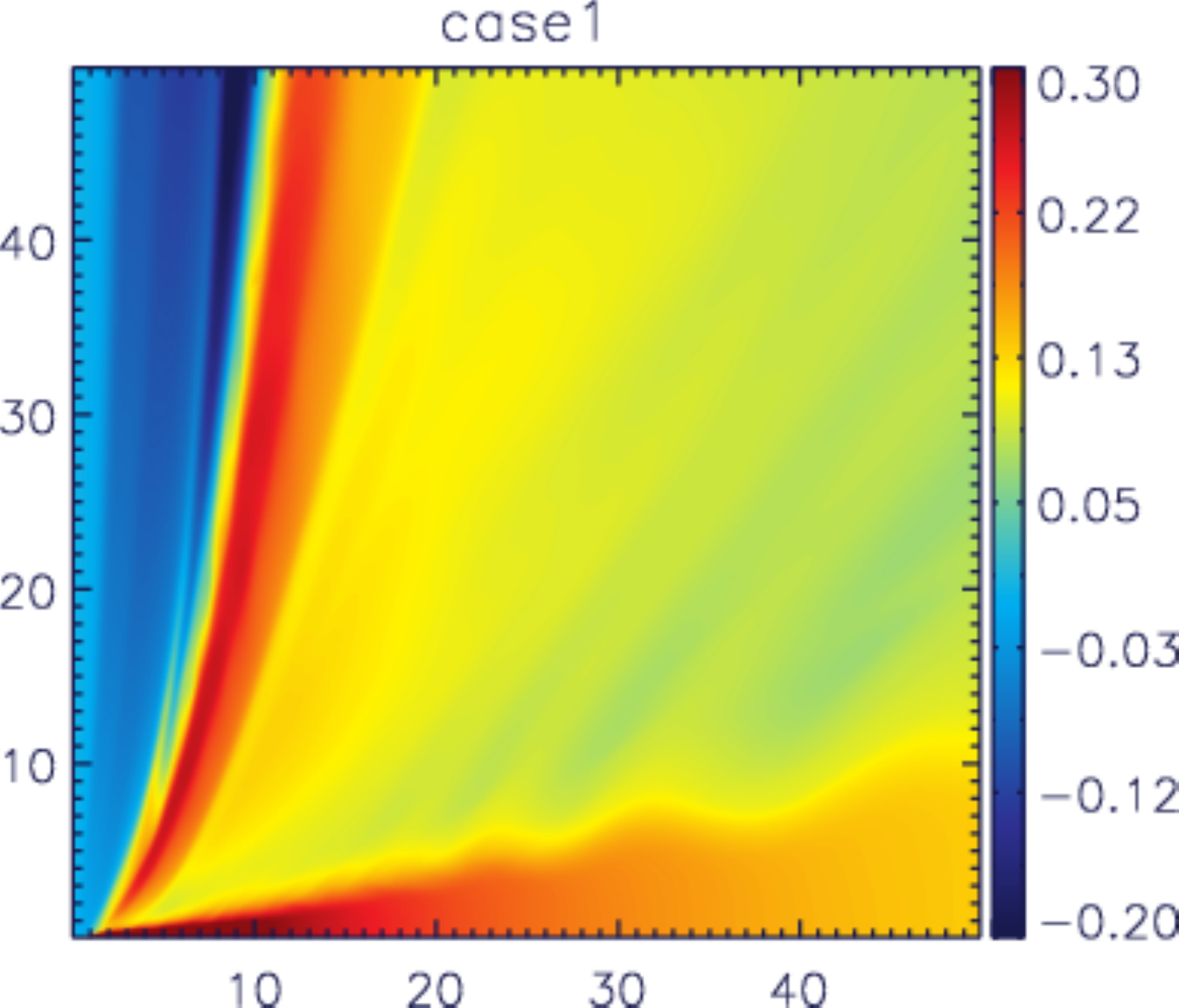}
\hspace{0.25cm}
\includegraphics[width=4cm]{\figurepath/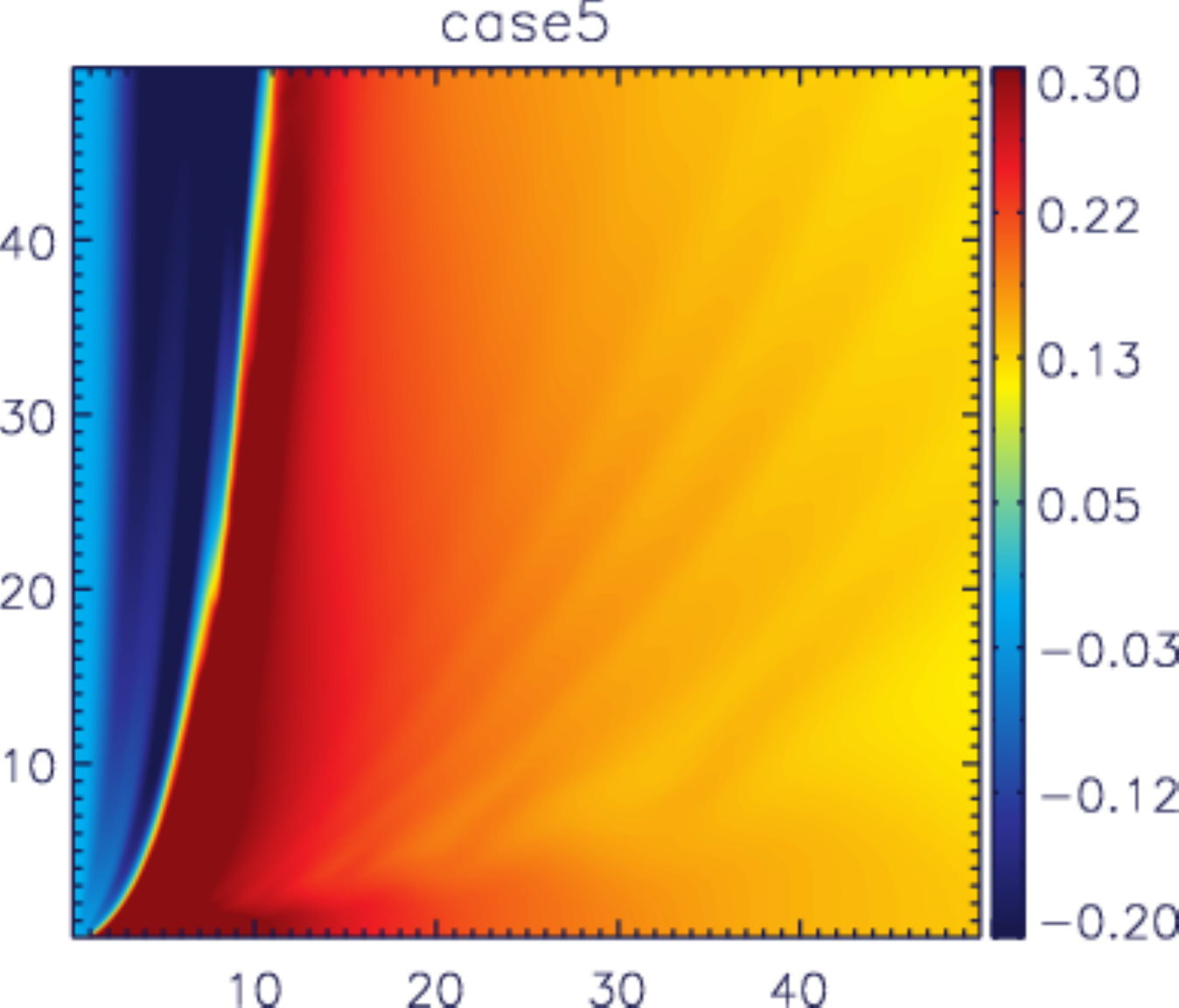}
\hspace{0.25cm}
\includegraphics[width=4cm]{\figurepath/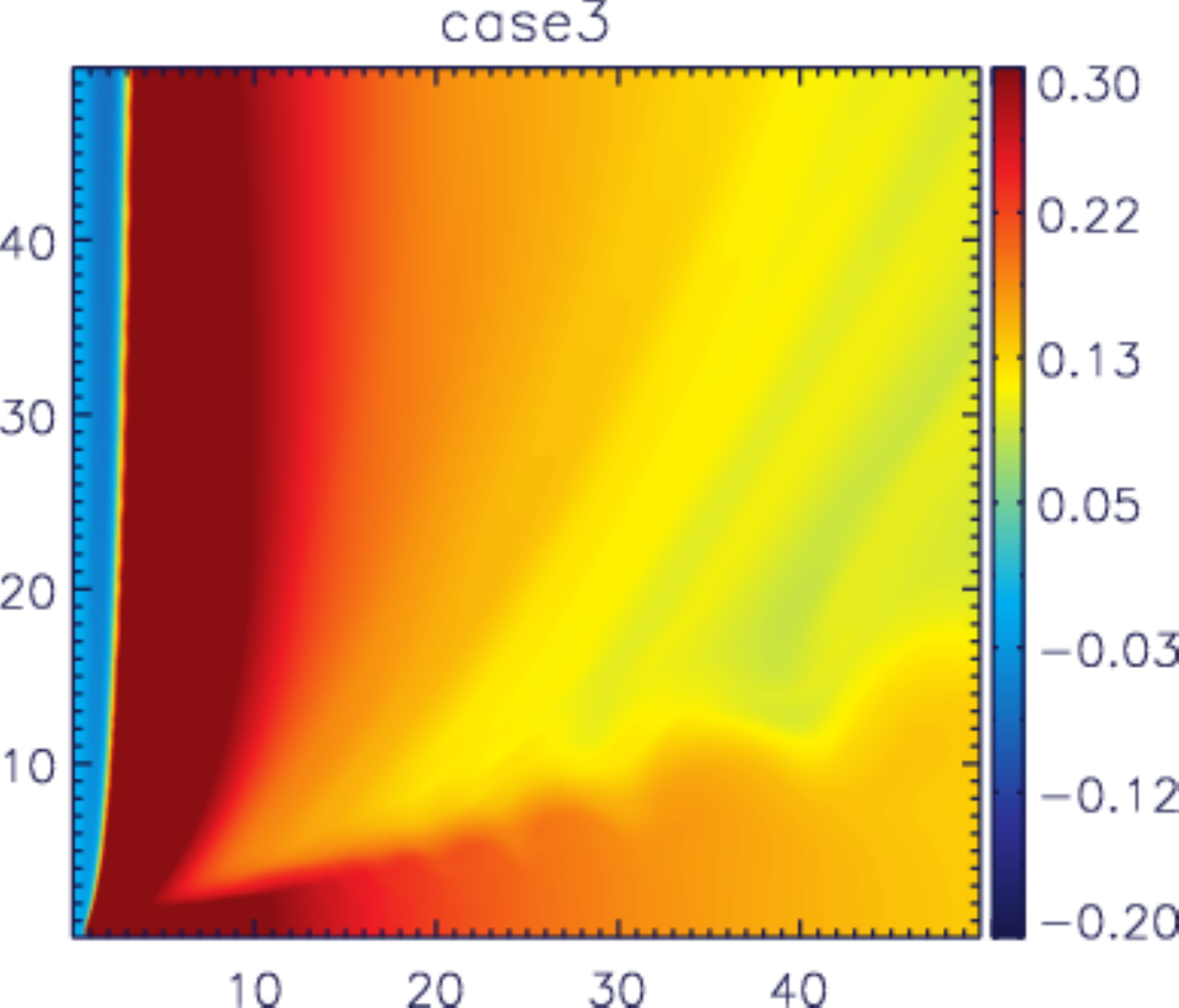}
\hspace{0.25cm}
\includegraphics[width=4cm]{\figurepath/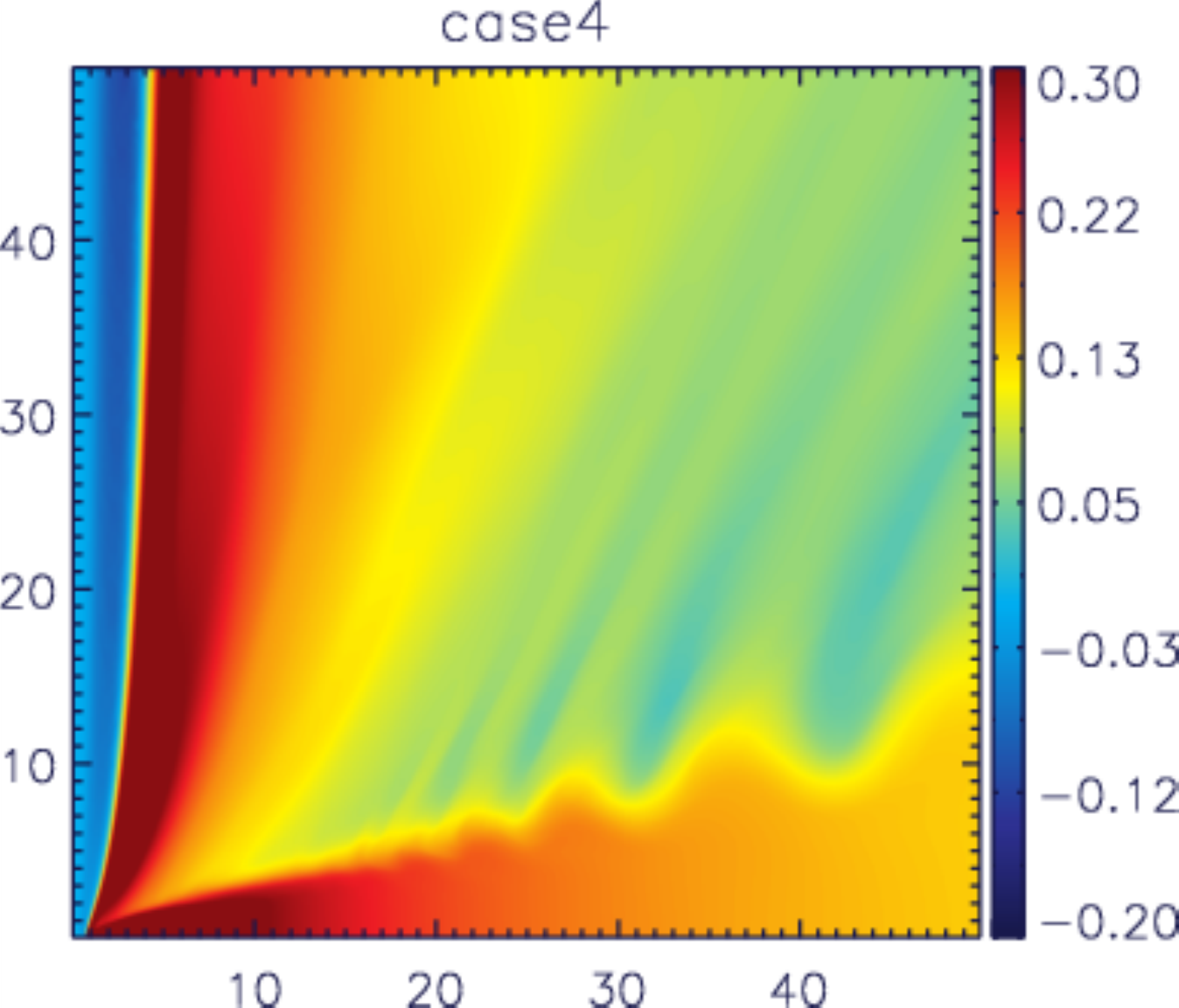}
\caption{Outflow rotation and toroidal magnetic diffusivity. 
Shown is the rotational velocity $v_\phi$ of the inner disk-jet system at t= 3000 for simulation runs with 
different toroidal magnetic diffusivity, respectively different anisotropy parameters $\chi$, 
i.e. cases 1, 5, 3, 4 (from left to right).} 
\label{fig:anisotropy}
\end{figure*}

\begin{figure}
\centering
\includegraphics[width=0.9\columnwidth]{\figurepath/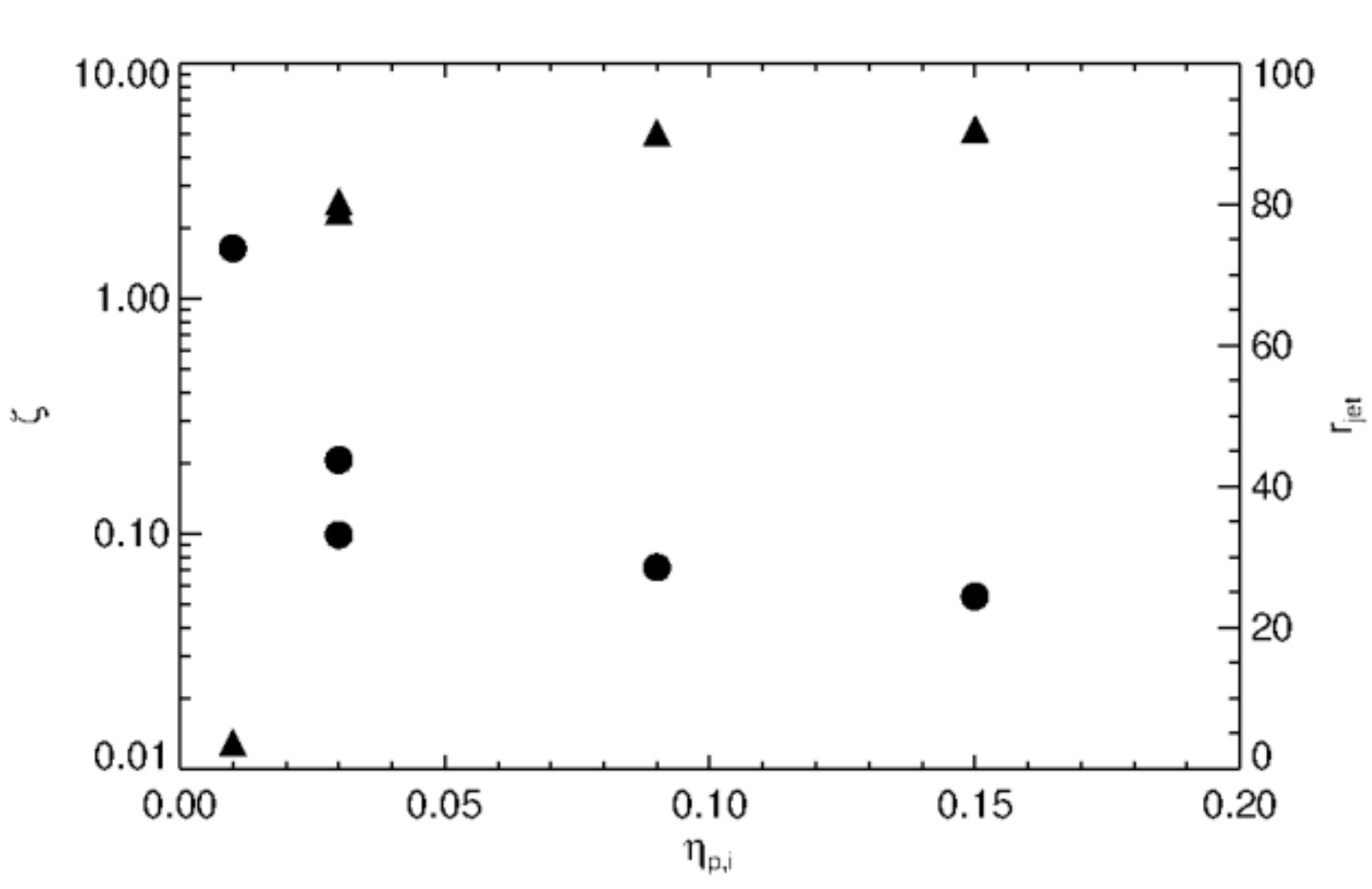}
\caption{Collimation degree and jet radius vs poloidal diffusivity. 
Shown is the collimation degree $\zeta$ (filled triangles) and the mass flux weighted jet radius $r_{\rm jet}$
(filled circles) over the poloidal diffusivity
for simulations with the same magnetic filed strength, but and different diffusivity,
 i.e. case 1 ($\eta_{\rm p,i}=0.03$), 
case 2 ($\eta_{\rm p,i}=0.01$), case 3 ($\eta_{\rm p,i}=0.15$), case 4 ($\eta_{\rm p,i}=0.09$), 
case 5 ($\eta_{\rm p,i}=0.03$). } 
\label{fig;eta_collimation}
\end{figure}

Next, we investigate a possible impact of anisotropy in the diffusivity-tensor, parametrized by $\chi$.  
We compare case 1 and case 5 with the same polodial diffusivity $\eta_{\rm p}= 0.03$,
but with case 5 having a lower diffusivity in toroidal direction.
By comparing their leading properties (see Tab.~\ref{tbl:cases}), we find them quite similar properties.
In particular, for both cases (with the same poloidal diffusivity) the ejection rates are also similar.
However, case 5 with less toroidal diffusivity has a higher accretion rate.

Anisotropy of magnetic diffusivity does also impact the rotation of the outflow.
It is believed that jets are rotating and the rotation is basically driven by the underlying disk rotation and 
the coupling between matter and field 
\citep{Bacciotti2002, Anderson2003,Coffey2004,Fendt2011}.
For a high toroidal diffusivity, the coupling is weak and, thus the acceleration in toroidal direction is 
weak as well.
With our simulations we can approve this concept.
Figure \ref{fig:anisotropy} shows the rotational velocity for the lower part of the jet. 
The outflow resulting from simulation case 1 with a nine times higher toroidal diffusivity shows a $50\%$ lower 
rotation rate than for case 5. 
Similarly, the outflows in simulation case 4 with a three times higher toroidal diffusivity shows a $50\%$ lower 
rotation rate than case 3\footnote{We cannot compare cases 1 and 4 or 5 and 4 directly, as, due to their different 
poloidal diffusivity these outflows have different mass fluxes}.

We observe a close correlation between the accretion rate $\dot{M}_{\rm acc}$ and angular momentum 
flux $(\dot{J}_{\rm kin}+\dot{J}_{\rm mag})$ from the disk.
In a system with higher angular momentum removal, a higher accretion rate is observed (Fig.~\ref{fig:ang-flux}), 
with the accretion rate in case 5 being larger than for case 1, and for case 1 larger than for case 3.
This confirms the common believe that in order to obtain higher accretion rates, a more efficient angular 
momentum removal is required.

To discuss the interrelation between collimation and diffusivity, we again
apply a collimation degree by comparing the directed mass fluxes (see Eq.~\ref{eq:angle}).
As a general result we find that outflows launched from disks of higher poloidal 
diffusivity $\eta_{\rm p}$ are more collimated.
This interrelation are displayed in Fig.~\ref{fig;eta_collimation}.

We see case 2 with the magnetic diffusivity of $\eta_{\rm p, i} = 0.01$ - maybe below a {\em critical} value.
evolves differently from the others and has both a very low degree of collimation
and an exceptionally large mass flux weighted jet radius.
The inner part of this outflow appears rather hollow (Fig.~\ref{fig:rho_parameterrun}).

The comparison of the (mass flux weighted) jet radii for these runs provide a similar picture.
We find an interrelation such that for increasing magnetic diffusivity the jet radius decreases
\footnote{For simulations with higher resolution (case 6 and  case 7), we adopt a smaller physical
grid size with $r_{\rm  out} = 50$, so we cannot compare the jet  radius properly.},
such that
$r_{\rm jet} {\rm (case\;3)} \leq r_{\rm jet} {\rm (case\;4)} < r_{\rm  jet}{\rm (case\;5)} < 
r_{\rm jet}{\rm (case\;1)} < r_{\rm jet}{\rm (case\;2)}$.

\begin{table}
\caption {Comparison between different diffusive scale  heights. Shown are some of the physical properties 
mentioned in Tab.~\ref{tbl:cases} 
for different diffusive scale height runs which have  been carried out up to t=2000.}
\begin{center}
 \begin{tabular}{cccccc}
\hline
\hline
\noalign{\smallskip}
                                  & case 1  & case 11 & case 12    &  case 13\\
$\epsilon_\eta$                   & 0.1     &  0.2    & 0.3       & 0.4   \\
$\Delta  r$                       & 0.064    & 0.064  & 0.064     & 0.064\\
$\Delta z$                        & 0.066   & 0.066 & 0.066      &  0.066  \\
$\eta_{\rm  p,\rm i}$             & 0.03     & 0.03  & 0.03       & 0.03\\
$\eta_{\phi,\rm i} $              & 0.09    & 0.09  & 0.09       &  0.09\\
$\chi$                            & 3        & 3     & 3      & 3\\
$\beta$                           & 10       &10     &  10    & 10 \\
\noalign{\smallskip}
\hline
\noalign{\smallskip}  
$r_{\rm jet,z=180}$              & 43.8    & 73.3     & 56.4   & 48.4  \\
\noalign{\smallskip}
$r_{\rm jet,z=60}$               & 25.01   & 45.8     & 39.2   & 34.4  \\
\noalign{\smallskip}
$r_{\rm l}$                      & 3.8     & 7.2      & 7.2    & 6.2     \\
\noalign{\smallskip} 
$v_{\rm  jet,z=170}$              & 0.46      & 0.35    &  0.33   &  0.41    \\
$v_{\rm jet,z=280}$              & 0.53     & 0.46    &  0.41   &  0.47    \\
\noalign{\smallskip}
\hline
$\dot{M}_{\rm acc}$             & 0.015  & 0.024     & 0.022  &  0.018   \\
\noalign{\smallskip} 
$\dot{M}_{\rm  ejec}$            & 0.005  &  0.015     & 0.015  & 0.01    \\
\noalign{\smallskip} 
$\dot{J} _{\rm  kin}$             &0.01    &  0.005     & 0.009  & 0.01      \\   
\noalign{\smallskip} 
$\dot{J}_{\rm mag}$            &0.033 & $0.015\pm 0.01$     & $0.02\pm  0.003$   & 0.028   \\
\hline
 \end{tabular}
\end{center}
\label{tbl:scale}
\end{table}

%----------------------------------------------------------------------------------------------------------- 
\subsection{Scale height of magnetic diffusivity}
\begin{figure*}
\centering
\includegraphics[width=8.cm]{\figurepath/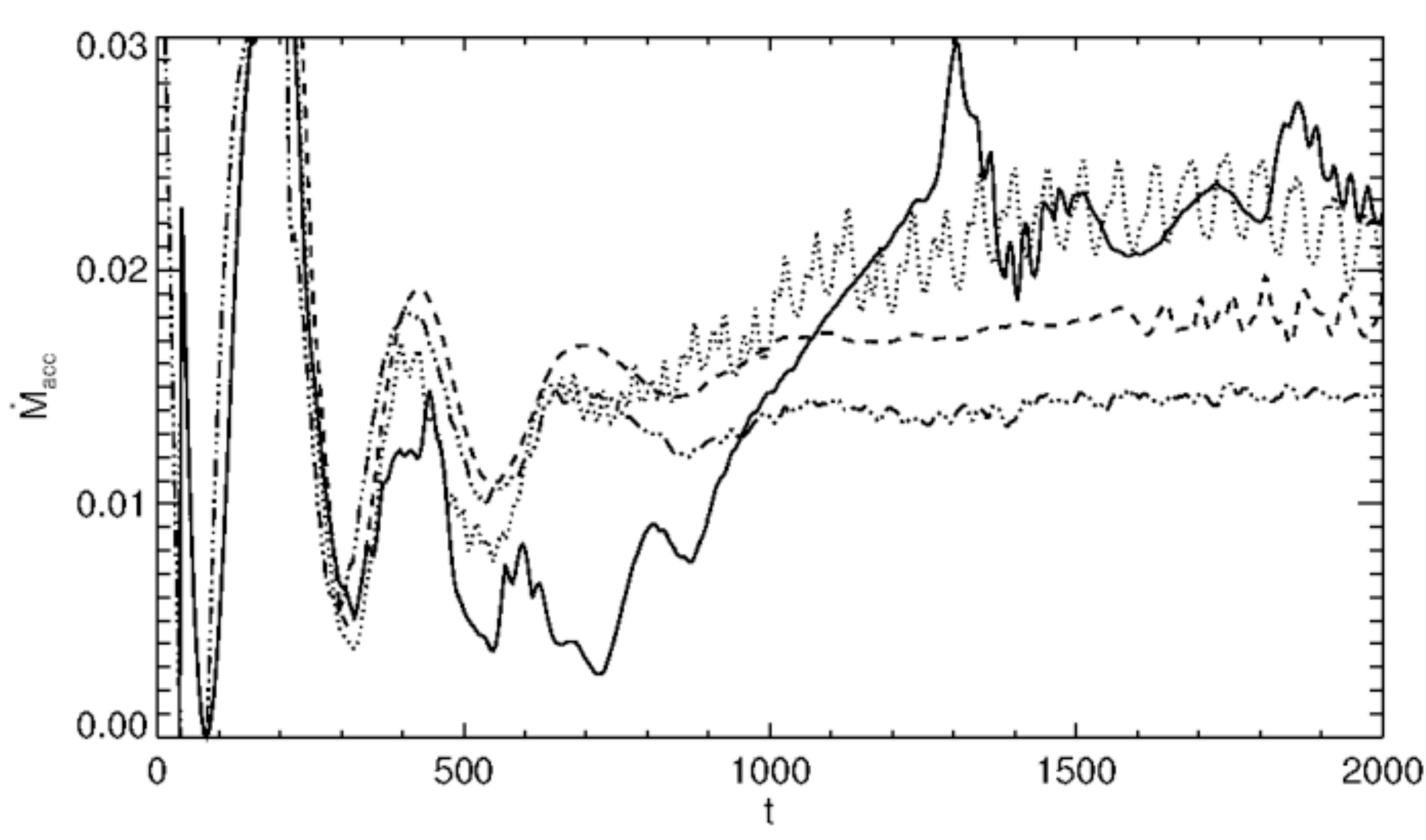}
\includegraphics[width=8.cm]{\figurepath/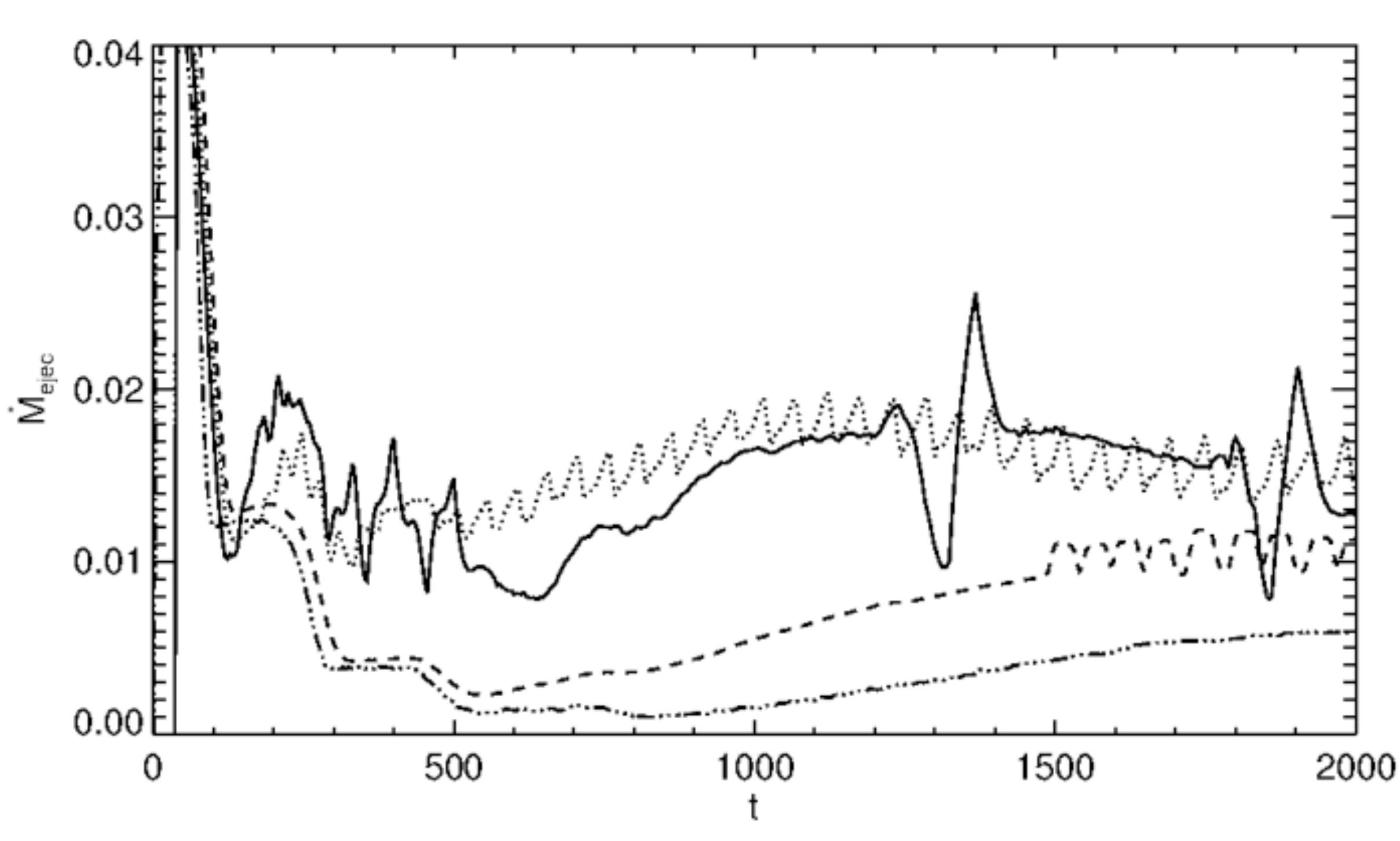}\\
\includegraphics[width=8.cm]{\figurepath/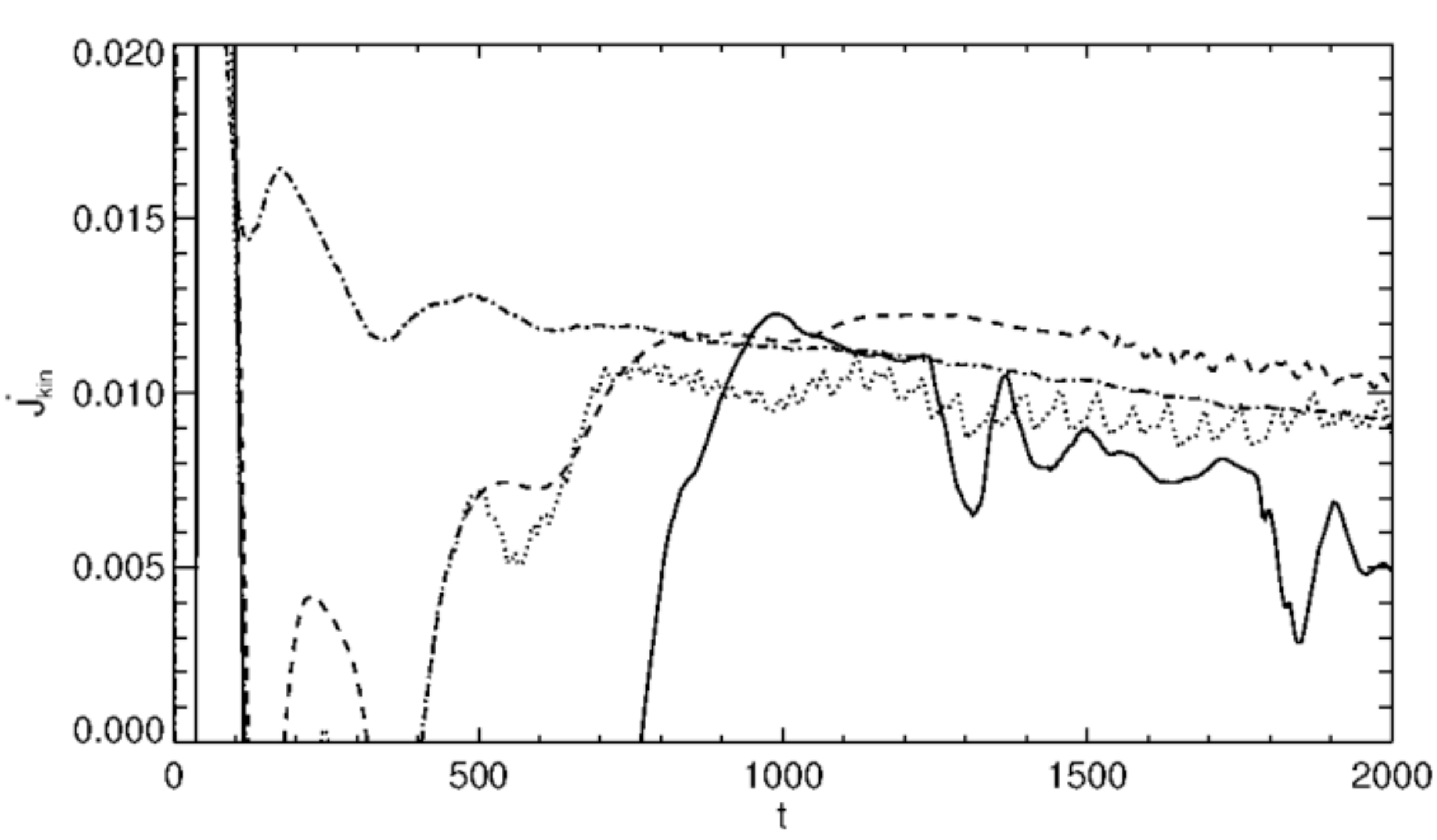}
\includegraphics[width=8.cm]{\figurepath/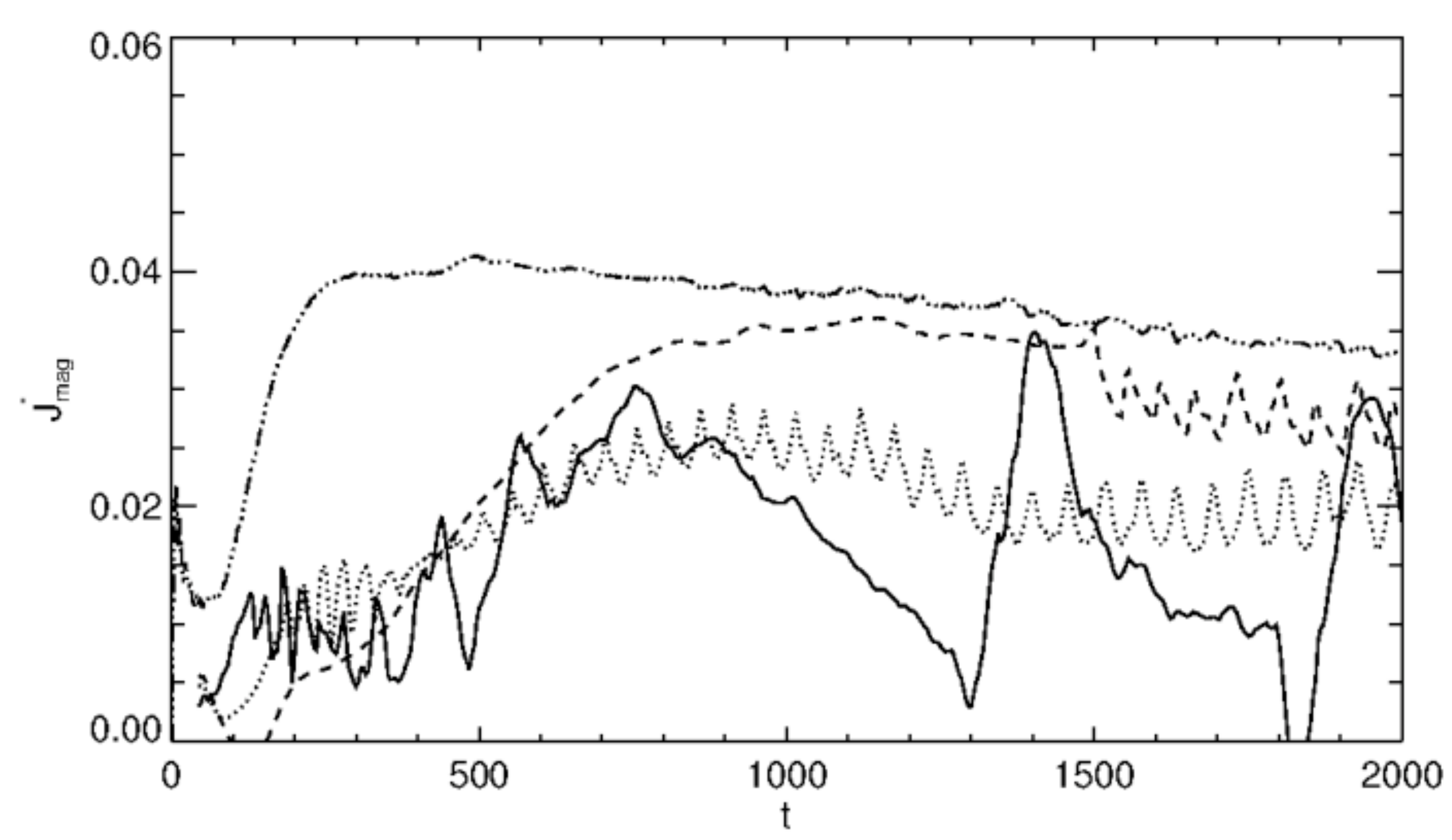}
\caption{Effect of the magnetic diffusivity scale height. 
Shown is the evolution of the accretion and ejection rate $\dot M_{\rm kin}, \; \dot M_{\rm mag}$ (top),
and the evolution of the kinetic and magnetic angular momentum flux $\dot J_{\rm kin}, \; \dot J_{\rm mag}$ (bottom),
for simulations applying a different diffusive scale height 
$H_{\eta} = 0.1$ (solid line), 0.2 (dotted line), 0.3 (dashed line), 0.4 (dot-dashed line).}
\label{fig:hstudy_scale}
\end{figure*}

We have argued above that the disk material which is launched to form an outflow is likely to be turbulent, 
so we may expect a magnetically diffusive disk wind above the disk surface.
Thus, since mass loading requires poloidal magnetic diffusivity, the jet launching
area can be extended into higher altitudes above the disk surface.
Furthermore, the jet launching area is extended into a domain where the plasma-$\beta$ is comparatively
low.
Taking into account both effects, one may therefore expect 
i) to launch jets from disks which would themselves be insufficiently magnetized for jet-driving, and also
ii) to launch outflow which have a rather low mass flux and subsequently higher terminal velocity.
In order to investigate this effects, we have therefore prescribed different height scales $H_\eta$ 
for the magnetic diffusivity in our simulations (see Eq.\ref{eq:magdiff}).
The physical properties measured for different diffusive scale heights are shown in Tab.~\ref{tbl:scale}.
In general, we find different stages in the temporal evolution of the mass flux (accretion and ejection) 
in the disk-jet system (Fig.~\ref{fig:hstudy_scale}).
For small diffusive scale heights $H_\eta$, a turbulent pattern and perturbations appear in the outflow
resulting in a more filamentary outflow.
It seems that these perturbation have larger amplitudes in case of a smaller diffusive scale height.
The amplitudes decrease when the diffusive scale height decreases. 
For $\epsilon_\eta=0.4$ they are damped completely.
We believe that these perturbations are generated by a physical process within the launching region
and that they are simply smoothed out by the magnetic diffusivity.
A further study is needed to clarify this issue.

For increasing scale height the accretion rate and ejection rate decrease, confirming the correlation 
derived above between the increasing magnitude of diffusivity and decreasing mass fluxes.
In other words, an increasing diffusivity scale height simply corresponds to a higher diffusivity with 
the consequences as discussed above.
Interestingly, all our simulations for different diffusivity scale heights converge to the same kinetic 
angular momentum flux in the outflow.
On the other hand, we find a trend of decreasing magnetic angular momentum flux with decreasing 
diffusivity scale height.

%===============================================================================================

\section{Comparison with previous numerical studies}
In this section we highlight some of our results discussing them in the context
of numerical studies published previously.
Our main goal was to perform a consistent parameter study to investigate a variety of physical 
effects involved in jet launching by using {\em one} code with {\em one} numerical setup.
Only this allows us to disentangle the leading effects involved in jet launching without the 
uncertainty introduced by interpreting results obtained with different codes 
(VAC, PLUTO, FLASH) or setups.

% 1.) 
We first stress the point that we apply a well defined, fixed-in-time {\em magnetic diffusivity}
  distribution. The magnetic diffusivity is an leading parameter which influences
  the whole disk-jet evolution \citep{Zanni2007}.
  Previous simulations \citep{Casse2002, Zanni2007, Tzeferacos2009,
  Murphy2010} have applied a diffusivity evolving in time. This might be considered 
  as more advanced, however, the previous authors do not provide details about time-evolution 
  of the diffusivity, thus intruducing some level of uncertainty. 
  Here, we have demonstrated how diffusion and advection of magnetic flux does change 
  the disk Alfv\'en speed (and thus the magnetic diffusivity if a time-dependent 
  $\eta(t) \sim v_{\rm A}(t)$ is applied) by a factor of ten.
  On the other hand, by applying different strength and scale height for the diffusivity
  we investigate the interrelation between magnetic diffusivity and the accretion and 
  ejection mass fluxes.

% 2.) 
  Similarly, we provide information about the spatio-temporal {\em evolution of the magnetization}
  (or plasma-$\beta$ in our case).
  \citet{Tzeferacos2009} have published a magnetization study, however, on a 
  smaller grid and for substantially shorter dynamical evolution. 
  In our paper we show in particular how the magnetization profile flares over 1000s of dynamical 
  time scales (Fig.~\ref{fig:log_beta_case1}). 
  Thus, along with the evolving outflow, the magnetization above the disk surface is substantially
  lower compared to the initial condition.
  
%3.)
We obtain {\em jet velocities} which seem to be smaller compared to previously published 
  values.
  For example, \citet{Zanni2007} detect asymptotic velocities of 1.5-1.8 times 
  the footpoint Keplerian speed for the field lines crossing the upper boundary.
  Simulations of jet formation from a fixed-in-time disk surface boundary
  (see e.g.~\citet{Ouyed1997, Fendt2002} ) give similar values. 
  In contrast, we report lower velocities of about 0.8 times the Keplerian speed at the
  inner disk radius. Note however, that our values result from a differently defined 
  measure of the jet velocity.
  We suggest that this mass-flux weigthed asymptotic speed - the average velocity 
  of the bulk mass flux - is more applicable to observations.
  The {\em maximum jet velocities} we observe in our simulations are higher and also up 
  to 1.8 time the Keplerian speed at the inner disk radius, which is consistent
  with the previous studies.
  We explicitely consider the {\em jet rotation} in our simulations a topic which has not been
  treated in previous launching simulations.

%4.) 
  We quantify the {\em jet collimation} degree by the directed mass fluxes.
  For the jet opening angle we find somewhat larger values than in previous studies. 
  We suppose that this results from the modified outflow condition we have applied 
  as we have seen that the original outflow condition lead to smaller opening angles.

% 5.)
We have run {\em long-lasting simulations} of up to 5000 dynamical time-scales
  $t_{\rm i} = r_{\rm i}/v_{\rm K,i}$.
  This allows us to i) reach a quasi steady-state situation of the simulation, but
  also to investigate the evolution beyond. 
  We find a slight but persistent change in the inflow-outflow dynamics on this very
  long time scales, such that the accretion rate slightly decreases.
  The main reason for this we see in the decrease of the disk mass for such long
  time scales. 
  However, the effect has an direct astrophysical application as it may be applied to 
  the long-term evolution of Classical TTS when the disk accretion in fact weakens
  after some $10^6$ years. 
  We note that previous simulations were substantially shorter, lasting until
  400 dynamical time steps $t_{\rm i} = r_{\rm i}/v_{\rm K,i}$ 
  \citep{Zanni2007, Tzeferacos2009} or 30 inner disk rotations
  \citep{Casse2004,Meliani2006}. 
  An exception is the simulation by \citet{Murphy2010} lasting for 
  more than 900 inner disk orbits.
  That paper focusses, however, on the launching physics and does not show the 
  large-scale evolution of the outflow.

%6.) 
  Together with the long time evolution, we have also have applied a {\em large grid size}.
  Compared to \citet{Tzeferacos2009} who investigate how magnetization affects launching
  by using a $40\times 120\,r_{\rm i}$ grid, our grid has about double size.
  \citet{Murphy2010} concentrate on the launching process of weakly magnetized
  disk and display only results on $40\times 40\,r_{\rm i}$ images (which are a subsets
  of a $280\times 840\,r_{\rm i}$).
  For YSO, our grid extension corresponds to about 28\,AU along the jet, well into
  the observable region. 
  We note that we have applied an equidistant grid for the magnetically diffusive disk 
  area which is consistent with the PLUTO code requirements. This is different to
  \citet{Murphy2010} who attach a scaled grid for the diffusive outer disk.

%-------------------------------------------------------------------------------------------------------
\section{Conclusions}
We have presented results of MHD simulations investigating the launching of jets and outflows 
from a magnetically diffusive disk in Keplerian rotation. 
The time evolution of the accretion disk structure is self-consistently taken into account.
The simulations are performed in axisymmetry applying the MHD code PLUTO.
The main goal of our simulations was to study how magnetic diffusivity (its magnitude
and distribution) and magnetization affect the disk and outflow properties,
such as mass and angular momentum fluxes, jet collimation, or jet radius.

Our grids extend to $(96 \times 288)$ inner disk radii with a resolution of $(0.064 \times 0.066)$,
respectively $(50 \times 180)$ inner disk radii with a higher resolution of $(0.025 \times 0.025)$.
An internal boundary (sink) is placed close to the origin absorbing the accreted mass and angular 
momentum.

We have prescribed a magnetic diffusivity in the disk based on an $\alpha$-prescription.
One of our parameters was the scale height of the magnetic diffusivity with the option to have 
it higher than the thermal scale height.
This can be motivated considering that it is the turbulent disk material which is loaded into 
the outflow, and that the turbulence pattern is swept along with the disk wind until it 
decays.
We have investigated disks carrying a magnetic flux corresponding to an initial plasma-beta 
ranging from 10 to 5000 at the inner disk radius.

As a general result we observe a continuous and robust outflow launched from the inner part of 
the disk, expanding into a collimated jet and is accelerated to super fast magneto-sonic speed. 
The key results of our simulations can be summarized as follows.

(1)
Concerning the {\em acceleration of the outflows},
our simulations confirm that the magneto-centrifugal acceleration mechanism 
is most efficient in the low plasma-$\beta$ regime,
while for weak magnetic fields the toroidal magnetic pressure gradient 
drives the ejected material.

We also confirm that the magnetocentrifugal mechanism also depends on the 
mass load in the outflow, as this mechanism works more efficiently for
outflows with low mass fluxes.
However, compared to the magnetic pressure driven outflows, jets in the 
magnetocentrifugal acceleration regime have usually higher mass fluxes.

In our simulations with very high plasma-$\beta$ we detect a highly
unsteady behaviour

(2) 
Efficient magneto-centrifugal driving which can accelerate jets to high 
kinetic energy relies on a strong coupling between magnetic field and the 
rotating disk, thus on a low diffusivity.
We find that it is the poloidal diffusivity $\eta_{\rm p}$ which mainly affects 
the driving of the outflow.
However, besides the coupling needed for acceleration, also the {\em launching of
material} depends on diffusivity. 
With increasing $\eta_{\rm p}$, the mass fluxes (both the accretion rate and 
the ejection rate) decrease.
Subsequently, the higher ejection rates result in a lower asymptotic outflow velocities.

(3)
We measure typical {\em outflow velocities} are in the range of $0.3 - 0.8$ times the inner disk
rotational velocity with the tendency that the mass fluxes obtained in magneto-centrifugally
driven outflow are substantially higher.
Here, we confirm the clear (inverse) correlation between jet velocity and mass load, as it
is well known from the literature.
Note that we have apply a mass-flux-weighted jet velocity which we find more applicable
to the observations. 
For the bulk mass flux we find lower velocities compared to other papers, which are
mostly dealing with the maximum speed obtained in the simulation. Our maximum velocities
are similar.
The relatively high speed of the outflows with low Poynting flux which are driven by
poloidal pressure gradient is unclear.

We find that the toroidal diffusivity affects the {\em outflow rotation} - a small toroidal 
diffusivity implies a larger jet rotational velocity. This has not shown before in simulations.

(4)  
We do not find a clear correlation between the {\em outflow collimation} and the magnetic
field strength. 
Also weakly magnetized outflows, which are driven by the magnetic pressure gradient, 
and which we find to be quite unsteady, show a high degree of collimation.
The question of collimation for the weakly magnetized outflows is not answered.

We find that outflows within the magneto-centrifugal-driving regime the flows ejected from
a weakly diffusive disks are only weakly collimated. 
Similarly, their jet radius (here defined as mass flux weighted radius) is larger in 
case of a lower poloidal magnetic diffusivity.

Follwoing the magnetic flux surface along the bulk mass flux from the asymtotic regime
to the launching area, we can defined the {\em launching area} of the outflow.
We find a size of the launching area from which the bulk of the mass flux originates
in the range between 3 and 8 inner disk radii or about 0.4 AU.

(5) 
Depending on the strength of magnetic diffusivity, the disk-jet structure may evolve into a steady-state.
We found that the cases with the strong field with $\beta_{\rm i} \sim 10$ and poloidal diffusivity
$\eta_{\rm p, i} \geq 0.03$ will reach a quasi steady state, 
confirming the literature.

(6)
The magnetic flux profile along the disk is subject to advection and diffusion. 
We find that the magnetization (or plasma-$\beta$) of disk and outflow may therefore substantially 
change during the time evolution. 
We have observed that the initial disk magnetization may change by a factor of 100.
This may have severe impact on the launching process and the formation of the outflow in the sense
that a rather highly magnetized disk may evolve into a weakly magnetized disk which cannot
drive strong outflows.
This issue has not been discussed before in the literature.

(7) 
For very long time scales the accretion disk changes its internal dynamics, as due to outflow ejection
and disk accretion the disk mass decreases. 
As a consequence, the accretion and ejection rates slightly decrease.
In order to compensate for this effect, we have applied a large outer disk radius providing a large mass 
reservoir for the inner jet-launching disk.

(8) 
For our simulations, we found that 10-50\% of the accreting plasma can be diverted into the outflow. 
For 
i) less diffusive disks, ii) a strong magnetic field, iii) a low poloidal diffusivity, 
or a iv) lower numerical diffusivity (resolution),
the mass loading into the outflow is increased - resulting in more massive jets.
We interpret as physical reason the more efficient extraction of angular momentum from the disk, 
due to the stronger matter-field coupling. 
Note that we do not consider in our simulations viscosity or the wealth of thermal effects 
which play an essential role for launching \citep{Casse2000}.

(9)
We found that jets launched in a setup with smaller diffusive scale height are more perturbed.
The same effect is seen in outflows launched from disks with weaker diffusivity.

(10)
We finally remark that it is essential to do long-term simulations covering thousands of rotational
periods in order to find a steady state situation of the accretion-ejection dynamics.
Our simulations run for 5000 dynamical time steps, corresponding to about 900 revolutions at the 
inner disk radius as adopted in our reference run.

In summary, we confirm the hypothesis that efficient magneto-centrifugal jet driving requires a strong 
magnetic flux (i.e. a low plasma beta), together with a large enough magnetic torque in order to produce
a powerful jet. 
In addition, the magnitude of (turbulent) magnetic diffusivity plays the major role in the ejection efficiency,
while the anisotropy in the diffusivity mainly affects the jet rotation.
Both results imply that the structure of the asymptotic jet is indeed governed by the properties 
of the accretion disk, here parametrized by the magnetization and magnetic diffusivity.
The mass ejection-to-accretion ratio along with the momentum and energy transfer rates from inflow to outflow
are essential properties for any feedback mechanism in star formation or galaxy formation scenarios and 
could only be derived from simulations resolving the inner region of the jet-launching accretion disk.

%///////////////////////////////////////////////////////////////////////////////////////////////////////////
%
\acknowledgements
We thank the Andrea Mignone and the PLUTO team for the possibility to use their code.
We acknowledge helpful and constructive criticism by two referees, which has improved the
presentation of the paper.
S.S. acknowledges the warm hospitality by the Max Planck Institute for Astronomy.
All simulations were performed on the THEO cluster of Max Planck Institute for Astronomy.
This work was financed partly by a scholarship of the Ministry of Science, Research, 
and Technology of Iran, and by the SFB 881 of the German science foundation DFG.

\appendix

\section{Units and normalization}
\label{app:units-normalization}
We normalize all variables, namely $r, \rho, v, B$ to their fiducial values at the inner disk radius.
We have adopted the following typical number values for 
a YSO of $M = 1\,M_{\odot}$ and an AGN of $M = 10^8\,M_{\odot}$.
A change in one of the system parameters accordingly changes the scaling of our simulation results
which can thus be applied for a variety of jet sources.
The inner disk radius $r_{\rm  i}$ is usually assumed to be a few radii of the central object,
\begin{eqnarray}
 r_{\rm i} & = & 0.028\,{\rm AU} \left( \frac{r_{\rm i}}{3 R_{\rm YSO}} \right)
                             \left( \frac{R_{\rm YSO}}{2 R_{\odot}} \right) \quad\quad {\rm (\rm YSO)} \nonumber \\
           & = & 10^{-4}\,{\rm pc} 
                             \left( \frac{r_{\rm i}}{10 R_{\rm S}} \right)
                             \left( \frac{M}{10^8 M_{\odot}} \right) \quad \quad (\rm AGN),
\end{eqnarray}
where $R_{\rm S} = 2GM/c^2$ is the Schwarzschild radius of the central black hole. The inner disk
radius is usually assumed to be located at the marginally stable orbit at $3\,R_{\rm S}$.
Since we apply the non-relativistic version of the PLUTO code, we cannot treat any relativistic 
effects. 
We therefore apply a scaling of $r_{\rm i} \simeq 10 R_{\rm S}$.
For simplicity and convienient comparison with the previous literature we apply $r_{\rm i} = 0.1\,$AU
for most comparisons concerning stellar sources.
The orbital velocity at the inner disk radius is
\begin{eqnarray}
 v_{\rm k,i} & = & 180\,{\rm km\,s^{-1}}\,
                 \left( \frac{M}{M_\odot} \right)^{1/2}
                 \left( \frac{r_{\rm i}}{3 R_{\rm YSO}} \right)^{-1/2} 
                 \left( \frac{R_{\rm YSO}}{2 R_{\odot}}  \right)^{-1/2}      \quad \quad {\rm (YSO)} \nonumber \\
             & = & 6.7\times 10^{4}\,{\rm km\,s^{-1}}
                 \left( \frac{r_{\rm i}}{10 R_{s}} \right)^{-1/2}
                 \left( \frac{M}{10^8 M_{\odot}}   \right)^{-1/2}    \quad \quad (\rm AGN).
\end{eqnarray}
For a $r_{\rm i} = 0.1\,$AU distance from a YSO the orbital speed is $v_{\rm k,i} = 94\,\rm km\,s^{-1}$.
The mass accretion rate is a parameter which is in principle accessible by observation.
Subject to the disk model applied, the observed disk luminosity can be related to an 
accretion rate. 
For a YSO, the accretion rate is typically of the order of $\dot{M}_{\rm acc} \simeq 10^{-7} {\rm M_{\odot} yr^{-1}}$,
providing a normalisation of the density $\rho_{\rm i}$ with 
$\dot M_{\rm i} = r_{\rm i}^2 \rho_{\rm i} v_{\rm K,i}$
Applying a length scale $r_{\rm i} = 0.1\,$AU and a velocity scale  $v_{\rm k,i} = 94\,\rm km\,s^{-1}$, 
we obtain
\begin{eqnarray}
 \dot{M}_{\rm i} & = & 10^{-5} {\rm M_{\odot} yr^{-1}} 
                      \left( \frac{\rho_{\rm i}}{10^{-10} \rm g\,cm^{-3}} \right)
                      \left( \frac{M}{M_{\odot}} \right)^{1/2} 
                      \left( \frac{r_{\rm i}}{0.1\,\rm AU} \right)^{3/2}   \quad\quad {\rm (YSO)} \nonumber \\
            & = & 10\,{\rm M_{\odot} yr^{-1}}
                       \left(\frac{\rho_0}{10^{-12} \rm g cm^{-3}}\right)
                        (\frac{M}{10^8 M_{\odot}})^{1/2}           \quad\quad (AGN)
 \end{eqnarray}
Normalized value of the magnetic field is obtained by considering the plasma-$\beta$ and the field strength at
the equator at the inner radius $B_{\rm i}= \sqrt{8 \pi P_{\rm i}/\beta_{\rm i} } $,

\begin{eqnarray}
 B_{\rm i} & = & 14.9 \left(\frac{\beta_{\rm i}}{10}\right)^{-1/2}
                  \left(\frac{\epsilon}{0.1}\right) \left(\frac{\rho_0}{10^{-10} \rm g cm^{-3}}\right)^{1/2}
                  \left(\frac{M}{M_\odot}\right)^{1/2} \left(\frac{r_{\rm i}}{\rm 0.1 AU}\right)^{-5/4}
                  \quad\quad {\rm G \quad (YSO) }\\
               & = & 1.06\times10^3 \left(\frac{\beta_{\rm i}}{10}\right)^{-1/2}\left(\frac{\epsilon}{0.1}\right) 
                  \left(\frac{\rho_0}{10^{-12} \rm g cm^{-3}}\right)^{1/2}
                  \left(\frac{r_{\rm i}}{\rm 10 R_{\rm s}}\right)^{-5/4} 
                  \quad\quad {\rm G \quad (AGN) }
\end{eqnarray}

\section{A modified radial outflow boundary condition}
An essential point of our setup is to impose an outflow boundary at the outer part of the domain 
in $r$-direction avoiding artificial collimation.
We have implemented a current-free outflow boundary condition which prevents spurious collimation
by Lorentz forces and which has been thoroughly tested in our previous papers
\citep{Porth2010,Porth2011, Vaidya2011}.
The boundary condition considers a vanishing toroidal electric current density in the ghost cells.
Therefore, the Lorentz force component perpendicular to the grid boundary vanishes.
The boundary constraints are applied on the poloidal and toroidal component of the electric current density,
$j_r = -\partial_z B_\phi $, $ j_z = r^{-1}\partial_r r B_\phi $, and
$j_\phi = \partial_z B_r-\partial_r B_z$, respectively.

In principle, this is considered for the grid ghost cells $(i_{\rm end}, j)$, 
adjunct to the domain boundary at $(i_{\rm end} + 1, j)$ (see the \S 3.1.1 in \citep{Porth2010}).
The transverse $ B_{\rm t}(i_{\rm end}, j + 1/2)$ and 
normal $B_{\rm n}(i_{\rm end} + 1/2, j ), B_{\rm n} (i_{\rm end} + 1/2, j + 1)$ 
magnetic field components in the domain,
together with the transverse field component $B_t (i_{\rm end} +1, j +1/2)$ of the first ghost zone, 
constitute a toroidal corner-centered electric current $I_\phi (i_{\rm end} + 1/2, j + 1/2)$.
With that a constraint for a current-free  $I_\phi = 0$ boundary condition can been implemented
numerically,
\begin{eqnarray}
 B_{\rm t}|_{i_{\rm end}+1,j+1/2} =  B_{\rm t}|_{i_{\rm end},j+1/2}
 + \frac{\Delta r}{\Delta z} \left[ B_{\rm n}|_{i_{\rm end}+1/2,j+1} - B_{\rm n}|_{i_{\rm end}+1/2,j}\right]. 
\label{app:outer-boundary}
\end{eqnarray}

%\bibliographystyle{apj}
%\bibliography{ms}
 % \input{mspdf.bbl}
\bibliographystyle{apj}

\end{document}